\documentclass[preprint, notrackchanges]{aastex631} % linenumbers
\usepackage{multirow, url, color}
\usepackage{bbding, amssymb} % symbols
\usepackage{ulem} % underline for emphasis
\usepackage{verbatim} % display LaTeX commands
\usepackage{enumerate}

\usepackage{amsmath} % \usepackage{txfonts}
\usepackage{booktabs} % Table Top/Mid/Bottom Rule
\usepackage{graphicx, subfigure}
\usepackage{amssymb}
\usepackage[fleqn]{mathtools}
\usepackage{mathrsfs}
\usepackage[makeroom]{cancel}
\def\gsim{\;\lower4pt\hbox{${\buildrel\displaystyle >\over\sim}$}\;}
\def\lsim{\;\lower4pt\hbox{${\buildrel\displaystyle <\over\sim}$}\;}
\def\grls{\;\lower4pt\hbox{${\buildrel\displaystyle >\over <}$}\;}

\newcommand{\dblind}[1]{{\color[rgb]{0,0,0} will be revealed after the double-blind review process.}}

\newcommand{\di}{\mathrm{d}}
\newcommand{\bs}[1]{\boldsymbol{#1}}
\makeatletter
\newcommand*{\rom}[1]{\expandafter\@slowromancap\romannumeral #1@}
\makeatother

\submitjournal{\apj}

\shorttitle{Simulation of the 2013 April 11 SEP Event}
\shortauthors{Liu et al.}
\begin{document}

\title{Physics-Based Simulation of the 2013 April 11 Solar Energetic Particle Event}

\author[0000-0002-2873-5688]{Weihao Liu} % Primary Investigator
\affiliation{Department of Climate and Space Sciences and Engineering, University of Michigan, Ann Arbor, MI 48109, USA}
\author[0000-0002-6118-0469]{Igor V. Sokolov} % Poisson Bracket Scheme & MFP
\affiliation{Department of Climate and Space Sciences and Engineering, University of Michigan, Ann Arbor, MI 48109, USA}
\author[0000-0003-3936-5288]{Lulu Zhao} % Shock & Simulation & Visualization
\affiliation{Department of Climate and Space Sciences and Engineering, University of Michigan, Ann Arbor, MI 48109, USA}
\author[0000-0001-9360-4951]{Tamas I. Gombosi} % Conceptualization & Comments
\affiliation{Department of Climate and Space Sciences and Engineering, University of Michigan, Ann Arbor, MI 48109, USA}
\author[0000-0001-9114-6133]{Nishtha Sachdeva} % Shock & Solar Wind & CME Study
\affiliation{Department of Climate and Space Sciences and Engineering, University of Michigan, Ann Arbor, MI 48109, USA}
\author[0000-0003-2865-1772]{Xiaohang Chen} % Shock & Visualization
\affiliation{Department of Climate and Space Sciences and Engineering, University of Michigan, Ann Arbor, MI 48109, USA}
\author[0000-0001-8459-2100]{G\'abor T\'oth} % Shock Surface & Comments
\affiliation{Department of Climate and Space Sciences and Engineering, University of Michigan, Ann Arbor, MI 48109, USA}
\author[0000-0002-3176-8704]{David Lario} % SEP Measurements & Comments
\affiliation{Heliophysics Science Division, NASA Goddard Space Flight Center, Greenbelt, MD 20771, USA}
\author[0000-0003-0472-9408]{Ward B. Manchester IV} % CME & Comments
\affiliation{Department of Climate and Space Sciences and Engineering, University of Michigan, Ann Arbor, MI 48109, USA}
\author[0000-0002-3787-1622]{Kathryn Whitman} % SEP Measurements: Particle Fluxes & Comments
\affiliation{Space Radiation Analysis Group, NASA Johnson Space Center, Houston, TX 77058, USA}
\affiliation{KBR, Houston, TX 77002, USA}
\author[0000-0002-0978-8127]{Christina M. S. Cohen} % SEP Spectrum and Fluence Interpretations & Comments
\affiliation{California Institute of Technology, Pasadena, CA 91125, USA}
\author[0000-0001-5191-1662]{Alessandro Bruno} % SEP Measurements: Particle Fluences & Comments
\affiliation{Heliophysics Science Division, NASA Goddard Space Flight Center, Greenbelt, MD 20771, USA}
\affiliation{Department of Physics, Catholic University, Washington, DC 20064, USA}
\author[0000-0001-9177-8405]{M. Leila Mays} % Shock-Capturing Tool & CCMC
\affiliation{Heliophysics Science Division, NASA Goddard Space Flight Center, Greenbelt, MD 20771, USA}
\author[0000-0003-2595-3185]{Hazel M. Bain} % Shock-Capturing Tool & SWPC
\affiliation{Cooperative Institute for Research In Environmental Sciences, University of Colorado Boulder, Boulder, CO 80309, USA}
\affiliation{NOAA Space Weather Prediction Center, Boulder, CO 80305, USA}

\correspondingauthor{Weihao Liu}
\email{whliu@umich.edu}

\begin{abstract}
Solar energetic particles (SEPs) can pose hazardous radiation risks to both humans and spacecraft electronics in space. Numerical modeling based on first principles offers valuable insights into the underlying physics of SEPs and provides synthetic observables for SEPs at any time and location in the inner heliosphere. 
In this work, we present a numerical scheme, which conserves the number of particles based on integral relations for Poisson brackets \citep{sokolov2023high}, to solve the kinetic equation for particle acceleration and transport processes. We implement this scheme within the Space Weather Modeling Framework, developed at the University of Michigan. In addition, we develop a new shock-capturing tool to study the coronal mass ejection-driven shock originating from the low solar corona. These methodological advancements are applied to conduct a comprehensive study of a historical SEP event on April 11, 2013. Multi-spacecraft observations, including \textit{SOHO}, \textit{SDO}, \textit{GOES} and \textit{ACE} near Earth, and \textit{STEREO-A/B}, are used for model--data comparison and validation. 
We show synthetic observables, including extreme ultraviolet and white-light images, proton time--intensity profiles, and energy spectra, and discuss their differences and probable explanations compared to observations. Our simulation results demonstrate the application of the Poisson bracket scheme with a particle solver to simulating a historical SEP event. We also show the capability of extracting the complex shock surface using our shock-capturing tool and understand how the complex shock surface affects the particle acceleration process. 
\end{abstract}
\keywords{Solar energetic particles (1491), Solar coronal mass ejection shocks (1997), Heliosphere (711), Space weather (2037), Computational methods (1965)}
% \keywords{Solar Energetic Particles ... See \url{https://astrothesaurus.org/concept-select/} when needed.}

\section{Introduction} \label{sec:Intro}
% SEP threats, sources, predictions, physics-based modeling, targets of this work
% Para 1. SEP events
Solar energetic particles (SEPs) consist of protons, heavier ions and electrons originating in association with solar eruptions. They are observed in energies ranging from suprathermal (a few keV per nucleon) to relativistic (a few GeVs per nucleon) energies \citep{reames1999particle, reames2021solar, klein2017acc}. Generally, SEP events can be classified into impulsive or gradual ones \citep{cane2006role, reames2013two}. Impulsive SEP events are believed to be associated with magnetic reconnection processes within solar flares and coronal jets. Their time--intensity profiles usually show a sudden onset followed by a fast decay with a duration typically less than 1 day \citep[e.g.,][]{nitta2006solar, mason2007he3, buvcik2020he3, lario2024rapid}. 
On the other hand, gradual SEP events are usually associated with shocks driven by coronal mass ejections (CMEs) and typically last for a few days \citep{kahler1978prompt, kahler1984associations, desai2016large}. 

CMEs can drive shock waves that have been identified in coronagraph images \citep{sime1987coronal, vourlidas2003direct} and are often observed \textit{in-situ} at 1 astronomical unit (au) and sometimes at larger heliocentric distances up to several au \citep{chen2011coronal, webb2012coronal, manchester2017physical}. As a shock wave propagates across the solar corona (SC) and through the interplanetary (IP) medium, it may continue to accelerate particles from the ambient solar wind plasma or remnants from previous events \citep[e.g.,][]{gopalswamy2002interacting, rouillard2011interpreting, luhmann2020icme}. 
The resulting energetic particles can then propagate through the SC and IP space, reaching Earth's location and posing hazardous radiation risks to both humans and spacecraft in space \citep[e.g.,][]{miroshnichenko2018retrospective, guo2021radiation, buzulukova2022space, cliver2022extreme}. Therefore, a better understanding of the acceleration and transport of SEPs and the capability to predict SEPs become critical to the human endeavor for deep space exploration. 

% Para 3. DSA mechanism for SEPs, power law, double power law)
Diffusive shock acceleration (DSA), also known as first-order Fermi acceleration \citep{fermi1949origin}, is believed to be the mechanism at shock fronts that produces energetic particles in many heliophysics and astrophysical systems \citep[e.g.,][]{axford1977acceleration, krymskii1977regular, bell1978acceleration1, bell1978acceleration2, blandford1978particle, blandford1987particle, jokipii1982particle, jokipii1987rate, armstrong1985shock, zank2000particle, petrosian2012stochastic}. Particles can be accelerated as they travel across a shock front with strong plasma compressions \citep[see Chapter 13.4.2 of][and the references therein]{gombosi1998physics}. This acceleration process can naturally lead to a universal power-law momentum distribution $f(p) \propto p^{-\gamma}$, where $f$ is the omnidirectional distribution function and $p$ denotes the magnitude of the particle momentum. The power-law index $\gamma$ depends only on the shock compression ratio, i.e., the ratio of the plasma downstream density to the upstream value \citep{drury1983introduction, jones1991plasma, melrose1993diffusive, sokolov2006diffusive, giacalone2008energy}. 
However, in SEP energy spectra, there is usually an exponential rollover \citep{ellison1985shock} or a double power-law feature \citep{band1993batse} with the rollover/break energy depending on the ion charge-to-mass ratio \citep[e.g.][]{cohen2005heavy, mewaldt2005proton, tylka2005shock, li2009shock, yu2022double}. Possible explanations suggested by, e.g., \cite{li2015scatter, zhao2016double, zhao2017effects} and \cite{kong2019acceleration} are the finite lifetimes and sizes of the shock for particle acceleration, as well as the particle transport processes. % particle diffusion, adiabatic cooling and magnetic focusing effects accompanying the transport processes, and the diffusion processes. 

% Para 4. Models: Role and Groups + State-of-the-art Physics-based Models
In order to investigate the underlying acceleration and transport mechanisms of SEPs, numerous models have been developed to predict SEP properties. These include empirical, machine-learning and physics-based approaches, as reviewed by \cite{whitman2023review}. Empirical and machine-learning SEP models are built upon the observational data and can offer quick predictions of SEP events. On the other hand, first-principles physics-based models consider the mechanisms that regulate the observed SEP properties and use different kinds of sophisticated computational techniques \citep[e.g.,][]{decker1988computer, ng1994focused, ng2003modeling, sokolov2004new, kota2005simulation, aran2006solpenco, luhmann2007heliospheric, zhang2009propagation, droge2010anisotropic, strauss2015aspects, hu2017modeling, zhang2017precipitation, borovikov2018toward, linker2019coupled, wijsen2019modelling, zhang2023data, palmerio2024improved, zhao2024solar}. These models leverage our current understanding of particle seed population, acceleration and transport in the SC and IP space and allow us to analyze the processes responsible for the properties associated with SEP events. Due to the dimensionality and stiffness of SEP simulations, these models are usually computationally expensive to obtain meaningful results and need much attention and effort in model validation and evaluation \citep{bain2023noaa, zheng2024overview}. Moreover, there are still challenges and open questions for the complete accurate modeling of SEP events as reported by \cite{anastasiadis2019solar}, such as the underlying physical mechanisms for particle acceleration \citep[e.g.,][]{giacalone2005efficient, giacalone2005particle, lee2012shock, verkhoglyadova2015theoretical, tsurutani2024review}, properties of the seed particle population injected into the acceleration process \citep[e.g.,][]{li2012twin, ding2015seed, zhuang2021successive, wijsen2023seed}, and the interaction of energetic particles with the turbulent magnetic field in the heliosphere \citep[e.g.,][]{giacalone2000small, zank2014particle, engelbrecht2019pitch, shalchi2020perpendicular}. In spite of high demands of computational resources and challenges of developing techniques to deliver meaningful results, the physics-based models remain attractive in the community, since these models are able to derive the shock properties and provide synthetic observables such as the time--intensity profiles and energy spectra of SEPs at any time and location of interest in the SC and inner heliosphere (IH). These synthetic observables can offer a unique insight to analyze the SEP events and interpret the underlying physics, advancing our knowledge of particle acceleration and transport processes. 

% Para 5. Paper structure
Our previous study in \cite{zhao2024solar} has demonstrated the capability of the SOlar wind with FIeld lines and Energetic particles (SOFIE) model as applied to predict historical SEP events. In this work, we advance the SOFIE model by introducing a newly developed shock-capturing tool and implementing a particle-number-conserving numerical scheme to simulate the acceleration and transport processes of SEPs. These methodological advancements have been applied to simulate a historical SEP event on April 11, 2013. 
The structure of the paper is as follows. In Section \ref{sec:Method}, we describe our numerical models in detail, including the magnetohydrodynamic (MHD) code to simulate the solar wind plasma, the CME flux rope initialization tool and the new SEP model setup. In Section \ref{sec:Event}, we provide an overview of the 2013 April 11 SEP event investigated in this work. In Section \ref{sec:Result}, we show our simulation results and the model--data comparisons for this event. We also analyze the synthetic observables and provide plausible explanations for their differences compared to observations. Conclusions are summarized in Section \ref{sec:Summary}.

\section{Methodology} \label{sec:Method}

In order to simulate SEPs with a physics-based model, we need to have modules simulating the background solar wind, CME generation and propagation, and the particle acceleration and transport processes. In this study, we employ the Space Weather Modeling Framework (SWMF\footnote{\url{https://github.com/SWMFsoftware}}) developed at the University of Michigan, which provides a high-performance computational capability to simulate the space weather environment from the upper solar chromosphere to the upper atmosphere of Earth and/or the outer heliosphere \citep{toth2005space, toth2012adaptive, gombosi2021sustained}. The SWMF has integrated various components that represent different physical domains of the space environment, each offering several models available. Our focus here is on the SC and IH components for three-dimensional (3D) global solar wind simulations, the Eruptive Event generator (EE) for the CME study, and the particle acceleration and transport model for SEPs. % Mark: Solve PDE /= Particle Tracker; simulating or designed to simulate 

\subsection{Background Solar Wind} \label{sec:Method1}
% 1. AWSoM-R model
The 3D global solar wind plasma is modeled by the Alfv\'en Wave Solar‐atmosphere Model(‐Realtime) \citep[AWSoM(-R),][]{van2010data, van2014alfven, sokolov2013magnetohydrodynamic, sokolov2021threaded, sokolov2022stream, oran2013global, gombosi2018extended}. 
The AWSoM-R is an Alfv\'en wave turbulence‐driven and self‐consistent solar atmosphere model and has been validated by comparing simulations and observations of both the \textit{in‐situ} macroscopic properties of the solar wind and the line‐of‐sight (LOS) appearance of the corona observed at different wavelengths \citep[e.g.,][]{jian2015validation, meng2015alfven, sachdeva2019validation, sachdeva2021simulating, sachdeva2023solar, van2019predictions, van2022improving, shi2022awsom, wraback2024simulating}. In AWSoM-R, the Block-Adaptive-Tree-Solarwind-Roe-Upwind-Scheme (BATS-R-US) code plays a critical role in solving the MHD equations that describe the plasma dynamics \citep{powell1999solution}. The steady-state solar wind solution is obtained with the local time stepping and the second-order shock-capturing scheme \citep{toth2012adaptive, gombosi2021sustained}. The inner boundary condition for the magnetic field is specified by solar magnetograms. In this study, we use the hourly updated synoptic magnetograms collected by the Global Oscillation Network Group of the National Solar Observatory \citep[NSO/GONG\footnote{\url{https://gong.nso.edu/data/magmap/} \label{url:gong}},][]{harvey1996global, hill2018global}. % Mark: boundary condition for/on xxx 

\begin{figure*}[tp!]
\centering {\includegraphics[width=0.97\hsize]{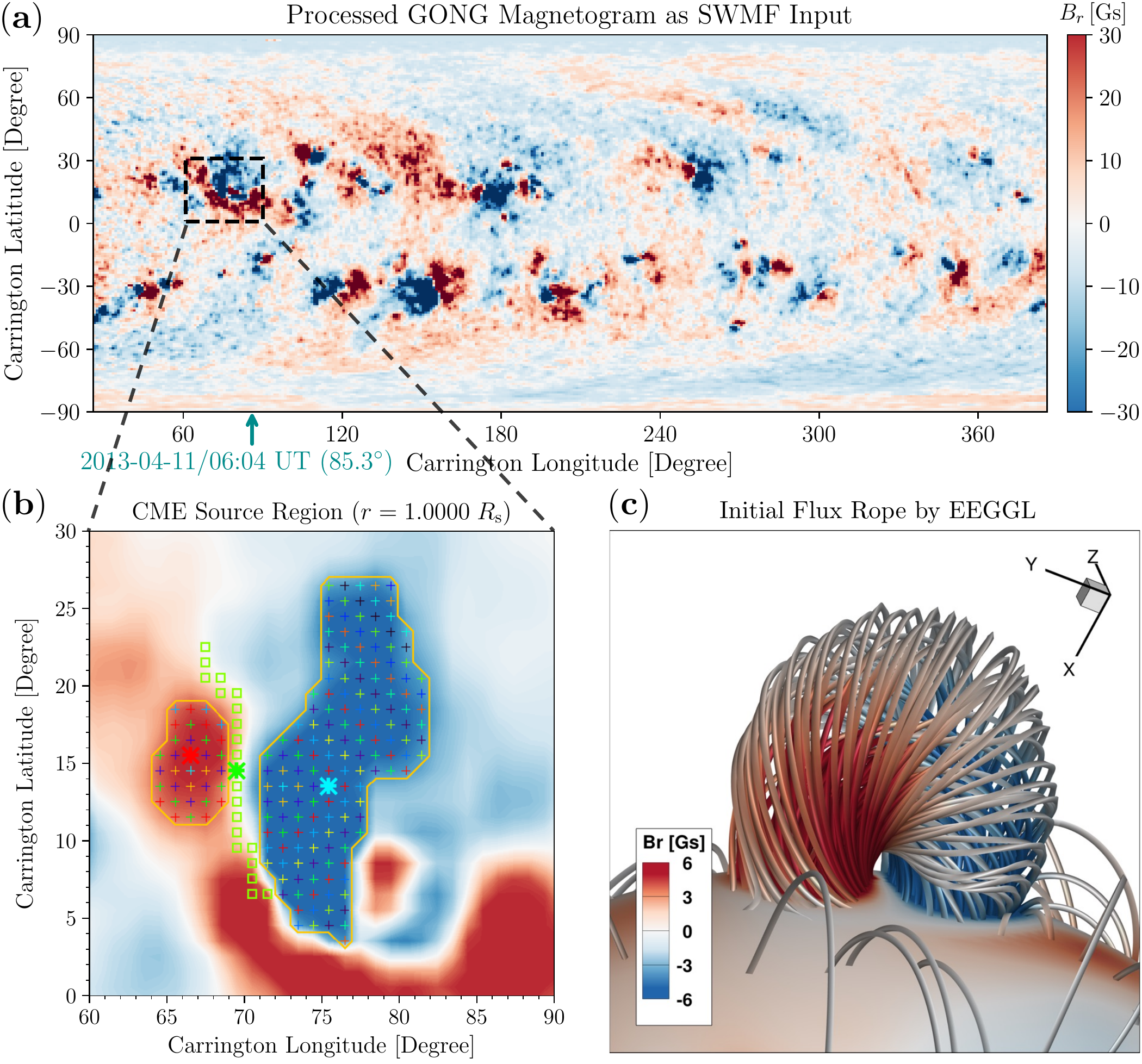}}
\caption{(a): Processed GONG magnetogram as of 06:04 UT on 2013 April 11, with the green arrow representing the Carrington longitude of Earth at that time. The magnetogram region $60^{\circ}$ eastward of the green arrow remains unchanged from the previous Carrington rotation. Weak magnetic fields in the original GONG magnetogram have been enhanced as described by Eq.~(\ref{eqn:BrMod}). The black dashed box shows the area of the active region where the CME flux rope originated. (b): Zoomed-in active region field at the inner boundary, $r=1.0000\;R_\mathrm{s}$, with the same color bar as used in panel (a). The red and blue asterisks indicate the chosen locations with the positive and negative magnetic polarity, respectively, and the green asterisk indicates the center of the configuration. A series of green squares denote the polarity inversion line. (c): The 3D topology of the flux rope initialized by EEGGL, superposed with the radial magnetic field of the AR adjusted to simulate the CME event on 2013 April 11.} \label{fig1:input} % The 60 degrees to the left of these arrows are regions which have not yet crossed the central meridian. 
\end{figure*}

Owing to the limitations of the observation geometry, there are significant uncertainties in the radial magnetic field measurements of the polar regions \citep[e.g.,][]{petrie2015solar, reiss2023progress}. In order to reduce this uncertainty and achieve better agreement of the global simulation results with observations, it is customary to modify the photospheric radial magnetic field in the polar regions \citep[e.g.,][]{nikolic2019solutions, sokolov2023titov, huang2024solar}. Specifically, the GONG-observed radial magnetic field, $B_r^\mathrm{GONG}$, used as the boundary condition at the heliocentric distance of 1 solar radii ($r=1\;R_\mathrm{s}$), is intensified in the weak-field regions using the following expression \citep{huang2024solar}: 
\begin{equation}
    \left.B_r\right|_{r=1\;R_\mathrm{s}} = \mathrm{sign}\left(B_r^\mathrm{GONG}\right) \times {\min}\left(3.75\left| B_r^\mathrm{GONG} \right|, \left| B_r^\mathrm{GONG} \right| + 5 \mathrm{\; Gs} \right). \label{eqn:BrMod}
\end{equation}
Figure \ref{fig1:input}(a) shows the processed GONG magnetogram recorded at 06:04 UT on 2013 April 11, used as the input for AWSoM-R in the SWMF. To obtain a 3D distribution of the strapping field configuration, the Potential Field Source Surface \citep[PFSS,][]{altschuler1969magnetic, schatten1969model} model with the source surface at $r=2.5\;R_\mathrm{s}$ is applied to express the intensified field as a series of spherical harmonics to the order of 180 in this study \citep{toth2011obtaining}. 

% 2. AWSoM-R parameters
In AWSoM-R, the coronal plasma is heated by the dissipation of two discrete turbulence populations that propagate parallel and antiparallel to the magnetic field \citep{sokolov2013magnetohydrodynamic, sokolov2021threaded, van2014alfven}. Using physically consistent treatments of wave reflection, dissipation, and heat partitioning between electrons and protons, AWSoM-R has been shown to simulate the SC plasma comparable to observations with three free parameters: the Poynting flux parameter for the energy input ($\left(S_\mathrm{A}/B\right)_{\odot}$), the correlation length for Alfv\'en wave dissipation ($L_\perp\sqrt{B}$) and the stochastic heating exponent and amplitude ($h_\mathrm{S}, A_\mathrm{S}$). Other parameters for the model setup are described in, e.g., \cite{sokolov2013magnetohydrodynamic, van2014alfven, van2022improving} and \cite{sachdeva2019validation}. The default settings for the free parameters are: $\left(S_\mathrm{A}/B\right)_{\odot} = 1.0\; \mathrm{MW\;m^{-2}\;T^{-1}}$, $L_\perp\sqrt{B} = 1.5\times10^5\; \mathrm{m\;T^{1/2}}$ and $(h_\mathrm{S},\, A_\mathrm{S}) = (0.21,\, 0.18)$, based on the studies of \cite{hollweg1986transition, fisk2001behavior, de2007chromospheric, chandran2011incorporating, sokolov2013magnetohydrodynamic, van2014alfven} and \cite{hoppock2018stochastic}. 
Recently, \cite{huang2024solar} performed uncertainty quantification of these free parameters and found that the parameters have a strong solar cycle dependence. In this simulation, the optimal Poynting flux parameter is $\left(S_\mathrm{A}/B\right)_{\odot} = 0.3\; \mathrm{MW\;m^{-2}\;T^{-1}}$ and the other two free parameters are set to default. 

% 3. Stream-aligned AWSoM-R
A validated background solar wind solution is essential for modeling the transport processes of energetic particles, as it provides the magnetic field configuration where particles propagate, thereby enabling the computation of the energetic particle properties observed by spacecraft at specific heliospheric locations \citep[e.g.,][]{hinterreiter2019assessing, jin2022assessing, zhao2024solar}. Nevertheless, current numerical solutions of the ideal or resistive MHD equations have struggled to reproduce the aligned interplanetary streamlines and magnetic field lines in corotating frames \citep[see the discussions in][]{kleimann2022structure, kennis2024magnetic}. 
One of the reasons for this discrepancy is the numerical reconnection across the heliospheric current sheet (HCS): the reconnected field is directed across the HCS, while the global solar wind streams along the HCS, thus resulting in ``V‐shaped" magnetic field lines and significant misalignment between the magnetic field lines and plasma streamlines \citep[e.g.,][]{brchnelova2022or, sokolov2022stream}. It is not feasible to follow the trajectory of particles in such ``V‐shaped" magnetic field lines, and thus streamlines are usually used instead \citep[e.g.,][]{young2021energetic}. Recently, \cite{sokolov2022stream} have introduced the stream-aligned MHD method that ``nudges" the magnetic field lines and plasma streamlines to restore their alignments. This stream-aligned AWSoM-R model has recently been validated in steady-state solar wind simulations \citep[e.g.,][]{wraback2024simulating, zhao2024solar}. In this study, we utilize the stream-aligned AWSoM-R model to obtain a steady-state solar wind plasma for CME and SEP propagation. % or ``through which CMEs and SEPs propagate" 

In the simulation, we use a block-adaptive 3D spherical grid in SC and a block-adaptive Cartesian cubic grid in IH, with an overlapping buffer grid that couples the solutions from SC over to IH. 
The computational domain in SC consists of grid blocks of $6 \times 8 \times 8$ cells (\textit{control volumes}). In heliocentric distance, $r$, the grid extends from 1.1 to $24\;R_\mathrm{s}$. Radial stretching is achieved using $\ln r$ as a coordinate instead of $r$. The smallest radial cell size is around $0.01\;R_\mathrm{s}$ near the Sun, to resolve the steep density and temperature gradients in the lower SC. The largest radial cell size in SC is approximately $0.4\;R_\mathrm{s}$. Inside $r = 1.7\;R_\mathrm{s}$, the angular resolution is $\sim1.4^{\circ}$; outside this radial range, the grid is coarsened by one level to $\sim2.8^{\circ}$. % Mark: Upper chrmosphere/transinsition region are solved by the thread model.
The computational domain in IH surrounds the spherical domain of SC and is composed of $8 \times 8 \times 8$ grid blocks, extending from 20 up to $500\;R_\mathrm{s}$. The cell size ranging from $\sim0.3\;R_\mathrm{s}$ near the inner boundary to $\sim20\;R_\mathrm{s}$ near the outer boundary. 
For both SC and IH, the adaptive mesh refinement \citep[AMR, see][and references therein]{berger1989local, gombosi2003adaptive, gombosi2004solution, van2011crash, toth2012adaptive} technique is performed to resolve the HCS. The grid resolution is increased by a factor of 2 along the path of the CME to resolve the CME structures. The total number of cells is on the order of 5 million in SC and 100 million in IH. 

\subsection{Eruptive Event Generator} \label{sec:Method2}

Within the steady-state solar wind domain, the CME flux rope is then modeled by the EE module in the SWMF, which has been extensively used and validated to model the CME initialization and propagation \citep[e.g.,][]{manchester2004eruption, manchester2004three, manchester2004modeling, manchester2006modeling, manchester2008three, manchester2014simulation, manchester2014flux, lugaz2005evolution, lugaz2005numerical, lugaz2007numerical, lugaz2013interaction, kataoka2009three, van2009breakout, jin2013numerical, shiota2016magnetohydrodynamic, jin2016numerical, jin2017chromosphere, jin2017data, alvarado2018suppression, chen2025decent}. 
Currently, there are a few different flux rope models embedded in the EE module, such as the breakout model \citep{antiochos1999model}, the flux-emergence model \citep[e.g.][]{manchester2004eruption} and analytical flux rope models including the Gibson-Low flux rope \citep[GL,][]{gibson1998time}, the Titov-D\'emoulin flux rope \citep[TD,][]{titov1999basic, roussev2003three, titov2014method, titov2022magnetogram, sokolov2023titov} and the STatistical InjecTion of Condensed Helicity (STITCH) initialization mechanism \citep{antiochos2013helicity, dahlin2022stitch}. The last two models can be used independently, or the STITCH model can be used to trigger the TD eruption. 

In this study, we employ the spheromak-type magnetic field configuration anchored to the inner boundary, adopting the Gibson-Low model for the initial condition to erupt and propagate as a CME.  %initiate (EEGGL does not initiate)
For this flux rope type, the defining parameters are specified by the Eruptive Event Generator Gibson-Low (EEGGL\footnote{Available on \url{https://github.com/SWMFsoftware} and \url{https://ccmc.gsfc.nasa.gov/analysis/EEGGL/}.}) model \citep[e.g.,][]{gibson1998time, borovikov2017eruptive, jin2017chromosphere, jin2017data}. 
The processed-GONG magnetogram shown in Figure \ref{fig1:input}(a), the active region (AR) location and the observed CME speed are used to calculate the GL flux rope parameters. Specifically, the \textit{STEREO} CME Analysis Tool \citep[StereoCAT\footnote{\url{https://ccmc.gsfc.nasa.gov/tools/StereoCat/} \label{url:STEREOCAT}},][]{millward2013operational, mays2015ensemble} can be utilized to calculate the CME speed with elimination of the projection effect by determining the CME's 3D trajectory through multi-spacecraft observations. These 3D kinematic properties of CMEs are included in the Space Weather Database Of Notifications, Knowledge, Information (DONKI\footnote{\url{https://kauai.ccmc.gsfc.nasa.gov/DONKI/search/} \label{url:DONKI}}) database. We adopt $675\;\mathrm{km\; s^{-1}}$ reported by DONKI as the accurate CME speed to calculate the flux rope parameters. Figure \ref{fig1:input}(b) shows the zoomed-in AR field with the chosen locations for the filament footpoints, as well as the polarity inversion line (PIL) identified by EEGGL. 
Note that the input magnetogram is smoothed by a $5\times5$ pixel window in EEGGL to reduce the complexity of the photospheric magnetic field configuration. 
Based on empirical features of pre-event conditions \citep[e.g.,][]{borovikov2017eruptive, jin2017data}, EEGGL can offer an efficient parameter setup, including the flux rope location, orientation, size and magnetic field strength, as detailed in Table \ref{tab:param}. With the force-imbalanced flux rope parameterized by EEGGL and inserted on top of the AR (see its 3D topology in Figure \ref{fig1:input}(c)), the CME propagation in SC and IH is then modeled with time-accurate simulations. 

\subsection{Particle Solver} \label{sec:Method3}

\subsubsection{Governing Equation} \label{sec:Method3.0}

As the SEP population likely forms out of the suprathermal tail of the solar wind, whose distribution is far from the Maxwellian \citep[e.g.,][]{pierrard2010kappa, kahler2019suprathermal, lario2019evolution}, we characterize SEPs by a canonical distribution function $F(\bs{r},\bs{p},t)$ of coordinates ($\bs{r}$), momentum ($\bs{p}$) and time ($t$), such that the number of particles, $\di N$, within the elementary volume, $\di^3\bs{r}$, is given by the following normalization integral: $\di N = \di^3\bs{r}\int \di^3 \bs{p}\; F(\bs{r},\bs{p},t)$. 
In a magnetized moving plasma, it is convenient to consider the distribution function at any given point, $\bs{r}$, in a frame of reference moving with the local plasma bulk velocity, $\bs{u}(\bs{r}, t)$. Also, we adopt the spherical coordinates ($p=|\bs{p}|, \mu = \bs{b}\cdot\bs{p}/p, \varphi$) in the momentum space with its polar axis aligned with the direction, $\bs{b}=\bs{B}/B$, of the magnetic field, $\bs{B}(\bs{r}, t)$. Herewith, $p$ indicates the magnitude of the particle momentum, $B$ denotes the magnetic field amplitude ($\left|\bs{B}(\bs{r}, t)\right|$), $\mu$ is the cosine value of the pitch angle and $\varphi$ is the phase angle of the particle Larmor gyration. The normalization integral in these new variables gives: 
\begin{equation}
    \di N = \di^3 \bs{r} \int_0^{+\infty} p^2\di p \int_{-1}^1 \di \mu \int_0^{2\pi} \di\varphi \; F(\bs{r},p,\mu,\varphi,t). \label{eqn:dN1}
\end{equation}

Using the canonical distribution function, one can define a gyrotropic distribution function, $\mathcal{F}(\bs{r}, p, \mu, t) = \frac{1}{2\pi} \int_0^{2\pi} \di\varphi\; F(\bs{r},p,\mu,\varphi,t)$, to describe the particle motion averaged over the phase of gyration around the magnetic field. The omnidirectional distribution function, $f(\bs{r}, p, t) = \frac{1}{2} \int_{-1}^1\di\mu\; \mathcal{F}(\bs{r}, p, \mu, t)$, is additionally averaged over the pitch angle. The normalization integrals in Eq.~(\ref{eqn:dN1}) becomes:
\begin{align}
\begin{aligned}
    \di N & = 2\pi \di^3 \bs{r} \int_0^{+\infty} p^2\di p \int_{-1}^1 \di \mu \; \mathcal{F}(\bs{r}, p, \mu, t) \\
    & = 4\pi \di^3 \bs{r} \int_0^{+\infty} p^2\di p \; f(\bs{r}, p, t). 
\end{aligned} \label{eqn:dN2}
\end{align}

% Mark: diffusion is the kind of particle motion and ``parallel diffusion'' is the above mentioned particle motion along the magnetic field 
The acceleration and transport of energetic particles in IP space is described by the focused transport equation \citep[e.g.,][]{northrop1963adiabatic, roelof1969propagation, skilling1971cosmic, isenberg1997hemispherical, kota1997energy, kota2000diffusion, kota2004cosmic, van2020primer}, which accounts for the effects of particle displacement along the magnetic field, drift in the inhomogeneous magnetic field, adiabatic heating or cooling, and adiabatic focusing and particle scattering by the magnetic turbulence, forming the kinetic equation for the gyrotropic distribution function $\mathcal{F}(\bs{r}, p, \mu, t)$:
\begin{align}
    \begin{aligned}
    & \frac{\partial \mathcal{F}}{\partial t} 
    + \underbrace{ \ \mu v \frac{\partial \mathcal{F}}{\partial s} \ }_{\text{Particle Streaming}} 
    + \underbrace{ \ \left(\bs{u} \cdot \nabla\right) \mathcal{F} \ }_{\text{Drift}} 
    + \underbrace{ \frac{\partial \mathcal{F}}{\partial p} \frac{\di p}{\di t} }_\text{Adiabatic Heating/Cooling} 
    + \underbrace{ \frac{\partial \mathcal{F}}{\partial \mu} \frac{\di \mu}{\di t} }_\text{Magnetic Focusing} \\ 
    & = \ \underbrace{ \frac{\partial}{\partial \mu} \left(D_{\mu\mu} \frac{\partial \mathcal{F}}{\partial \mu}\right) }_\text{Scattering} 
    + \underbrace{Q}_\text{Additional Source/Sink}, 
    \end{aligned} \label{eqn:FocusedEqn}
\end{align}
in which $s$ is the distance along the magnetic field, $D_{\mu\mu}$ is the pitch-angle diffusion coefficient, and $Q$ denotes the additional acceleration source or sink terms. In the diffusive limit, where the distribution function is assumed to be isotropic, the focused transport equation reduces to the Parker transport equation \citep{parker1965passage}:
\begin{align}
    \begin{aligned}
    \frac{\partial f}{\partial t}  
    + \underbrace{ \ \left(\bs{u} \cdot \nabla\right) f \ }_{\text{Drift}} 
    - \underbrace{\frac{1}{3}\left(\nabla \cdot \bs{u}\right) \frac{\partial f}{\partial \ln p}}_\text{Adiabatic Heating/Cooling} 
    = \underbrace{\nabla \cdot\left(\overset{\leftrightarrow}{\kappa} \cdot \nabla f\right)}_\text{Diffusion} 
    + \underbrace{Q}_\text{Additional Source/Sink}, 
    \end{aligned} \label{eqn:ParkerEqn}
\end{align}
where $\overset{\leftrightarrow}{\kappa}=D_{\parallel}\bs{bb}$ is the tensor of parallel diffusion along the magnetic field, $D_{\parallel}$ is the parallel spatial diffusion coefficient. The term proportional to the divergence of $\bs{u}$ accounts for the adiabatic cooling for $\nabla\cdot\bs{u}>0$, or the first-order Fermi acceleration in compression or shock wave fronts for $\nabla\cdot\bs{u}<0$ \citep{fermi1949origin}. As the transport equation in Eq.~(\ref{eqn:ParkerEqn}) captures the effects of interplanetary magnetic field (IMF) and IP plasma properties on the SEP acceleration and transport processes, we use Eq.~(\ref{eqn:ParkerEqn}) for the SEP numerical modeling in this study. 

\subsubsection{M-FLAMPA} \label{sec:Method3.1}

In the SWMF, the Multiple Field-Line-Advection Model for Particle Acceleration \citep[M-FLAMPA,][]{sokolov2004new, borovikov2018toward, borovikov2019toward} has been developed to simulate the particle acceleration and transport processes, where the particles are accelerated at the shocks driven by CMEs through the first-order Fermi acceleration mechanism \citep{fermi1949origin}. With no loss in generality, M-FLAMPA reduces a 3D problem of particle propagation in the IMF to a multitude of simpler 1D problems of the particle transport along a single line of the IMF. As the simulation begins, AWSoM-R and M-FLAMPA run simultaneously. At each time step, the time-evolving magnetic field lines, as well as the plasma properties, are extracted from the AWSoM-R solutions, along which the particle distribution function is solved \citep{borovikov2015efficient, borovikov2018toward}. Moreover, as proposed by \cite{sokolov2004new}, novel mathematical approaches are applied to the extracted magnetic field lines to sharpen the shock wave front, thus enhancing the efficiency of the DSA process. 

In M-FLAMPA, the particles are assumed to couple with the magnetic field lines. The particle motions consist of the displacement of the particle's guiding center along the IMF line, and the joint advection of both the guiding center and the IMF line together with the plasma where the magnetic field is frozen. Mathematically, this method employs the Lagrangian coordinate, $\bs{x}_L$, which stays with the advecting fluid elements in space \citep{landau1987fluid}. Herewith, the partial time derivative at the constant Lagrangian coordinate, $\bs{x}_L$, and the time, $\tau$, is denoted as $\frac{\di}{\di t}$ or $\frac{\partial}{\partial \tau}$, while the notation $\frac{\partial}{\partial t}$ denotes the partial time derivative at the constant Eulerian coordinate, $\bs{x}$, with the relations: $\frac{\partial}{\partial \tau} = \frac{\di}{\di t}=\frac{\partial}{\partial t} + \bs{u}\cdot\nabla$. Certain terms in Eq.~(\ref{eqn:ParkerEqn}) can be expressed in terms of Lagrangian derivatives and the spatial derivative along magnetic field lines, combining with the plasma motion equations. Eq.~(\ref{eqn:ParkerEqn}) can be eventually rewritten as \citep{borovikov2018toward, borovikov2019toward, sokolov2023high}:
\begin{equation}
    \frac{\partial f}{\partial \tau} = \frac{\di f}{\di t} = - \dfrac{1}{3}\frac{\mathrm{D}\ln\rho}{\mathrm{D}t} \frac{\partial f}{\partial \ln p} + \nabla \cdot\left(\overset{\leftrightarrow}{\kappa} \cdot \nabla f_0\right) + Q, \label{eqn:ParkerEqnL}
\end{equation}
where $\rho$ denotes the mass density of the plasma.

In addition, the Strang splitting method \citep[e.g.,][]{strang1968construction, macnamara2016operator} is applied in M-FLAMPA to split the advection and diffusion terms, in order to solve Eq.~(\ref{eqn:ParkerEqnL}) efficiently. Here, we implement the high-resolution Poisson bracket scheme for advections \citep{sokolov2023high} and use the theoretical derivations for the diffusion coefficient based on the quasi-linear theory \citep[QLT,][]{jokipii1966cosmic} and the turbulent magnetic field \citep[see][and references therein]{li2003energetic, sokolov2004new, borovikov2019toward}. More details are described in Sections \ref{sec:Method3.2} and \ref{sec:Method3.3}.

\subsubsection{Implementation of the \texorpdfstring{\cite{sokolov2023high}}{Sokolov et al. (2023)} Poisson Bracket Scheme} \label{sec:Method3.2}

% Introduction: Numerical scheme in SEP acceleration and transport processes
In order to simulate the fluxes of shock-accelerated SEPs, we solve the kinetic equation throughout the whole computational domain, including the shock wave region, with the DSA mechanism in the heart of our SEP model. In this case, it is important to use a particle-number-conserving scheme. Otherwise, the prediction for SEP fluxes may be contaminated by fake particle productions or disappearances due to approximation errors at high spatial gradients near the shock wave front.

In classical mechanics \citep{landau1959course}, the Poisson bracket for the distribution function, $F\left(\bs{r}, \bs{p}, t\right)$, is introduced as:
\begin{equation}
    \left\{F; H \right\} \equiv \sum\limits_{\ell} \left\{F; H \right\}_{q_\ell,\, p_\ell} = \sum\limits_{\ell} \left(\frac{\partial F}{\partial q_\ell} \frac{\partial H}{\partial p_\ell} - \frac{\partial H}{\partial q_\ell} \frac{\partial F}{\partial p_\ell}\right). \label{eqn:defPB}
\end{equation}
Here, $H$ represents the Hamiltonian function, $p_\ell$ and $q_\ell$ are the canonical coordinates for momentum and position, respectively, and $\ell$ denotes the $\ell^\mathrm{th}$ degree of freedom. Along the Hamiltonian trajectory, where $\frac{\di q_\ell}{\di t} = \frac{\partial H}{\partial p_\ell}, \ \frac{\di p_\ell}{\di t} = -\frac{\partial H}{\partial q_\ell},\ \forall \ell$, the time evolution of the distribution function is governed by the Poisson bracket with the Hamiltonian function:
\begin{equation}
\frac{\di F}{\di t} = \frac{\partial F}{\partial t} + \left\{ F; H \right\} = 0, \label{eqn:Liouville}
\end{equation}
This fundamental conservation law, known as the \textit{Liouville theorem} \citep{liouville1838note}, states that the distribution function remains constant along the Hamiltonian trajectory. 

Based on the integral relations for Poisson brackets, \cite{sokolov2023high} has developed a computationally efficient scheme for solving kinetic equations using the finite volume method. This newly developed Poisson bracket scheme conserves the number of particles, and possesses the total-variation-diminishing \citep[TVD, e.g.,][]{sokolov2006tvd, krivodonova2021tvd, toth2023total} property with second order of accuracy in space, thus ensuring high-resolution numerical results. With the Poisson bracket scheme, Eq.~(\ref{eqn:ParkerEqn}) can be reformulated as:
\begin{equation}
    \frac{B}{\delta s}\left\{f_{j,k}; \frac{\delta s}{B}\frac{p^3}{3} \right\}_{\tau,\, p^3/3} = \frac{B}{\delta s} \frac{\partial}{\partial s_L} \left( \frac{D_{\parallel}}{B\delta s} \frac{\partial f_{j,k}}{\partial s_L} \right), \label{eqn:ParkerEqnPB}
\end{equation}
where $f_{j,k}\left(s_L,\frac{p^3}3,\tau\right)$ is the omnidirectional distribution function along the field line that has been initially traced through the grid point, $\bs{x}_{j,k}$, back to the inner boundary of the SC domain and outer boundary of the IH domain. These grid points are uniformly spaced in longitude and latitude with the index $j$ and $k$, respectively, on the spherical surface at $r=2.5\;R_\mathrm{s}$. Herewith, $\delta s=\di s/\di s_L$, where $\di s$ is the element of the length introduced above (cf. Eq.~(\ref{eqn:FocusedEqn})), $s$, along the magnetic field line, and $\di s_L$ is the mesh size in the Lagrangian coordinate, $s_L$. More details on the derivations of Eq.~(\ref{eqn:ParkerEqnPB}) can be found in Section 4 of \cite{sokolov2023high}. 

By the Strang splitting method \citep{strang1968construction, macnamara2016operator}, at each time step, we first solve the advection equation in the phase space:
\begin{equation}
    \left\{f_{j,k}; \frac{\delta s}{B}\frac{p^3}{3} \right\}_{\tau,\, p^3/3} = 0, \label{eqn:ParkerEqnPBLHS}
\end{equation}
where the Hamiltonian function is $H \equiv \frac{\delta s}{B}\frac{p^3}{3}$, with $\tau$ and $p^3/3$ being the two canonical coordinates as we solve the time-accurate transport equation for SEPs. 

Note that the Poisson bracket scheme is applicable to various types of kinetic equations that can be formulated in terms of Poisson brackets. It has been demonstrated in \cite{sokolov2019integral} that the focused transport equation (cf. Eq.~(\ref{eqn:FocusedEqn})) can be formulated into multiple Poisson brackets, showing the potential to study the pitch-angle dependence in testing cases using the Poisson bracket scheme. In this work, we solve the Parker transport equation (cf. Eq.~(\ref{eqn:ParkerEqn})) for the omnidirectional distribution function, notated as $f(\bs{r}, p, t)$, as a first implementation of the Poisson bracket scheme in the M-FLAMPA. More sophisticated numerical models that take into account the pitch-angle dependence for the distribution function will be investigated in the future. 

\subsubsection{Particle Diffusion} \label{sec:Method3.3}

The interaction between the energetic protons and turbulent magnetic fields is modeled by diffusion along time-evolving magnetic field lines. Following Eq.~(\ref{eqn:ParkerEqnPBLHS}), within each time step, the transport equation is subsequently solved for spatial diffusion along each field line in M-FLAMPA:
\begin{equation}
    \frac{\partial f_{j,k}}{\partial \tau} = \frac{B}{\delta s} \frac{\partial}{\partial s_L} \left( \frac{D_{\parallel}}{B\delta s} \frac{\partial f_{j,k}}{\partial s_L} \right), \label{eqn:ParkerEqnPBRHS}
\end{equation}
where the spatial diffusion coefficient along the magnetic field, $D_{\parallel}$, can be derived in the usual manner from the scattering integral with respect to the particle pitch angle, $D_{\mu\mu}$ \citep[e.g.,][]{jokipii1966cosmic, earl1974diffuse, lee1982coupled, lee1983coupled}:
\begin{align}
    D_{\parallel} & = \frac{v^2}{8}\int_{-1}^{1} \frac{\left(1-\mu^2\right)^2}{D_{\mu\mu}} \;\di\mu, \label{eqn:calcDxx} \\
    D_{\mu\mu} & = \frac{\pi\omega_{ci}k}{2B^2/\mu_0} \left(1-\mu^2\right) \sum\limits_{+,\,-}I_{\pm}(k), \label{eqn:calcDmumu}
\end{align}
in which $v=\left|\bs{v}\right|$ denotes the proton speed, $\mu_0$ is the vacuum permeability, and $\omega_{ci} = \frac{eB}{m}$ is the cyclotron frequency of protons, with $e$ being the proton charge. $I_{\pm}(k)$ denotes the spectral energy density of turbulent waves, which propagate parallel ($I_{+}$) and anti-parallel ($I_{-}$) to the magnetic field. The wave number taken at $k=\frac{\omega_{ci}}{v\left|\mu\right|}$ satisfies the resonance condition \citep{lee1982coupled, borovikov2019toward}. %Herewith, we consider only the high-energy particles with $\left|\bs{v} - V_\mathrm{g} \bs{b}\right| \gg V_\mathrm{A}$, where $V_\mathrm{g}$ is the group speed of the turbulence wave and $V_\mathrm{A}$ is the Alfv\'en speed. 
% We assume that the left- and right-polarized waves are balanced in the Alfv\'en wave turbulence, and that the total wave energy at $k$ equals $I_{+}(k)$ for waves propagating parallel to the magnetic field and $I_{-}(k)$ for those propagating anti-parallel \citep{borovikov2019toward}. 
Both turbulent wave spectra are assumed to follow Kolmogorov's power law with an index of $-5/3$ \citep{kolmogorov1941local, zakharov2012kolmogorov}. At the current stage, our model does not account for the contributions of the self-generated wave turbulence by energetic particles. More details on the assumptions, considerations and derivations of the model can be found in \cite{sokolov2009transport} and \cite{borovikov2019toward}. Finally, the parallel diffusion coefficient, $D_{\parallel}$, can be expressed in terms of the mean free path (MFP), $\lambda_{\parallel}$, and the proton speed \citep[e.g.,][]{sokolov2004new}:
\begin{equation}
    D_{\parallel} = \frac{1}{3} \lambda_{\parallel} v, \label{eqn:expressDxx}
\end{equation}
with different treatments of $\lambda_{\parallel}$ in the upstream and downstream regions of the shock in M-FLAMPA. In the upstream region of the shock, the MFP can be approximated as \citep{li2003energetic, li2005mixed, zank2007particle}: 
\begin{equation}
    \lambda_{\parallel} = \lambda_0 \,\left( \frac{r}{1 \mathrm{\; au}} \right)\, \left(\frac{pc}{1 \mathrm{\;GeV}}\right)^{\frac{1}{3}}, \label{eqn:lambdaxxUp}
\end{equation}
in which $\lambda_0$ is the mean free path for $1\;$GeV particles at $1\;$au and it is a free parameter in the model. 
With the relativistic relations between the proton speed, momentum and kinetic energy, the parallel diffusion coefficient, $D_{\parallel}$ in Eq.~(\ref{eqn:expressDxx}), can be expressed as:
\begin{equation}
    D_{\parallel} = \frac{1}{3} c\, \lambda_0 \,\left( \frac{r}{1 \mathrm{\; au}} \right)\, \left[ \frac{E_k\left(E_k+2E_{\mathrm{p}0}\right)}{\left(1 \mathrm{\;GeV}\right)^2} \right]^{\frac{1}{6}} \cdot \left[ \frac{E_k\left(E_k+2E_{\mathrm{p}0}\right)}{\left(E_k+E_{\mathrm{p}0}\right)^2} \right]^{\frac{1}{2}}, \label{eqn:DxxUp}
\end{equation}
where $c$ denotes the speed of light, $E_k$ denotes the kinetic energy of energetic protons, and $E_{\mathrm{p}0}=m_{\mathrm{p}}c^2= 938.1$ MeV is the rest proton energy, with $m_{\mathrm{p}}$ being the rest proton mass. It can be seen from Eq.~(\ref{eqn:DxxUp}) that the upstream parallel diffusion coefficient approximately follows $D_{\parallel} \propto r \cdot E_k^{2/3}$ for the protons in the energy range of keV to MeV. 

\begin{figure*}[tp!]
\centering {\includegraphics[width=0.97\hsize]{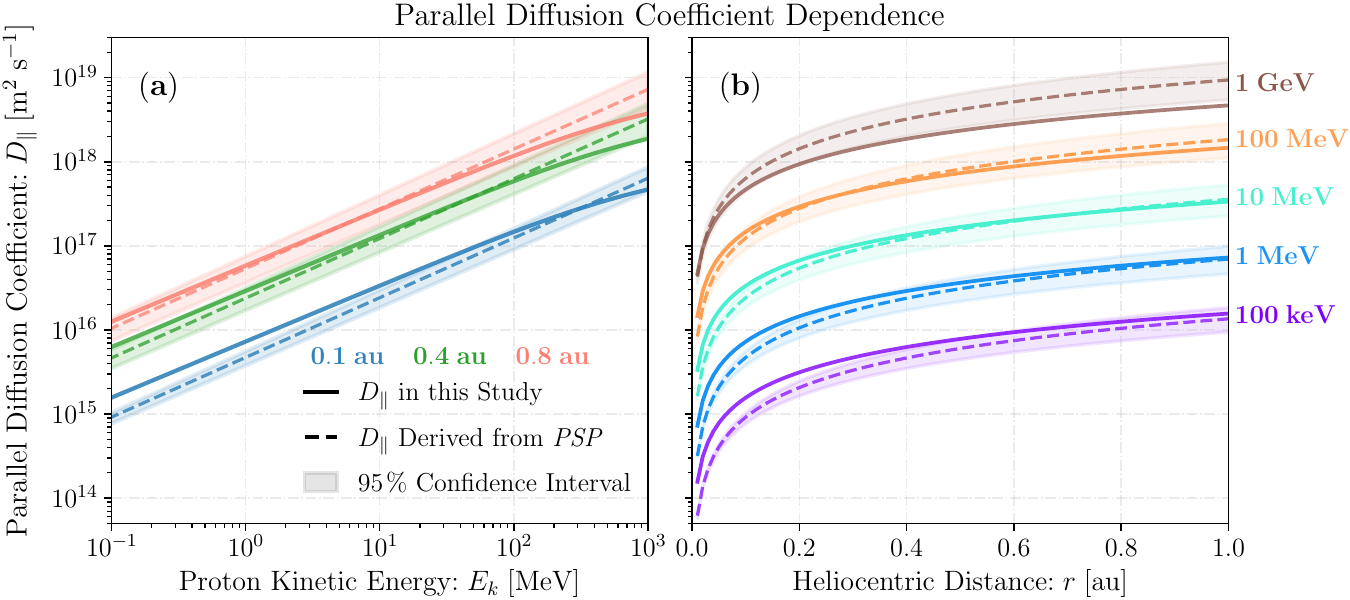}}
\caption{Parallel diffusion coefficient in the upstream shock region used in this study (solid lines) in comparison with that derived by \cite{chen2024parallel} from the \textit{PSP} observations (dashed lines, with the 95\% confidence interval plotted as shaded areas). Left panel (a): Parallel diffusion coefficient dependence on the proton kinetic energy at the heliocentric distance of $0.1$, $0.4$ and $0.8\;\mathrm{au}$, plotted in blue, green and red, respectively. Right panel (b): Parallel diffusion coefficient dependence on the radial distance for the proton kinetic energy of $100\;$keV, $1\;$MeV, $10\;$MeV, $100\;$MeV and $1\;$GeV, plotted in purple, blue, cyan, orange and brown, respectively.} \label{fig2:DxxUp}
\end{figure*}

Note that different parameter configurations of the MFP and the parallel diffusion coefficient may lead to different results \citep[e.g.,][]{kecskemety2009decay, zhao2016double, zhao2024solar}. In order to justify the validity of the adopted choice for $D_{\parallel}$, we refer to a recent study by \cite{chen2024parallel}, which examines the power spectrum density of the magnetic turbulence measured by the \textit{Solar Wind Electrons, Alphas, and Protons} \citep[SWEAP,][]{kasper2016solar} and FIELDS \citep{bale2016fields} instruments on board the \textit{Parker Solar Probe} \citep[\textit{PSP},][]{fox2016solar}. Based on the \textit{PSP} observations in its Orbits 5–13, \cite{chen2024parallel} derived an empirical formula of the parallel diffusion coefficient for 100 keV to 1 GeV energetic protons in the inner heliosphere: 
\begin{equation}
    D_{\parallel} = \left(5.16\pm1.22\right)\times10^{14} \,\left(\frac{r}{1 \mathrm{\; au}}\right)^{1.17\pm0.08}\, \left(\frac{E_k}{1 \mathrm{\;keV}}\right)^{0.71\pm0.02} \; \left[\mathrm{m^2\;s^{-1}}\right]. \label{eqn:DxxPSP}
\end{equation}
With a similar index simplified from Eq.~(\ref{eqn:DxxUp}) for the keV-to-MeV protons, i.e., $D_{\parallel} \propto r \cdot E_k^{2/3}$, we compare the dependence of the diffusion coefficient adopted in M-FLAMPA on the heliocentric distance (see Figure \ref{fig2:DxxUp}(a)) and the proton energy (see Figure \ref{fig2:DxxUp}(b)) with those derived by \cite{chen2024parallel} from the interplanetary turbulence level as observed by \textit{PSP}. In this work, we take $\lambda_0$ in Eq.~(\ref{eqn:lambdaxxUp}) to be $0.3\;$au, which is consistent with the results of \cite{chen2024parallel}. As shown in Figure \ref{fig2:DxxUp}, the comparison demonstrates a perfect agreement within the 95\% confidence interval. 

In the downstream region of the shock, the diffusion coefficient is calculated self-consistently through the total Aflv\'en wave intensity obtained from the MHD simulation. Following Eqs.~(\ref{eqn:calcDxx})–(\ref{eqn:calcDmumu}), we introduce a minimum wave number, $k_0$, below which the turbulence level becomes negligible. In this way, the downstream MFP is derived by only considering $I_{\pm}(k)$ for $k\geqslant k_0$ in Eq.~(\ref{eqn:calcDmumu}), corresponding to sufficiently small spatial scales. With the derivations shown in \cite{borovikov2019toward}, we have: 
\begin{equation}
    \lambda_{\parallel} = \frac{81}{7\pi} \left(\frac{B}{\delta B}\right)^2 \frac{r_{\mathrm{L}0}^{1/3}}{k_0^{2/3}} \cdot \left(\frac{pc}{1 \mathrm{\;GeV}}\right)^{\frac{1}{3}}, \label{eqn:lambdaxxDn} 
\end{equation}
in which $\delta B$ is the turbulent field strength, and $r_{\mathrm{L}0} = \frac{1\;\mathrm{GeV}}{ceB}$ is the Larmor radius for the particle momentum being 1 GeV$/c$. Herewith, we consider
\begin{equation}
    k_0 = \frac{2\pi}{L_{\max}(r)}, \label{eqn:setk0}
\end{equation}
with the maximum spatial scale in the turbulence, $L_{\max}(r) = 0.4\,r$ \citep[e.g.,][and references therein]{borovikov2019toward, tenishev2022application}, which gives a comparable magnitude of the MFP in the shock downstream and upstream as shown in Figure \ref{fig14:lxxUpDn} below. 
Besides, in order to compensate for the eroded width of the shock wave front due to the finite mesh size in the MHD simulations ($\sim0.1\;R_\mathrm{s}$), the parallel diffusion coefficient for the low-energy particles is artificially enhanced to 
\begin{equation}
    D_{\parallel} = {\max}\left\{D_{\parallel},\, D_{\min}\right\}, \label{eqn:DxxLow}
\end{equation}
where $D_{\min}=0.1\;R_\mathrm{s}\times10^5\;\mathrm{m\;s^{-1}}$, as used in \cite{sokolov2004new} and \cite{borovikov2018toward}. 

Using Eqs.~(\ref{eqn:calcDxx})–(\ref{eqn:DxxLow}), the diffusion equation in Eq.~(\ref{eqn:ParkerEqnPBRHS}) can be solved along each individual magnetic field line. It is important to note that this approach does not account for perpendicular diffusion due to the field line random walk and particle decoupling from field lines, which still remains a subject of active research and discussions within the community \citep[e.g.,][]{laitinen2013energetic, laitinen2016solar, laitinen2018effect, shalchi2019field, shalchi2021field, chhiber2021random} and will be implemented in our model in the future.

\subsubsection{Particle Injection} \label{sec:Method3.4}

Given the established dynamics that governs particle acceleration and transport, the next essential component of our model is the injection of particles into the shock acceleration process. In our simulations, the boundary condition at low energies for the omnidirectional distribution function is based on the assumption of a suprathermal tail extending from the plasma thermal energy to the injection energy or the equivalent momentum \cite[see also][]{sokolov2004new}. This tail is assumed to follow a power law, $f\propto p^{-5}$, commonly observed in the solar wind \citep[e.g.,][and references therein]{gloeckler2003ubiquitous, fisk2006common, fisk2008acceleration}:
\begin{equation}
    f(p_\mathrm{inj}) = \frac{c_i}{2\pi} \frac{n_\mathrm{p}}{\left(2m_{\mathrm{p}}k_\mathrm{B}T_\mathrm{p}\right)^{3/2}} \left( \frac{\sqrt{2m_{\mathrm{p}}k_\mathrm{B}T_\mathrm{p}}}{p_\mathrm{inj}} \right)^5, \label{eqn:DistInj}
\end{equation}
where $k_\mathrm{B}$ is the Boltzmann constant, $n_\mathrm{p}$ and $k_\mathrm{B}T_\mathrm{p}$ denote the ambient plasma density and temperature in energy units calculated from the AWSoM-R simulation, respectively. Here, $p_\mathrm{inj}$ is the injection momentum corresponding to the injection energy, which is set to be 10 keV at any location of the shock wave front \citep[e.g.,][]{ellison1990first, giacalone2007acceleration}. Also, the amplitude of the injected particles is determined by the so-called injection coefficient, $c_i$, which indicates the fraction of suprathermal protons and can be derived from:
\begin{equation}
    4\pi \int\limits_{\sqrt{2m_{\mathrm{p}}k_\mathrm{B}T_\mathrm{p}}}^{+\infty}\, p^2f(p) \; \di p = c_i n_\mathrm{p}. \label{eqn:injcoef}
\end{equation}
The injection coefficient, $c_i$, is assumed to be 1 in the simulations. In order to match the observation, a scaling factor, 1.2, is used to scale up the calculated particle flux \citep{zhao2024solar}. A scaling factor greater than 1 indicates that there are actually more seed particles injected into the shock acceleration than what is calculated by Eq.~(\ref{eqn:DistInj}). Since the wave turbulence self-generated by the streaming protons \citep[e.g.,][]{ng1994focused, vainio2003generation, treumann2009fundamentals} is not included in the simulation, the acceleration and transport of energetic protons remain unaffected by such a scaling factor \citep{zhao2024solar}. 

\begin{table*}[ht!]
\begin{center}
\caption{Key input parameters of the AWSoM-R, EEGGL and M-FLAMPA models in the SWMF for this study.} \label{tab:param}
\setlength\tabcolsep{20pt}{
\begin{tabular}{clcccc}
\hline\hline
    Model & Parameter & Value \\ 
\hline % \cline{2-3} \\ \multirow{LINE}{*}{CONTENT}
AWSoM-R & Poynting flux parameter ($\left(S_\mathrm{A}/B\right)_{\odot}$) & $0.3\;\mathrm{MW\;m^{-2}\;T^{-1}}$ \\
 & Correlation length for dissipation ($L_\perp\sqrt{B}$) & $1.5\times10^5\;\mathrm{m\;T^{1/2}}$ \\
 & Stochastic heating exponent ($h_\mathrm{S}$) & $0.24$ \\
 & Stochastic heating amplitude ($A_\mathrm{S}$) & $0.18$ \\
\hline % \cline{2-3}
EEGGL & CME speed\tablenotemark{$^\mathrm{a}$} & $675\;\mathrm{km\; s^{-1}}$ \\
 & Type of the inserted flux rope & GL \\
 & Selected AR positive pole location\tablenotemark{$^\mathrm{b}$} & $\left(66^{\circ},\;\! 16^{\circ}\right)$ \\
 & Selected AR negative pole location\tablenotemark{$^\mathrm{b}$} & $\left(75^{\circ},\;\! 14^{\circ}\right)$ \\
 & Flux rope radius & $0.53\;R_\mathrm{s}$ \\
 & Flux rope stretching & $0.60\;R_\mathrm{s}$ \\
 & Flux rope height & $0.73\;R_\mathrm{s}$ \\
 & Flux rope magnetic field strength & $15.0\;\mathrm{nT}$ \\
\hline
M-FLAMPA & Diffusion coefficient free parameter ($\lambda_0$) & $0.3\;\mathrm{au}$ \\
 & Injection momentum spectral index & $-5$ \\
 & Injection scaling factor & $1.2$ \\
\hline
\end{tabular}}
\end{center}
\tablenotetext{\text{a}}{ This CME speed is re-evaluated and reported in the DONKI database. } %calculated with the effect of 2D projections eliminated. }
\tablenotetext{\text{b}}{ These locations are given as the Carrington longitude and latitude. }
\end{table*}
To summarize, Table \ref{tab:param} recapitulates the key input parameters used for this study, as described through Section \ref{sec:Method}. 

\section{The 2013 April 11 SEP Event: Overview} \label{sec:Event}

% Overview about Lario 2014 paper
The SEP event on 11 April 2013, was one of the widespread SEP events observed in solar cycle 24 \citep[e.g.,][]{richardson201425, dresing2014statistical, gopalswamy2015high, paassilta2018catalogue}. \cite{lario2014solar} studied this event using observations from multiple spacecraft, including the \textit{Solar TErrestrial RElations Observatory Ahead/Behind} \citep[STA/STB, described in][]{kaiser2008stereo}, the \textit{SOlar and Heliospheric Observatory} \citep[\textit{SOHO}, described in][]{domingo1995soho}, the \textit{Advanced Composition Explorer} \citep[\textit{ACE}, described in][]{stone1998advanced}, the \textit{Geostationary Operational Environmental Satellite} \citep[\textit{GOES}, described in][]{menzel1994introducing} and the \textit{Wind} spacecraft \citep[described in, e.g.,][]{harten1995design}. 
They also analyzed the corresponding solar sources of this event by examining extreme ultraviolet (EUV) wave observations and white-light (WL) coronagraph images from STA, STB, and the Atmospheric Imaging Assembly \citep[AIA,][]{lemen2012atmospheric} on board the \textit{Solar Dynamics Observatory} \citep[\textit{SDO},][]{pesnell2012solar}. Furthermore, this particular event shows high ratios of Fe/O abundances observed by particle detectors on board STB and \textit{ACE} (see Figure 4 in \cite{lario2014solar} and more details in \cite{cohen2014longitudinal}).  

% Flare
In this event, the filament eruption that triggered the CME responsible for this SEP event has been investigated by multiple studies \citep[e.g.,][]{vemareddy2015full, joshi2016formation, kwon2017investigating, palmerio2018coronal, fulara2019kinematics, kilpua2019forecasting, pan2022sigmoid}. On 11 April 2013, an M6.5 X-ray flare erupted from the NOAA AR 11719, located at Stonyhurst heliographic latitude $+9^{\circ}$ and longitude $-12^{\circ}$ (N09E12) as viewed from Earth. The soft X-ray emission began at 06:55 UT and peaked at 07:16 UT. Both STB and \textit{Wind} observed type III radio bursts associated with this eruption from the highest frequencies that the instruments can detect ($\sim$16 MHz) starting at about 06:58 UT, while STA observed the type III burst only at frequencies below 1 MHz starting around 07:00 UT. 

% CME
An associated CME was then observed by the C2 coronagraph of the \textit{Large Angle and Spectrometric Coronagraph} \citep[LASCO,][]{brueckner1995lasco} on board \textit{SOHO} at 07:24 UT, and by the COR1 coronagraph \citep{thompson2003cor1} of the \textit{Sun Earth Connection Coronal and Heliospheric Investigation} \citep[SECCHI,][]{howard2008sun} telescopes on board both STA and STB at 07:54 UT \citep{cohen2014longitudinal}. The LASCO observations \citep[e.g., Section 3.4 of][]{joshi2016formation} indicate that this is a moderately fast halo CME, with the plane-of-sky CME speed of $861\;\mathrm{km\; s^{-1}}$ as reported in the Coordinated Data Analysis Web (CDAW) CME catalog\footnote{\url{https://cdaw.gsfc.nasa.gov/CME_list/} \label{url:CDAW}} \citep{yashiro2004catalog, gopalswamy2009soho}. The plane-of-sky CME speed is also reported as $668\;\mathrm{km\; s^{-1}}$ and $590\;\mathrm{km\; s^{-1}}$ by STA/SECCHI and STB/SECCHI, respectively, in the corresponding catalogs\footnote{\url{https://secchi.nrl.navy.mil/cactus/}} \citep{robbrecht2009automated}. By combining these multi-point spacecraft observations, the CME speed is estimated to be over $1000\;\mathrm{km\; s^{-1}}$, as reported by \cite{park2017dependence}. However, \cite{mays2015ensemble} and \cite{dumbovic2018drag} found that such a value can lead to a too early arrival of the CME at Earth in numerical modeling. The re-evaluated CME speed is $675\;\mathrm{km\; s^{-1}}$, as listed in the DONKI database. 
A type II radio burst was observed by both STB and \textit{Wind} starting at 07:10 UT and ending around 15:00 UT in the range of frequencies from 10 MHz to 200 kHz, as reported by \cite{lario2014solar} and \cite{park2015study}.

\begin{figure*}[tp!]
\centering {\includegraphics[width=0.98\hsize]{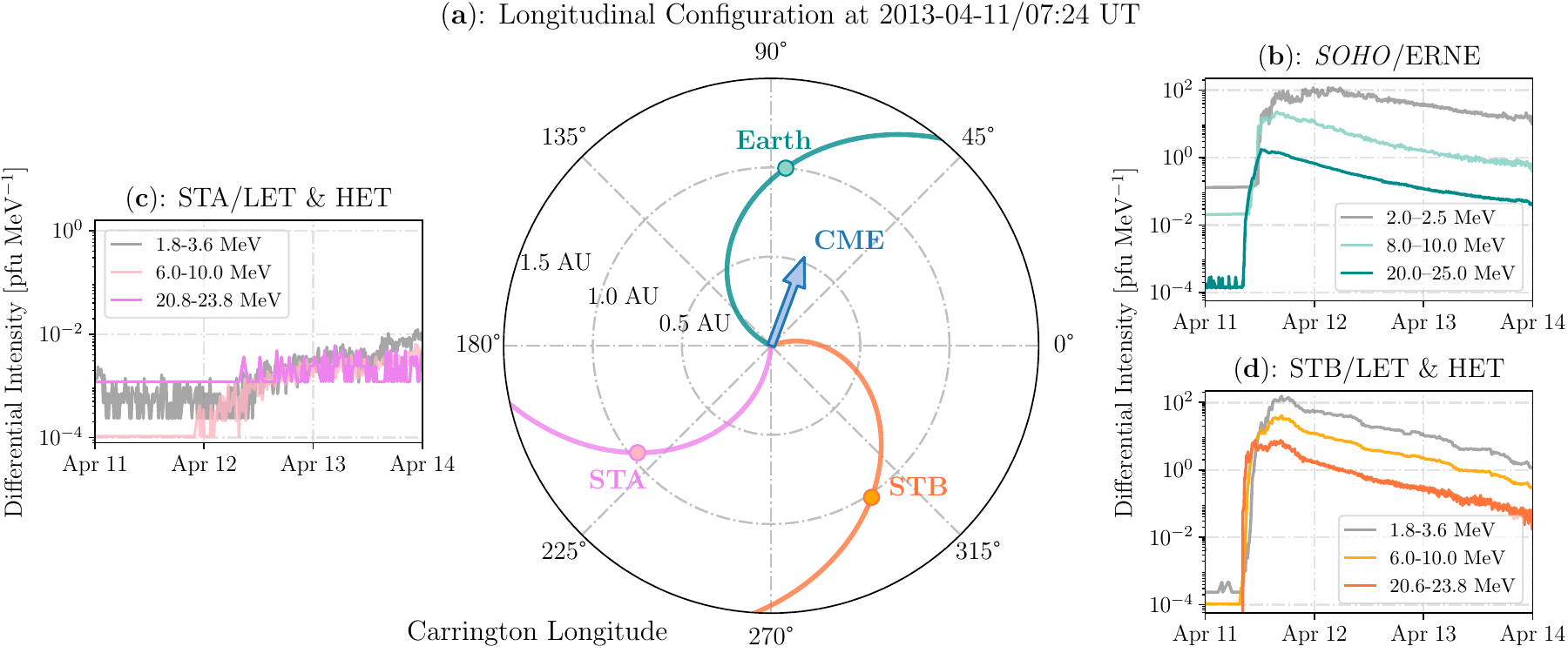}} 
\caption{Overview of the 2013 April 11 SEP Event. (a): Longitudinal configuration of Earth, STA and STB locations (circles) with the nominal IMF lines connecting them to the Sun, as taken from the Solar-Mach tool \citep{gieseler2023solar} and viewed from the north pole, plotted in the heliographic rotating (HGR) coordinates in green, pink and orange, respectively. The blue arrow shows the CME orientation for this event. (b)(c)(d): Measurements of energetic particles by \textit{SOHO}/ERNE near Earth, STA/LET and HET, and STB/LET and HET respectively.} \label{fig3:overview}
\end{figure*}

\begin{table*}[tp!]
\begin{center}
\caption{Solar wind and IMF parameters at different locations.} \label{tab:satprop}
\setlength\tabcolsep{15pt}{
\begin{tabular}{cccccccc}
\hline\hline
    Location & $\phi_\mathrm{C}\;$[$^{\circ}$] & $\theta_\mathrm{C}\;$[$^{\circ}$] & $r\;$[au] & $U_\mathrm{sw}\;[\mathrm{km\; s^{-1}}]$ & $\phi_\mathrm{F}\;$[$^{\circ}$] & $\Delta\phi_\mathrm{F,\,AR}\;$[$^{\circ}$]\\ 
 \hline Earth  & 85.3   & $-5.9$    & 1.00  & 363   & 154.0 & 84.5      \\
 \hline STA    & 218.7  & 7.2       & 0.96  & 514   & 264.7 & $-164.8$  \\
 \hline STB    & 303.5  & 2.3       & 1.02  & 327   & 21.8  & $-47.7$   \\ 
\hline % \cline{2-3} \\ \multirow{LINE}{*}{CONTENT}
\end{tabular}}
\end{center}
% \tablenotetext{}{NOTEs.}
\tablecomments{The list of observation locations included in this study, along with key parameters: the Carrington longitude ($\phi_\mathrm{C}$), the Carrington latitude ($\theta_\mathrm{C}$) and the heliocentric distance ($r$); the \textit{in-situ} solar wind bulk speed ($U_\mathrm{sw}$); the Carrington longitude of the magnetic footpoint following the nominal IMF ($\phi_\mathrm{F}$); and the angular distance between the magnetic footpoint derived from the nominal IMF and the source region from which the CME originates ($\Delta\phi_\mathrm{F,\,AR}$). Here, negative and positive values represent the eastward and westward direction toward the footpoints, respectively.}
\end{table*}

% Description: Figure 3(a), how this (location, field line, cf. Table 2) is done
Figure \ref{fig3:overview}(a) shows the longitudinal locations of Earth, STA and STB, as well as the corresponding nominal IMF lines shortly before the CME eruption on 2013 April 11, as viewed from the north pole. The exact observation locations in the HGR coordinate system are provided in Table \ref{tab:satprop}. Here, the nominal IMF lines assume a Parker spiral field with a constant solar wind speed \citep{parker1958dynamics} and an analytical solution taken from the Solar-Mach tool\footnote{\url{https://solar-mach.github.io/} \label{urk:solarmach}} \citep{gieseler2023solar}. Here, we estimate the solar wind speed ($U_\mathrm{sw}$) by averaging the \textit{in-situ} plasma measurements from the NASA Goddard Space Flight Center (GSFC) OMNI dataset\footnote{\url{https://omniweb.gsfc.nasa.gov/} \label{url:OMNI}} \citep{king2005solar} over a 12-hour window prior to the eruption. The resulting solar wind speed is approximately $363\; \mathrm{km\; s^{-1}}$ at Earth, $\gtrsim500\; \mathrm{km\; s^{-1}}$ at STA, and $327\; \mathrm{km\; s^{-1}}$ at STB, giving the nominal magnetic footpoint separations of $110^{\circ}$–$130^{\circ}$ between pairs of observation locations, as listed in Table \ref{tab:satprop}. 

% Description: Figure 3(a) for CME part, and what we infer from locations of CME and observations
In Figure \ref{fig3:overview}(a), the orientation of the CME flux rope derived from EEGGL and inserted at 07:24 UT, is marked by a blue arrow. The flux rope is placed above AR 11719, centered at $\left(69.5^{\circ},\;\! 14.5^{\circ}\right)$ in Carrington longitude and latitude, as shown in Figure \ref{fig1:input}(b). In Table \ref{tab:satprop}, we show the angular distance between the nominal magnetic footpoint and the parent AR from which the CME originates ($\Delta\phi_\mathrm{F,\,AR}$). These values demonstrate that the magnetic footpoint of STB is the closest to AR 11719 among the three observers, followed by Earth, and that STA is the farthest. Note that the coronal magnetic field can be highly complex and that this estimation can be limited. Alternative methods for calculating these angular distances are presented in Section 2 of \cite{lario2014solar}. 

Figure \ref{fig3:overview}(b)–(d) shows the energetic particle time--intensity profiles measured by (b) the \textit{Energetic and Relativistic Nuclei and Electron} instrument \citep[ERNE,][]{torsti1995energetic, valtonen1997energetic} on \textit{SOHO}; (c) the \textit{Low Energy Telescope} \citep[LET,][]{mewaldt2008low} and \textit{High Energy Telescope} \citep[HET,][]{von2008high} on STA; and (d) LET and HET on STB. We choose 3 energy channels for each spacecraft as shown in Figure \ref{fig3:overview}(b)–(d), as representative of particle measurements\footnote{Hereafter, ``pfu" refers to the particle flux unit, defined as $1\;\mathrm{pfu} = 1\;\mathrm{Count \; cm^{-2}\; s^{-1}\; sr^{-1}}$.} at low, intermediate and high energies, respectively. It can be found that 
\begin{itemize}
    \item[1.] In Figure \ref{fig3:overview}(b)(d), the onset phase of the SEP event appears sharper at STB compared to Earth, especially in the higher energy channels (20.6–23.8 MeV in STB/HET, and 20.0–25.0 MeV in \textit{SOHO}/ERNE). A more detailed comparison of the onset phase of Earth and STB can be found in Figure 5 of \cite{lario2014solar}. %, where particle measurements from the \textit{3D Plasma and Energetic Particle} instrument \citep[3DP,][]{lin1995three} on board \textit{Wind}, the \textit{Electron, Proton, and Alpha Monitor} \citep[EPAM,][]{gold1998electron} on board \textit{ACE} and the \textit{Solar Electron and Proton Telescope} \citep[SEPT,][]{muller2008solar} on both STA and STB are shown and analyzed. 
    Moreover, the low-energy channels (1.8–3.6 MeV in STB/LET, and 2.0–2.5 MeV in \textit{SOHO}/ERNE) in Figure \ref{fig3:overview}(b)(d) shows that the SEP fluxes at STB decay more quickly than at Earth. 
    \item[2.] While noticeable SEP fluxes are observed at both Earth and STB, there is only a slight enhancement of particle measurements at STA, as shown in Figure \ref{fig3:overview}(c). This disparity at STA is consistent with the large longitudinal distance between the source AR and the nominal magnetic footpoint of STA (also see Table \ref{tab:satprop}). 
\end{itemize}
In the following, we will show the numerical simulation results of this event using the models described in Section~\ref{sec:Method} and compare them with the measurements of observed particles. 

\section{Results and Discussion} \label{sec:Result}

In this section, we present the results for each component of the model. In Section \ref{sec:Result1sw}, we show the results of steady-state solar wind simulations for a 27-day period centered at 2013-04-11/06:04 UT. With this steady-state solar wind solution, the flux rope is inserted within AR 11719 from which the CME erupts. We show the CME initial state and evolution, and compare the synthetic WL images with observations in Section \ref{sec:Result2cme}. In Section \ref{sec:Result3shk}, we describe our newly developed shock-capturing tool and illustrate how the shock surface is identified and its evolution in the low solar corona. In Section \ref{sec:Result4sep}, we highlight our M-FLAMPA SEP simulation results and compare them with \textit{in-situ} particle measurements. 

\subsection{Steady-State Solar Wind} \label{sec:Result1sw}

\begin{figure*}[tp!]
\centering {\includegraphics[width=0.98\hsize]{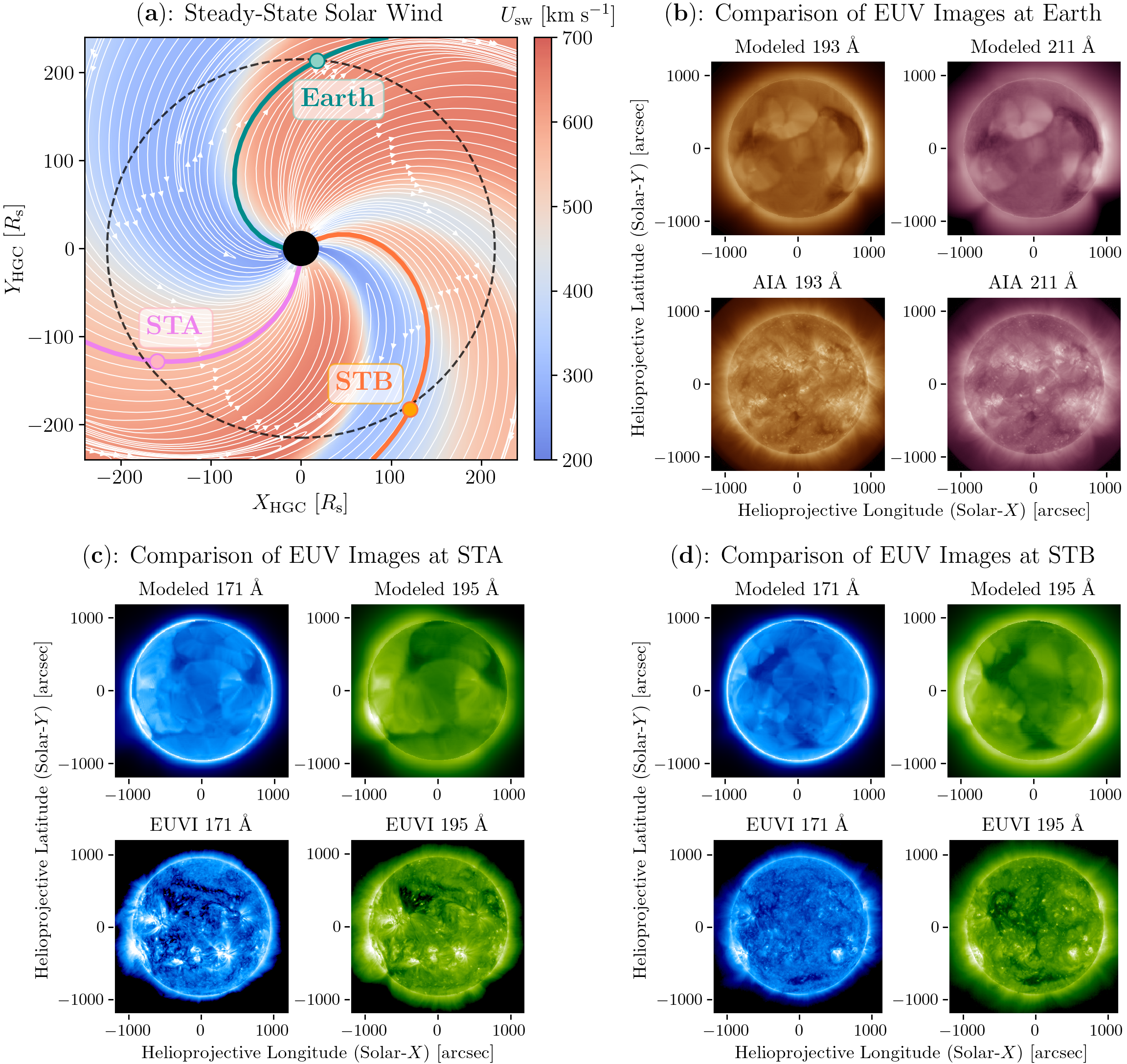}} % png eventually
\caption{Steady-state simulation results by AWSoM-R solving the stream-aligned MHD equations. (a): Background solar wind speed in the solar equatorial plane, with the Carrington Heliographic (HGC) coordinates used. The white curves with arrows indicate the magnetic field lines, and those connecting to Earth, STA and STB are plotted in green, pink and orange, respectively. The black dashed circle represents the $1\;$au heliocentric distance. The black solid circle at the center is the lower boundary of IH in our simulations ($20\;R_\mathrm{s}$). (b)(c)(d): Comparison of the EUV images at Earth, STA and STB, respectively. The modeled images at 193\;\AA\ and 211\;\AA\ wavelengths are compared with those observed by \textit{SDO}/AIA near Earth, and images at 171\;\AA\ and 195\;\AA\ wavelengths are compared with those observed by EUVI on board STA and STB, respectively. Helioprojective longitude and latitude are shown in the modeled and observed images for spatial references.} \label{fig4:backgroundsw}
\end{figure*}

Taking the processed GONG magnetogram shown in Figure \ref{fig1:input}(a) and the parameters listed in Table \ref{tab:param} as inputs, the stream-aligned AWSoM-R model calculates the 3D solar wind plasma properties. The solar wind speed in the solar equatorial plane in IH is shown in Figure \ref{fig4:backgroundsw}(a), plotted with the locations of Earth, STA and STB, as well as the magnetic field lines connecting them to the inner boundary of IH. As seen in Figure \ref{fig4:backgroundsw}(a), Earth and STA are located in regions with relatively fast solar wind ($>500\; \mathrm{km\; s^{-1}}$), while the solar wind is relatively slow at STB ($\simeq300\; \mathrm{km\; s^{-1}}$), comparable to the values in Table \ref{tab:satprop}. We also plot other magnetic field lines in the equatorial plane as white curves with arrows, demonstrating the alignment of the magnetic field and the solar wind plasma stream (see Section \ref{sec:Method1} and \cite{sokolov2022stream}). 

We use the simulated steady-state solar wind electron density and temperature to synthesize the LOS EUV images, which are compared with the multi-wavelength EUV observations\footnote{\url{https://sdac.virtualsolar.org/cgi/search} \label{url:vso}} from \textit{SDO}/AIA and the \textit{Extreme Ultraviolet Imager} \citep[EUVI,][]{wuelser2004euvi} on board STA and STB. The comparisons between simulations and observations are shown in Figure \ref{fig4:backgroundsw}(b)–(d), corresponding to 193\;\AA\ and 211\;\AA\ bands for \textit{SDO}/AIA, and 171\;\AA\ and 195\;\AA\ for STA/EUVI and STB/EUVI. For each comparison, the top row shows the model-synthesized LOS EUV images, and the bottom row shows the observation results. The key findings from the EUV image comparisons in Figure \ref{fig4:backgroundsw}(b)–(d) are the following: 
\begin{enumerate}
    \item[1.] The simulation results exhibit reasonable consistency with the observations in matching the relative brightness on a global scale, capturing the positions of major coronal holes (CHs) and ARs. This agreement indicates that the stream-aligned AWSoM-R model is able to reproduce the overall 3D structure of the density and temperature in the low solar corona. 
    \item[2.] %Notably, open fluxes from the CHs are simulated reasonably well. 
    The CHs in the northern hemisphere are captured in simulations from the STA and STB views, and the narrow CH close to the south pole is also reproduced in simulations from the STB view, as shown in Figure \ref{fig4:backgroundsw}(c)(d). Although the simulated CHs in the northern hemisphere are visible from Earth's point of view, they are relatively darker in comparison to \textit{SDO}/AIA observations in Figure \ref{fig4:backgroundsw}(b). As discussed in \cite{sachdeva2023solar}, solar CHs contain small-scale, closed field-line loops and magnetic flux that add to their brightness. In contrast, the numerical simulation often lacks these small-scale features, leading to darker CHs in the synthetic images. 
    \item[3.] As shown in Figure \ref{fig4:backgroundsw}(b)–(d), the stream-aligned AWSoM-R model reproduces the bright ARs on the west limb from the Earth view, and the ARs on both the west and east limbs from the STA and STB views. However, the small-scale ARs in the center of the EUV images observed are either partially visible or not present in the model results. This discrepancy can be attributed to a combination of factors, including the uncertainties of observational data prepared for the synoptic map, the evolution of all ARs, particularly near the solar maximum, as well as the order of accuracy in the modeling scheme \citep[e.g.,][and references therein]{bertello2014uncertainties, gombosi2021sustained, sachdeva2021simulating, sachdeva2023solar}. 
    % It is worth noting that April 2013 is near the solar maximum in solar cycle 24. Our steady-state simulation is performed with a synoptic magnetogram, which assumes the AR is shifted in location but with no changes in strength, whereas the observations are updated for particular timestamps. Therefore, it is reasonable that the model cannot reproduce all the dynamical and time-dependent solar activities at specific time \citep[e.g.,][]{sachdeva2019validation}. 
\end{enumerate}

Despite the discrepancies in some fine structures, such as CHs and ARs, the model demonstrates generally good agreement in terms of the overall brightness, as well as the spatial location and scale of these features, indicating high simulation performance in capturing the global structure in the low corona \citep{downs2010toward, sachdeva2019validation, sachdeva2021simulating}. 
We also note that the source surface radius of the PFSS model is set to be $2.5\;R_\mathrm{s}$ in this work, as described in Section \ref{sec:Method1}. Adjusting the outer boundary of the PFSS model may lead to an improved background solar wind solution \citep[e.g.,][and references therein]{lee2011coronal, huang2024adjusting}, which is important in modeling the propagation of CMEs and SEPs. 
Overall, the comparisons shown in Figure \ref{fig4:backgroundsw}(b)–(d) validate the synthesized EUV observables and suggest the readiness of steady-state solar wind solutions for subsequent simulations. 

\subsection{CME Eruption and Propagation} \label{sec:Result2cme}

\begin{figure*}[tp!]
\centering {\includegraphics[height=0.8\textheight]{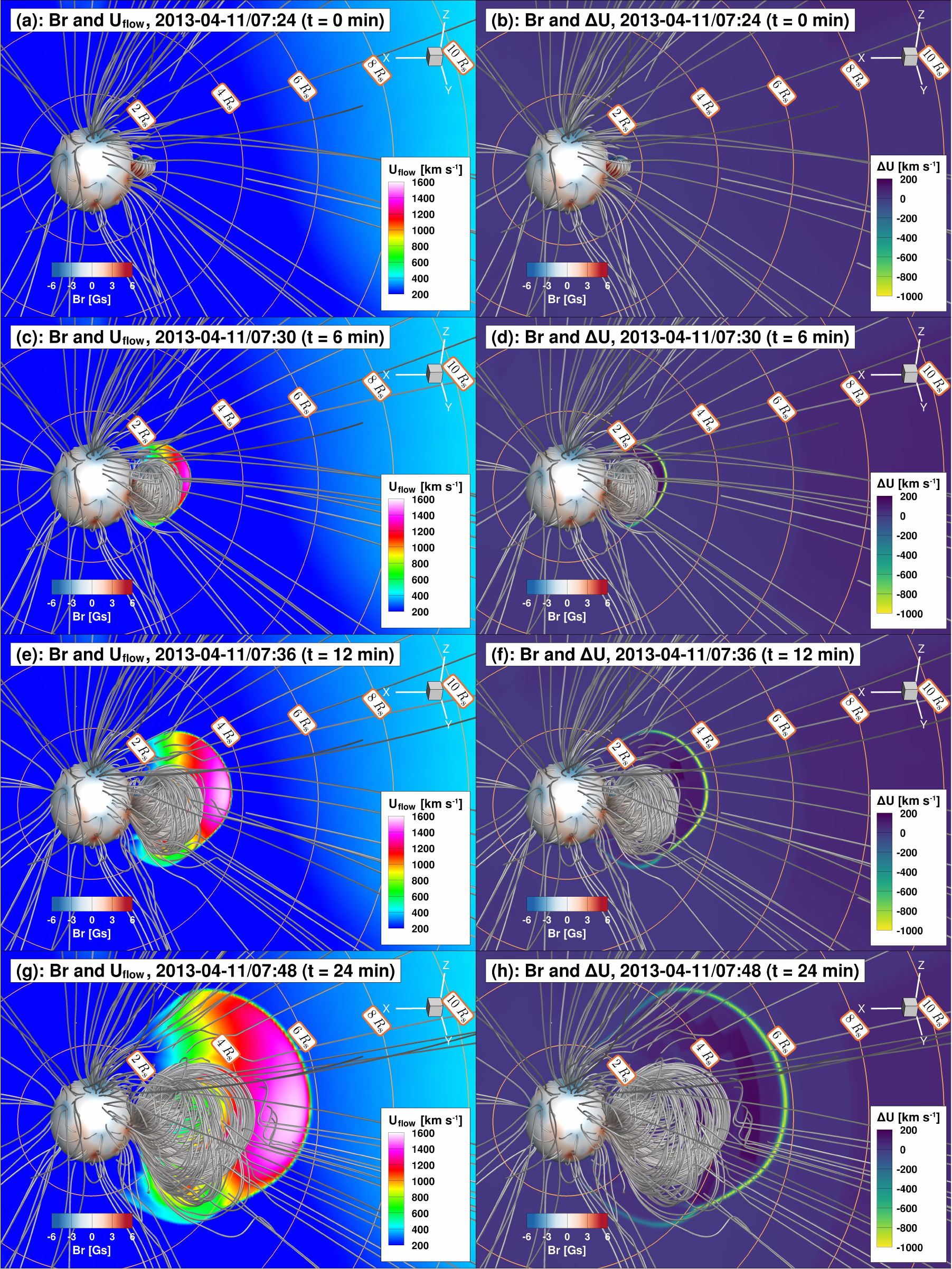}} 
\caption{Evolution of the flux rope, the flow speed ($U_\mathrm{flow}$) and the $\Delta U = \Delta x\, \nabla\cdot \bs{u}$ value (divergence of the velocity field times the cell size) in SC. In each panel, 3D topology of multiple magnetic field lines are shown. The concentric circles represent the heliocentric distance in the contour slice, plotted every $2\;R_\mathrm{s}$ with the values written on each circle. HGR coordinates are used with the system rotated such that the negative $X$-axis points toward Earth. Panel (a) shows the initial flux rope at the solar surface plotted with the radial magnetic field strength ($B_r$) plotted on the solar surface while the flow speed apears on the $x-z$ plane in SC. Panel (b) is similar to panel (a) but plotted with the $\Delta U$ value in SC. Panels (c)(d), (e)(f) and (g)(h) are similar to panels (a)(b) but at 6, 12 and 24 minutes after the CME eruption, respectively.} \label{fig5:3dplot}
\end{figure*}

\begin{figure*}[tp!]
\centering {\includegraphics[width=0.95\hsize]{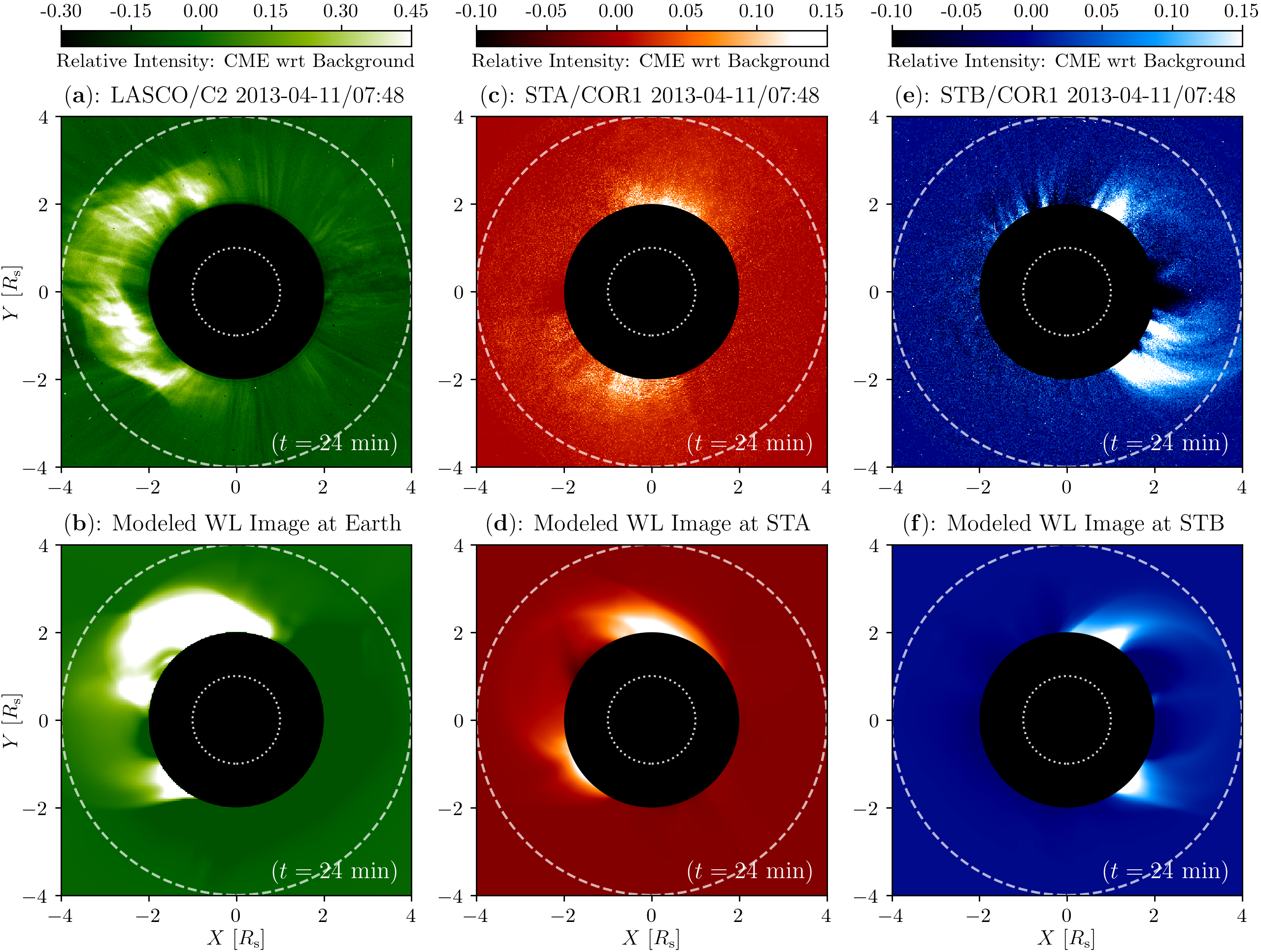}} % png eventually
\caption{Comparison of the LASCO/C2 (left column), STA/COR1 (middle column) and STB/COR1 (right column) WL images at 24 minutes after the CME eruption. Upper panels are observations, and lower panels are the corresponding model-synthesized WL images. The color scale shows the relative intensity changes of WL total brightness with respective to the solar wind before the eruption. In each image, the radius of the inner white dotted circle, the black solid circle and the outer dashed white circle are at $r=1.0$, 2.0 and $4.0\; R_\mathrm{s}$, respectively.} \label{fig6:WLimage}
\end{figure*}

After obtaining steady-state solar wind solutions, a force-imbalanced GL magnetic flux rope \citep{gibson1998time} and its entrained plasma are placed on top of the parent AR (AR 11719 for this event). After the insertion, we do not change the velocity of the initial states to drive self-similar evolution. Figure \ref{fig5:3dplot} presents the time evolution of the CME in 3D, at $t = 0$, 6, 12 and 24 minutes from the top panels to the bottom. In each panel, the magnetic field lines are plotted in the solid lines, and the blue-to-red colors represent the radial magnetic field strength on the $1.1\;R_\mathrm{s}$ sphere and along the field lines. A plane cut depicts the distribution of the plasma speed in the left column of Figure \ref{fig5:3dplot} and the $\Delta U = \Delta x\, \nabla\cdot\bs{u}$ value for the divergence of the velocity field times the cell size in the right column of Figure \ref{fig5:3dplot}. The color bars for different parameters are shown at the bottom of each panel. 
In Figure \ref{fig5:3dplot}(c)(e)(g), we can see that the evolution of the flux rope starts with a rapid acceleration to a speed greater than $1200\;\mathrm{km\; s^{-1}}$ in the low corona. The fast propagation of the flux rope drives a fast-mode MHD shock ahead of it, corresponding to the interface between the magenta and blue regions. As the flux rope propagates outward, it interacts with the background magnetic field, changing the field topology. The interaction is evident in the bent field lines downstream of the shock, as shown in panels (e)–(h) of Figure \ref{fig5:3dplot}. 

Here, we compare the model-synthesized WL images with observations\textsuperscript{\ref{url:vso}} to validate the CME flux rope propagation direction. In our model, the synthetic WL images are created by integrating the Thomson-scattered light along the LOS that comprises the image \citep[e.g.,][]{hayes2001deriving, morgan2006depiction}. As illustrated in Figure \ref{fig6:WLimage}, we compare the model-synthesized WL images with those captured by LASCO/C2 (Figure \ref{fig6:WLimage}(a)(b)), STA/COR1 (Figure \ref{fig6:WLimage}(c)(d)) and STB/COR1 (Figure \ref{fig6:WLimage}(e)(f)) coronagraphs. C2 has a field of view (FOV) from $2.0$ to $6.0\;R_\mathrm{s}$, and COR1 from $1.5$ to $4.0\;R_\mathrm{s}$. Therefore, in Figure \ref{fig6:WLimage}, we limit the FOV to $4.0\;R_\mathrm{s}$. In each panel of Figure \ref{fig6:WLimage}, the inner white dotted circle, the black solid circle and the outer dashed white circle show the radius at 1.0, 2.0 and $4.0\;R_\mathrm{s}$, respectively. The color scale shows the relative intensity changes of the WL total brightness with respect to the steady-state corona solar wind. Here we list our key findings by comparing the WL images, and propose the possible explanations for some of them:
\begin{enumerate}
    \item[1.] In the LASCO/C2 view, the core structure of the CME propagates eastward, as shown in Figure \ref{fig6:WLimage}(a)(b). The observation shows the CME to be symmetric along the equator, while the synthetic image shows the northern part of the CME to be brighter than the southern part. 
    % The envelope of the observed CME appears to be symmetric with respect to the solar equator. However, the northern part of the CME is brighter than the southern part in the model-synthesized WL image, demonstrating an asymmetric shape. 
    We examine the plasma properties in our simulations and find a high-density region ahead of the flux rope, which can slow down the CME propagation, thus contributing to this asymmetry \citep{zhao2024solar}. 
    \item[2.] In the STA/COR1 view, we primarily see the traces of a structure propagating toward the far side of the Sun in Figure \ref{fig6:WLimage}(c)(d). The noisy white dots in the east part of the images may indicate the CME propagation direction; nonetheless, the brightness changes are not very pronounced because of the location separation, as shown in Table \ref{tab:satprop} and Figure \ref{fig3:overview}(a). In our simulations, we basically reproduce these structures, including the propagation direction and some weakly intensified brightness in the east, as illustrated in Figure \ref{fig6:WLimage}(d). 
    \item[3.] In the STB/COR1 view, the CME propagates westward with a nearly symmetric structure with respect to the solar equator. By comparing Figure \ref{fig6:WLimage}(e)(f), our model-synthesized WL image aligns with the observation except for some slight differences in the brightness in the south. This difference is likely related to the high-density region ahead of the flux rope in the simulation, which also affects the symmetry of the CME core structure from the point of view of LASCO/C2.  
\end{enumerate}
Furthermore, we note that these images are influenced by significant projection effects, which can complicate the interpretation of the CME structure and brightness distribution \citep[e.g.,][and references therein]{temmer2009cme, temmer2023cme}. 

\subsection{Shock Wave Front} \label{sec:Result3shk}

Because of the role played by CME-driven shocks in particle acceleration processes, CME-driven shock simulations are essential to calculate particle acceleration in the modeling of SEP events \citep[e.g.,][]{mikic2006introduction, lee2012shock, manchester2017physical}. 
% Studies in the past have tried to reconstruct the shock surface, derive the shock properties, and find the magnetic connectivity to investigate the role of shocks in SEP events \citep[e.g.,][]{smith1990mhd, lario1998energetic, ontiveros2009quantitative, shen2011three, kwon2014new, bain2016shock, rouillard2016deriving, plotnikov2017magnetic, kouloumvakos2019connecting, maguire2020evolution}. A feasible way to obtain more information about the CME-driven shock is to capture the self-consistent shock from MHD simulations. Recently, satisfying agreements of shocks in simulations and observations have been found among several state-of-the-art models \citep[e.g.,][]{torok2018sun, downs2021validation, ding2022modeling, jin2022assessing, jin2024exploring}, which indicates the capability of constructing realistic CME-driven shocks from current numerical models. In our model, we develop a shock-capturing tool embedded in the MHD simulations to better understand of the time evolution of the properties of the shock structure. 
Studies in the past have tried to reconstruct the shock surface and derive its properties, which can be categorized into two broad types: (1) using observational data such as EUV, WL coronagraphs and radio observations from which it is possible to infer the shock surface and its properties \citep[e.g.,][]{ontiveros2009quantitative, kwon2014new, rouillard2016deriving, plotnikov2017magnetic, zucca2018shock, kouloumvakos2019connecting, maguire2020evolution}; and (2) employing physics-based simulations to model the shock surface and its evolution \citep[e.g.,][]{smith1990mhd, lario1998energetic, shen2011three, torok2018sun, downs2021validation, ding2022modeling, jin2022assessing}. 
While observational studies provide valuable constraints on shock properties, they are often limited by projection effects and assumptions about the applied shock geometry. On the other hand, simulations offer a more feasible way to study the evolution of shocks \citep[e.g.][]{manchester2005coronal}. 
In our model, we develop a shock-capturing tool embedded in the MHD simulations, which enables shock identification with high spatial resolution in 3D and demonstrates the refined structure and complexity of the shock front, thus distinguishing it from many existing methods listed above. 

% Grid and scheme for the shock-capturing tool
In addition to the grid initialized as described in Section \ref{sec:Method1}, in order to resolve the fine structures around the shock surfaces, the grid block resolution is refined by a factor of 2 at locations where the ion thermal pressure jump between neighboring cells exceeds 2.0. Moreover, AWSoM-R incorporates a second-order shock-capturing scheme with slope limiters to enhance the accuracy of MHD parameters simulated near the shock front \citep{toth2012adaptive, gombosi2021sustained}. 

% Criteria
In our model, $\Delta U = \Delta x\, \nabla\cdot\bs{u}$ is used as the criterion to extract the shock surface. The divergence of the velocity field is negative at the shock front and scales with the inverse of the shock width that is proportional to the local mesh size $\Delta x$. In essence, $\Delta U$ indicates the flow speed jump across one grid cell. In our shock-capturing tool, the shock surface is extracted along radial lines using a longitude-latitude grid with an angular resolution of $0.5^{\circ}$. Along each radial line, the shock wave front is identified by the smallest value of $\Delta U$. Taking into account the fluctuations for the modeling tool and realistic structures in the system, a threshold of $\Delta U$ needs to be specified. In our simulation, the threshold is $\Delta U_\mathrm{t}=-120\;\mathrm{km\; s^{-1}}$, as illustrated in Figure \ref{fig8:shockprop}(a). If the minimum $\Delta U$ is less than the threshold, the radial distance of the surface and the shock associated parameters are saved; otherwise, the shock surface is not recognized. 

\begin{figure*}[tp!]
\centering {\includegraphics[width=0.95\hsize]{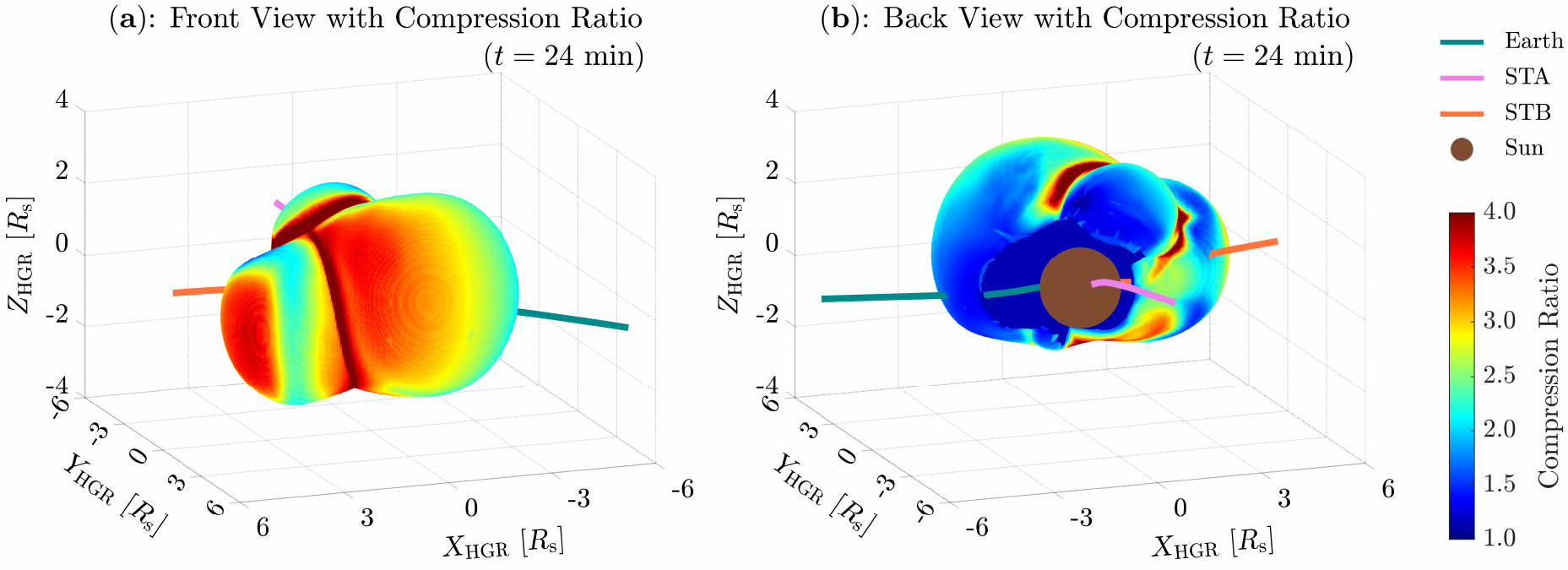}} \\
\vspace{0.5em}
{\includegraphics[width=0.95\hsize]{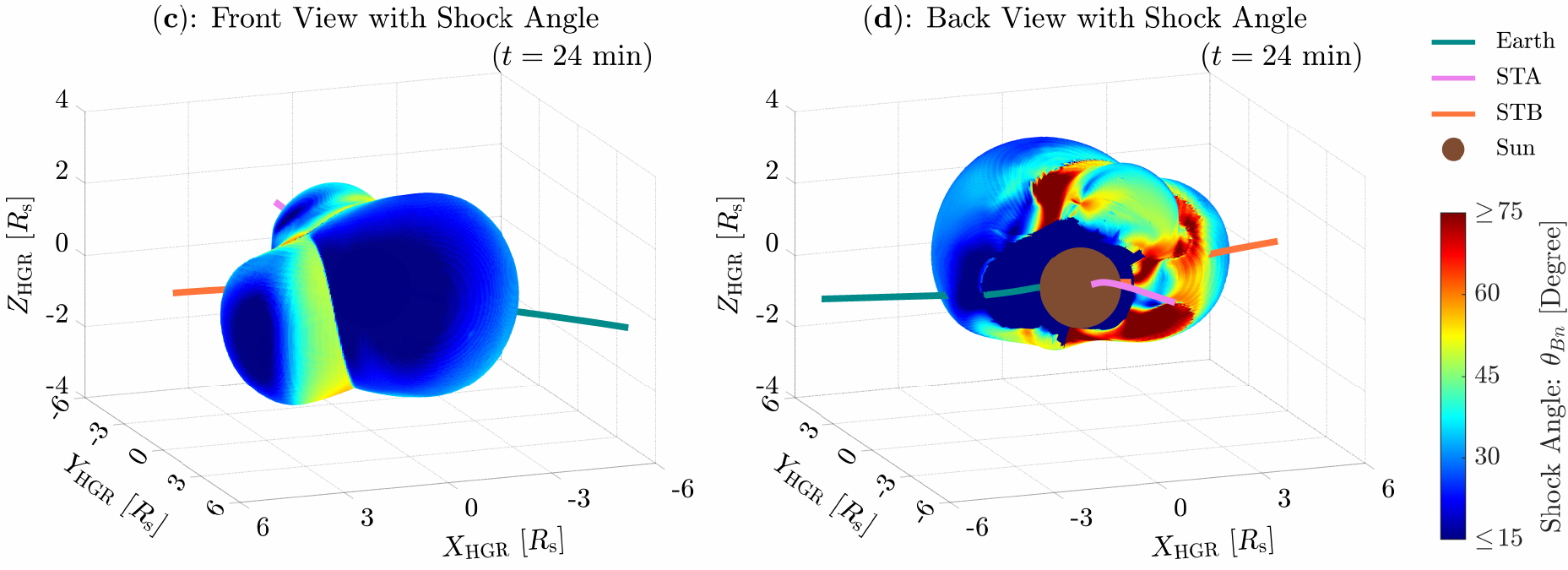}} \\
\vspace{0.5em}
{\includegraphics[width=0.95\hsize]{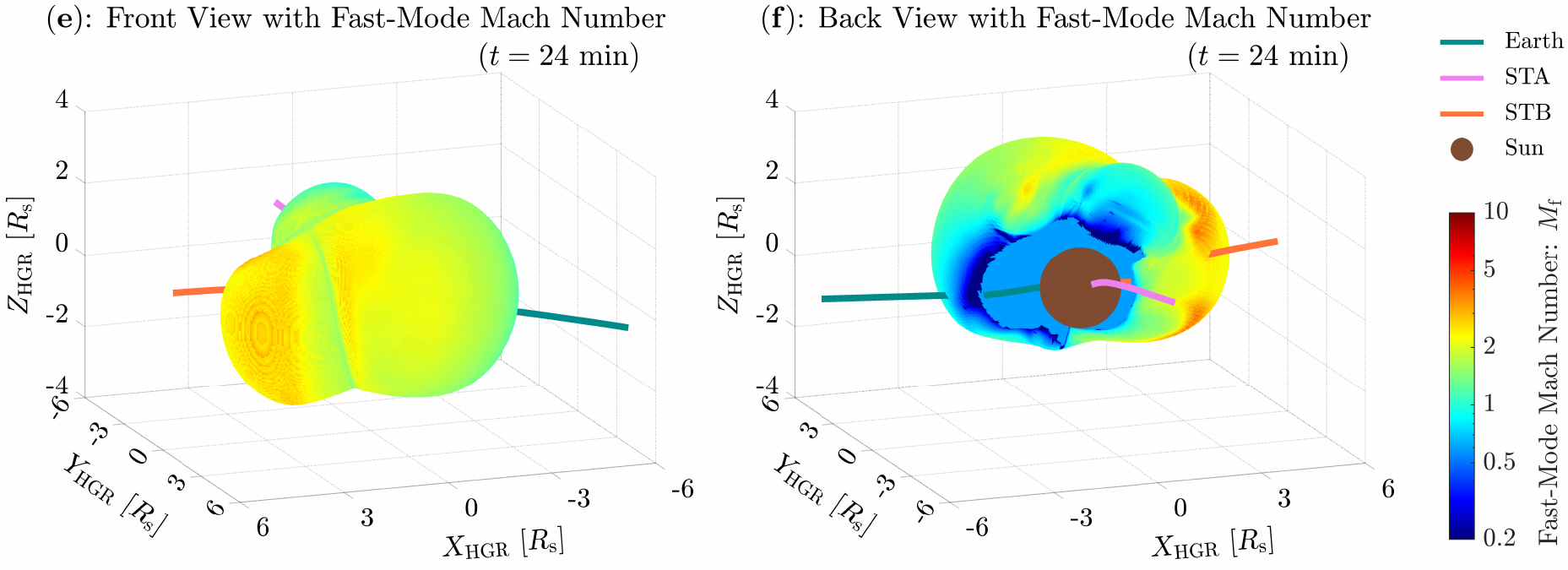}}
\caption{Extracted shock surface at 24 minutes after the flux rope eruption. HGR coordinates are used. In each panel, field lines connecting to Earth, STA and STB are plotted in green, pink and orange, respectively. The brown sphere in the center represents the Sun ($r=1\;R_\mathrm{s}$) in panels (b), (d) and (f). Panel (a) shows the front view of the shock surface colored with the shock compression ratio. The front view is defined as the view from above the AR. Panel (b) is similar to panel (a) but from the back view of the shock surface, which is 180-degree rotating about the $Z$-axis. Panels (c)(d) and (e)(f) are similar to panels (a)(b) but colored with the shock angle ($\theta_{Bn}$) and the fast-mode Mach number ($M_{\text{f}}$), respectively.} \label{fig7:shocksurf}
\end{figure*}

\begin{figure*}[tp!]
\centering {\includegraphics[width=0.8\hsize]{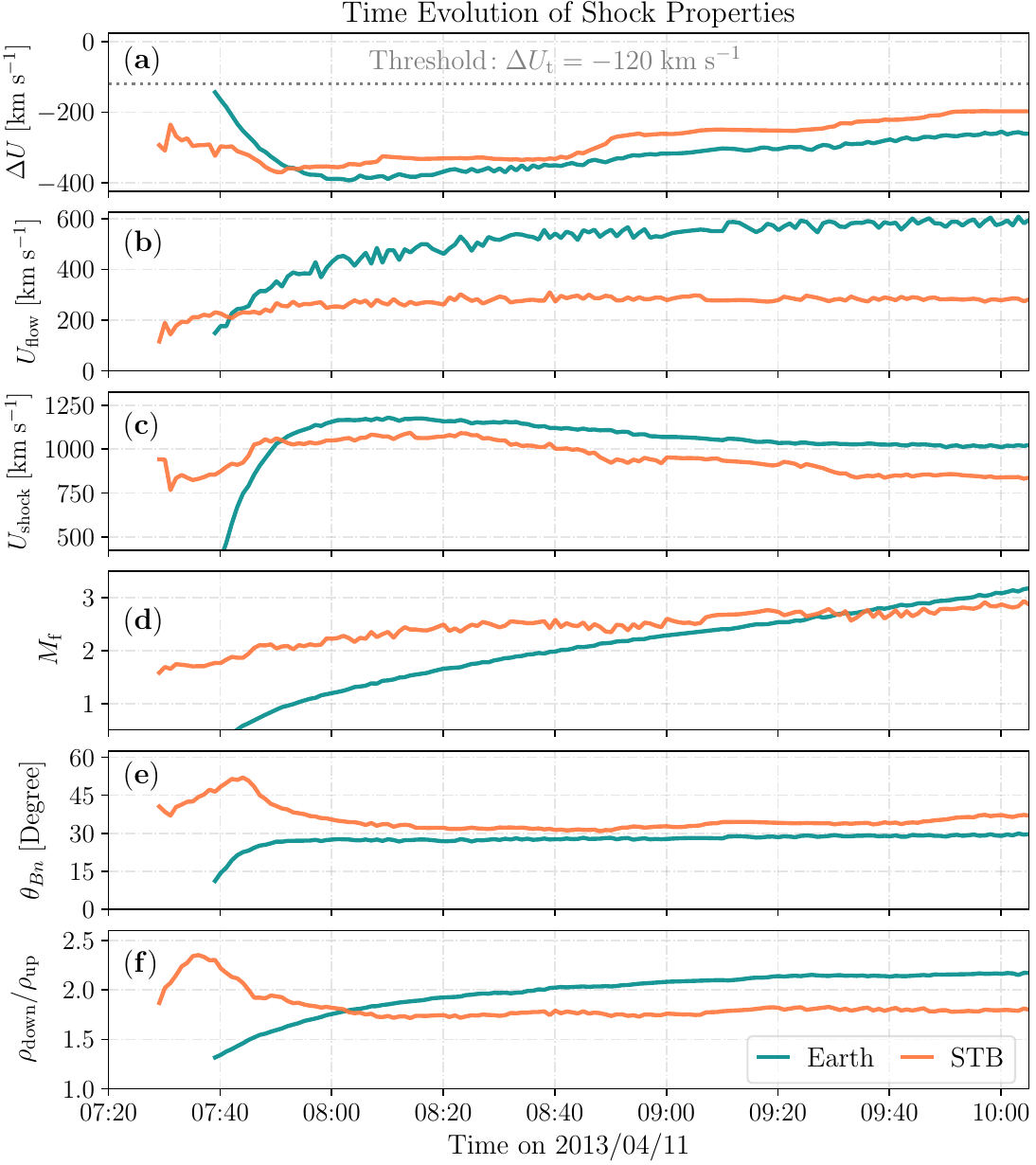}}
\caption{Time evolution of the shock properties at the intersection point of the shock surface with field lines connecting to Earth and STB, plotted in green and orange, respectively. Six parameters are presented from the top panel to the bottom: (a) The $\Delta U$ value, plotted with the gray dotted horizontal line for the shock identification threshold $\Delta U_\mathrm{t}=-120\;\mathrm{km\; s^{-1}}$ used in this study; (b): Upstream solar wind speed ($U_\mathrm{flow}$); (c) Shock speed ($U_\mathrm{shock}$); (d) Fast-mode Mach number ($M_{\text{f}}$); (e) Shock angle ($\theta_{Bn}$); (f) Density compression ratio ($\rho_\mathrm{down}/\rho_\mathrm{up}$).} \label{fig8:shockprop}
\end{figure*}

Figure \ref{fig7:shocksurf} shows the shock surface extracted at $t = 24\;$minutes after the CME eruption from the front (left column) and back (right column) views. Here, the front view refers to the view from above AR 11719, and the back view is with a 180-degree rotation about the $Z$-axis. We show the shock wave front colored by the compression ratio, the shock angle and the Mach number for the fast magnetosonic wave in panels (a)(b), (c)(d) and (e)(f) of Figure \ref{fig7:shocksurf}, respectively. We can see that the 3D shock surface is non-uniform and consists of three spherical regions, due to the deformation of the flux rope in its interaction with the inhomogeneous background solar wind (see Figure \ref{fig5:3dplot}(g)(h)). Distinct variations in the shock properties can be found across small distances on the shock surface, especially near the interfaces of different spherical regions. Detailed calculations of the upstream shock normal, the shock angle, the shock speed and the fast-mode Mach number are included in Appendix \ref{sec:appendA}. 

The connection point of each observer on the shock surface is determined by tracing the magnetic field line from the location of the observer back to the shock surface in the 3D magnetic field solutions of AWSoM-R. 
Hereafter in this paper, when referring to the observer-shock front magnetic connection, we use the concept of ``cobpoint", short for the \textit{Connecting with the OBserver POINT}, which is first explicitly considered in modeling by \cite{heras1995three}. 
In Figure \ref{fig7:shocksurf}, we show the field lines in the low solar corona at $t = 24$ minutes after the CME eruption, which connect to Earth, STA and STB, plotted in green, pink and orange, respectively. At $t = 24$ minutes, Earth and STB are magnetically connected to the shock, while STA is not. As illustrated in Figure \ref{fig7:shocksurf}, Earth is connected to the weak part of the shock with a compression ratio of about 1.5. The shock is quasi-parallel, with a $\theta_{Bn}$ being about $30^{\circ}$, and the fast-mode Mach number is around 1.0. STB is connected to a stronger part of the shock with a compression ratio of about 2.0. The $\theta_{Bn}$ is about $45^{\circ}$, indicating an oblique shock, and the fast-mode Mach number is around 2.0. 

With a one-minute cadence, we trace the field lines and examine the shock properties at the cobpoint. Figure \ref{fig8:shockprop} illustrates the time evolution of the cobpoint properties corresponding to Earth and STB, plotted in green and orange, respectively. Properties of the STA-related cobpoint are not shown because STA is not magnetically connected to the shock. The properties displayed include the criterion we use to identify the shock surface (the $\Delta U$ value), the upstream flow speed ($U_\mathrm{flow}$), the shock speed ($U_\mathrm{shock}$), the fast-mode Mach number ($M_\text{f}$), the shock angle ($\theta_{Bn}$) and the density compression ratio ($\rho_\mathrm{down}/\rho_\mathrm{up}$)\footnote{Hereafter, $\left\langle\cdots\right\rangle_\mathrm{down}$ and $\left\langle\cdots\right\rangle_\mathrm{up}$ refer to the parameter in the shock downstream and upstream, respectively.} from top to bottom panels in Figure~\ref{fig8:shockprop}. In Figure \ref{fig8:shockprop}, it is shown that STB is magnetically connected to the shock surface around 5 minutes after the flux rope erupted, while Earth is connected to the shock surface around 15 minutes after the flux rope eruption. Moreover, STB is connected to a stronger shock in the first hour, whereas the shock to which the Earth is connected is very weak. Therefore, more effective particle acceleration is expected at the STB-connected shock than Earth at the beginning of the event, explaining the fact that we see a more prompt onset at STB while the onset is more gradual at Earth, as shown in Figure 5 of \cite{lario2014solar} with a zoomed-in time axis. Later, the shock region which Earth is connected to becomes slightly stronger than STB, with a higher compression ratio and higher fast-mode Mach number, as shown in Figure \ref{fig8:shockprop}(d)(f). 

\subsection{SEP Fluxes and Evolutions} \label{sec:Result4sep}

In M-FLAMPA, 648 magnetic field lines are extracted and the distribution functions of energetic particles along each individual magnetic field line are solved. These 648 magnetic field lines are uniformly initialized on the $r = 2.5\;R_\mathrm{s}$ sphere that covers $360^{\circ}$ in longitude and $\pm85^{\circ}$ in latitude, as described in Section \ref{sec:Method3.2}. 
Once the time-accurate simulation begins, that is, after the flux rope is placed on top of the AR, AWSoM‐R and M‐FLAMPA run concurrently. M-FLAMPA extracts the plasma parameters from AWSoM-R simulations every 2 minutes and calculates the acceleration and transport processes of particles. 
In the following sections, we show the simulation results and compare them with observations. 

\subsubsection{{\it Z}=0 Plane Cut and Radial Distribution} \label{sec:Result4sep0_zcut}

\begin{figure*}[tp!]
\centering {\includegraphics[width=0.95\hsize]{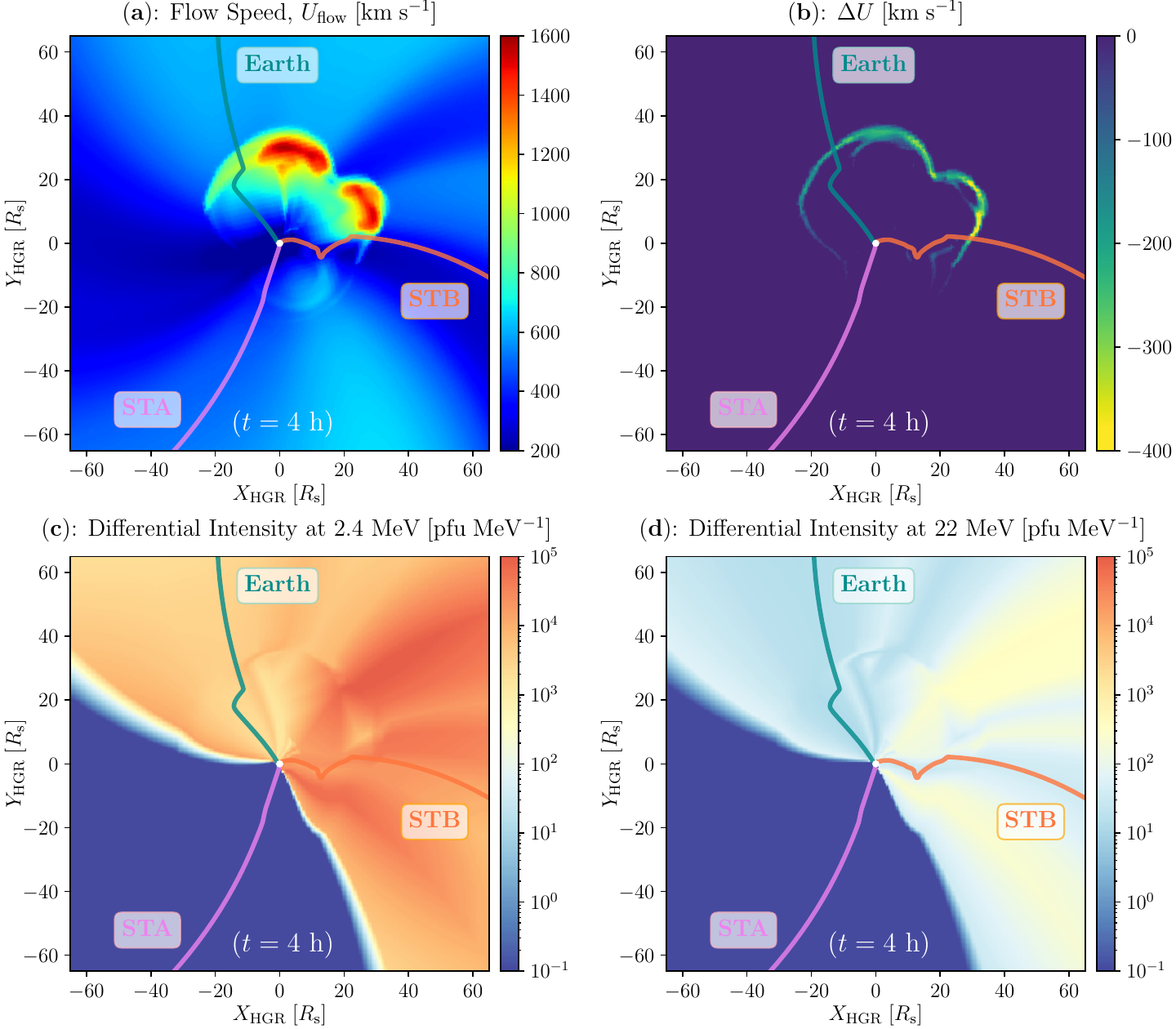}}
\caption{Simulation results in the solar equatorial plane ($Z_\mathrm{HGR}=0$, viewed from the north pole) at 4 hours after inserting the flux rope. In each panel, HGR coordinates are used, and the FOV is set as $-65\;R_\mathrm{s}\leqslant X_\mathrm{HGR},\, Y_\mathrm{HGR}\leqslant 65\;R_\mathrm{s}$. Field lines connecting to Earth, STA and STB are plotted in green, pink and orange, respectively, with corresponding labels next to the field lines. The white solid circle at the center represents the Sun ($1\;R_\mathrm{s}$). Panel (a) shows the flow speed. Panel (b) shows the $\Delta U$ value. Panels (c) and (d) show the energetic proton differential intensity at 2.4 and 22 MeV, respectively, with colors saturated if the intensity falls beyond the range from $10^{-1}$ to $10^{5}\;\mathrm{pfu\; MeV^{-1}}$.} \label{fig9:z0Cut}
\end{figure*}

Figure \ref{fig9:z0Cut} shows a snapshot of simulation results in the solar equatorial plane where $Z_\mathrm{HGR}=0$ as seen from above the north pole, at 4 hours after the flux rope eruption. The magnetic field lines connecting to Earth, STA and STB are plotted in green, pink and orange curves, respectively. In Figure~\ref{fig9:z0Cut}, panels (a) and (b) show the plasma speed and $\Delta U$ (see Section~\ref{sec:Result3shk}). Panels (c) and (d) show the calculated differential intensity of protons at 2.4 MeV and 22 MeV. Those two energy channels are chosen to compare with the \textit{SOHO} and STB observations, as discussed below in Section~\ref{sec:Result4sep2_time}. 
Since STA is not magnetically connected to the shock surface as seen in panels (a) and (b), and the perpendicular diffusion is not incorporated in this work, STA does not detect any particle enhancement. In addition, the longitudinal dependence of the particle intensity in panels (c) and (d) can be explained by the non-uniformity of the shock surface as seen in panels (a) and (b). 

\subsubsection{2D Spherical Distribution} \label{sec:Result4sep1_2D}

\begin{figure*}[tp!]
\centering {\includegraphics[width=0.95\hsize]{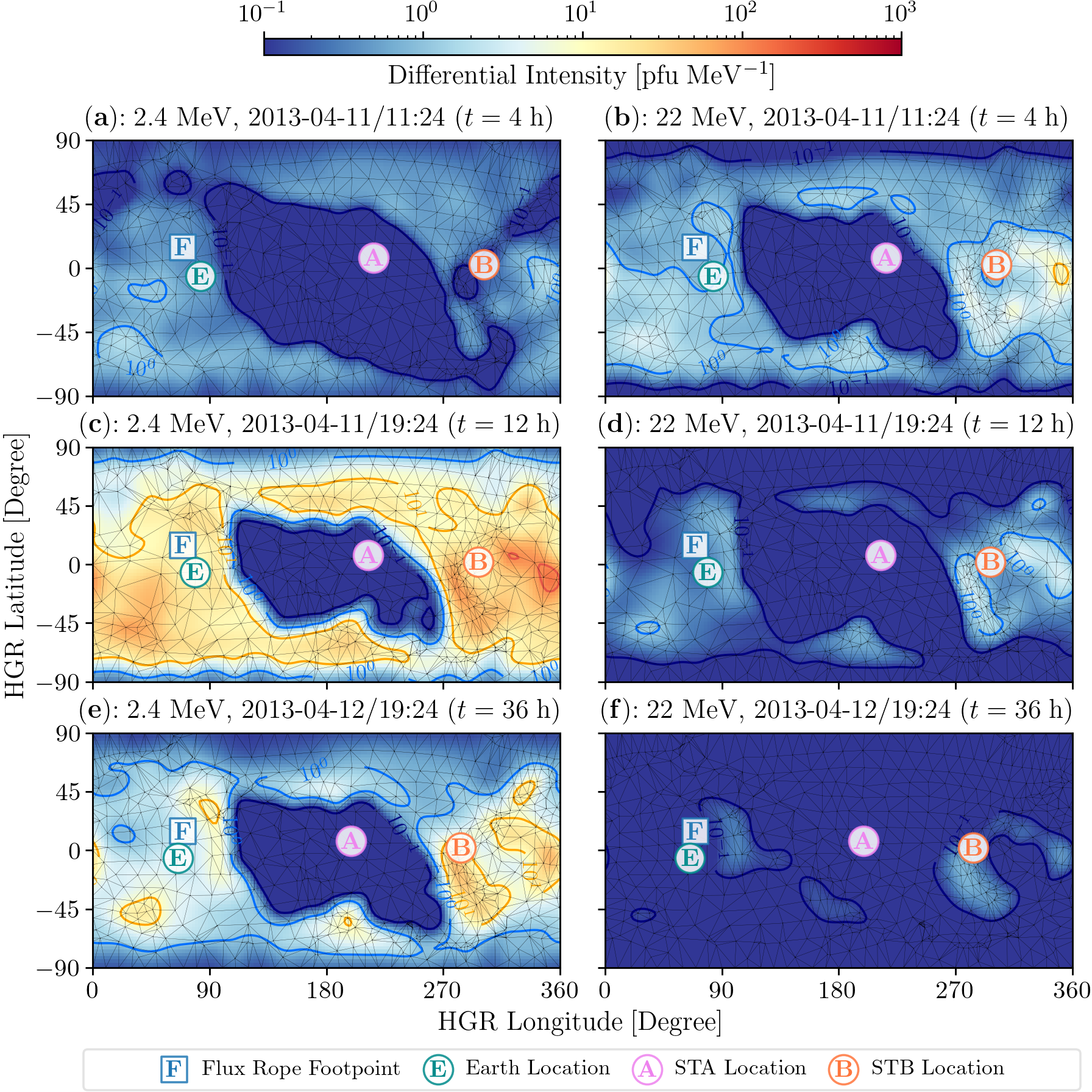}} % png eventually
\caption{The 2D distribution of the differential intensity at 2.4 MeV (left column) and 22 MeV (right column) on the $1\;$au sphere is shown, at 4 hours (upper row), 12 hours (middle row) and 36 hours (lower row) after the CME eruption. In each panel, $x$-axis and $y$-axis are the HGR longitude and latitude. The flux rope footpoint on the solar surface is marked as ``F" in a square in blue, and the locations of Earth, STA and STB are plotted as ``E", ``A" and ``B", in a circle in green, pink and orange, respectively.} \label{fig10:contour}
\end{figure*}

% Overview
Figure \ref{fig10:contour} shows the two-dimensional (2D) distribution of the energetic proton differential intensity on a logarithmic scale at 4 hours (top row), 12 hours (middle row) and 36 hours (bottom row) after the launch of the CME flux rope. Two energy channels are shown, 2.4 MeV (left column) and 22 MeV (right column), corresponding to lower- and higher-energy protons. The $x$- and $y$-axes represent the HGR longitude and latitude on a sphere at $1\;$au. Earth, STA and STB locations are marked by letters ``E", ``A" and ``B" within green, pink and orange circles, respectively. The location of the inserted flux rope on the Sun is marked as a blue square with the letter ``F", showing the relative location of observers with respect to the flux rope (also see Figure \ref{fig3:overview}(a)). A similar plot but for $\geqslant10$ MeV integral flux and with a different computational scheme is shown in Figure 5 of \cite{zhao2024solar}. 

% Triangulation
Note that the traced 648 magnetic field lines in M-FLAMPA are evenly distributed on the $2.5\;R_\mathrm{s}$, but not evenly distributed over the sphere at $1\;$au because of the inhomogeneous magnetic fields in the simulation domain. Therefore, we apply the Delaunay triangulation method \citep{delaunay1934sphere, lee1980two} to construct a skeleton of the sphere at $1\;$au, which uses a set of points to effectively divide the plane into multiple triangular cells and tends to avoid the formation of narrow or sliver triangles. In Figure \ref{fig10:contour}, the vertices indicate where the field lines intersect with the $1\;$au sphere, and the edges illustrate the skeletal representation of the $1\;$au sphere derived via Delaunay triangulation. With the skeleton and differential intensity values at each vertex, we interpolate the intensity across the entire $1\;$au sphere. In each panel of Figure \ref{fig10:contour}, the contours are plotted to show the structure of the distribution function. 

Furthermore, comparing the distributions at energies of 2.4 MeV (left column) with 22 MeV (right column) in Figure \ref{fig10:contour}, we see that the higher-energy protons generally arrive at the $1\;$au surface earlier than the lower-energy protons. We can also observe the phases of increase, peak and decay for the differential intensity distribution at 2.4 MeV in Figure \ref{fig10:contour}(a)(c)(e). 

% Distribution in longitude
Distinct variations in energetic proton intensities can also be found across longitudes and latitudes in Figure \ref{fig10:contour}. For instance, in all 3 time slices (4, 12 and 36 hours after the flux rope eruption), the differential intensity around STA is orders of magnitude lower than that at Earth and STB, if any. This agrees with the observations in Figure \ref{fig3:overview} and is consistent with the simulation results shown in Figure \ref{fig9:z0Cut}, which shows no SEP fluxes in the regions near STA due to the lack of magnetic connectivity (see Section \ref{sec:Result3shk}) and the absence of perpendicular diffusion in our model. Hereafter, we focus on the particles observed at Earth and STB. 
Since the IMF follows the Parker spiral in general \citep[e.g.,][]{xie2019statistical, zhao2019stat}, the SEP flux typically concentrates around $40^{\circ}$–$80^{\circ}$ east of the flux rope location, which also depends on the corona and IMF configurations \citep[e.g.,][]{lario2006radial, richardson201425, richardson2018prediction, paassilta2018catalogue}. As shown in Figure \ref{fig10:contour}(b)(c), the peak intensity at both energies occurs between $315^{\circ}$ and $360^{\circ}$ in our simulations, which is $70^{\circ}$–$115^{\circ}$ east of the flux rope location. 
Note that in our simulation, we assume a uniform injection coefficient throughout the shock front, that is, independent of the shock obliquity, as described in Section \ref{sec:Method3.4}. Therefore, the 2D distribution of the energetic particles reflects the collective effects of the shock strength, as well as the ambient plasma density and temperature. 

\subsubsection{Time--Intensity Profiles} \label{sec:Result4sep2_time}

\begin{figure*}[tp!]
\centering {\includegraphics[width=0.98\hsize]{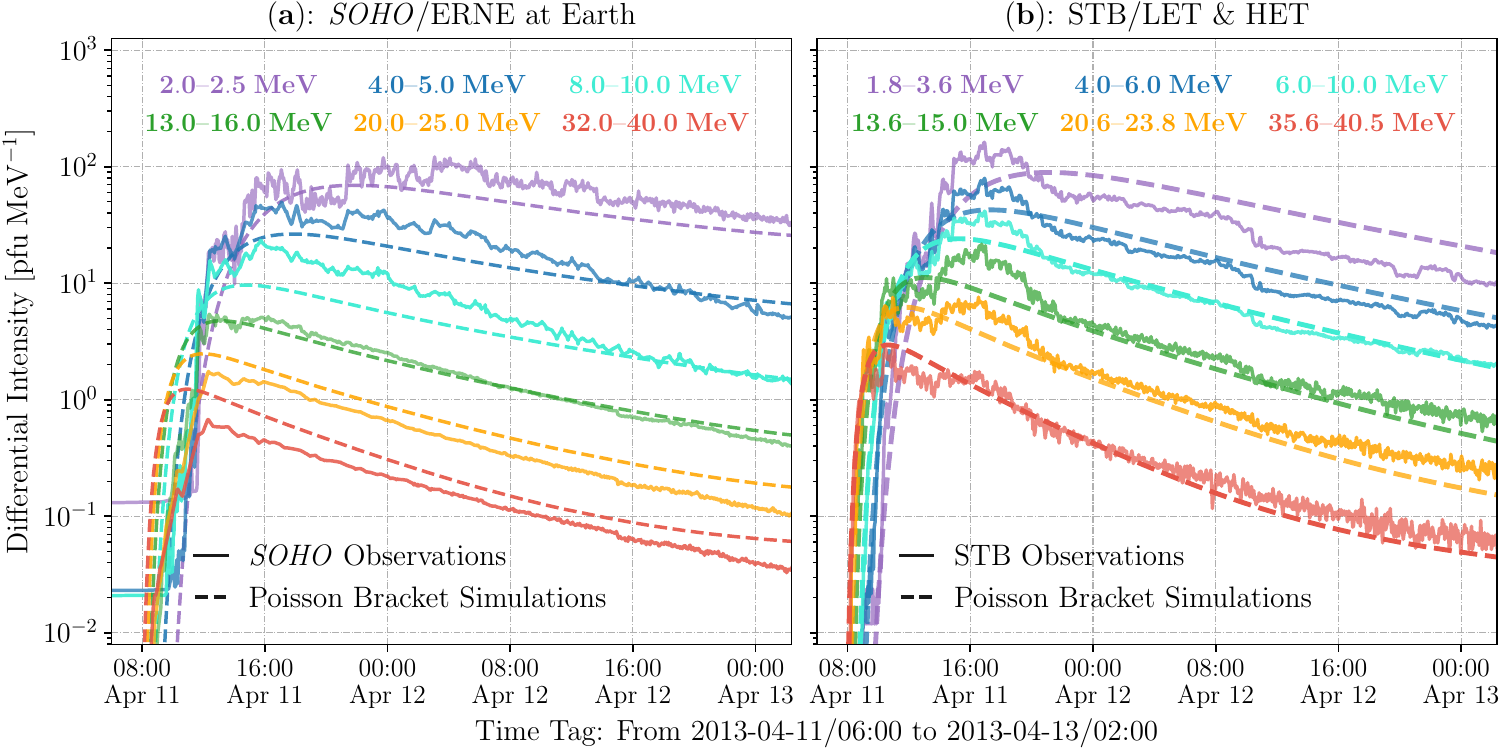}} % V3: w/n STA; V4: w/ STA
\caption{Comparison of the time intensity profiles between simulation results and observations across six energy channels that range from $\sim$2 MeV to $\sim$40 MeV. These energy channels are chosen to match with each particle instrument and they are slightly different for each instrument. The observations are plotted in solid curves, while the calculated intensities are plotted in dashed curves. Panel (a) shows the comparison with \textit{SOHO}/ERNE observations and panel (b) shows the comparison with STB/LET and HET. The ranges of $x$-axis and $y$-axis are the same for panels (a) and (b).} \label{fig11:dfluxtime}
\end{figure*}

% Descriptions of Time Profiles
With the triangulation method described in Section \ref{sec:Result4sep1_2D}, we interpolate the differential intensities at Earth and STB. Each panel of Figure \ref{fig11:dfluxtime} presents the calculated differential intensities across six energy channels, which we compare with particle measurements from \textit{SOHO}/ERNE, and STB/LET and HET, respectively. These six energy channels for each spacecraft are chosen between $\sim$2 MeV and $\sim$40 MeV and are shown in different colors. In Figure \ref{fig11:dfluxtime}, the simulation results using the Poisson bracket scheme are plotted in dashed lines, while the observational data are presented in solid lines. 
Overall, our model reproduces the time profiles across the six energy channels plotted for both \textit{SOHO} and STB, with discrepancies within roughly half an order of magnitude. The calculated intensities are slightly higher at STB with a faster decay phase than at Earth. Given that the scaling factor is set to be the same in the entire simulation domain, this discrepancy between STB and Earth is due to both the number of particles injected into the shock system and the different time-evolving shock properties at the cobpoint related to STB and Earth, as illustrated in Figures \ref{fig7:shocksurf}, \ref{fig8:shockprop} and \ref{fig10:contour}. 

% Details: Onset and Decay, for each energy channel in SOHO/ERNE and STB/LET & HET
In addition, for both \textit{SOHO} and STB, the onset times in the simulations agree with the observations in all six energy channels and a clear velocity dispersion is shown. 
For \textit{SOHO}/ERNE, the onset phases are comparable between simulations and observations in the two intermediate energy channels (8.0–10.0 and 13.0–16.0 MeV). 
However, in the two higher-energy channels (20.0–25.0 and 32.0–40.0 MeV), the SEP peak intensities arrive about 2 hours earlier in simulations than in observations. In the two lower-energy channels (2.0–2.5 and 4.0–5.0 MeV), the SEP peak intensities arrive slightly later in simulations than in observations. These differences may be due to the uncertainties in modeling the evolution of the Gibson-Low flux rope and the shock properties to which \textit{SOHO}/ERNE is magnetically connected. 

In terms of the time profiles in STB/LET and HET, there is an irregular structure in particle measurements around 17:00 UT on April 11 in Figure \ref{fig11:dfluxtime}(b), which is not reproduced by the simulations. In addition to this short-term irregularity, the onset phases in simulations are comparable to those in observations across the six energy channels, except for the lower-energy channel (1.8–3.6 MeV), in which the calculated peak intensities are about 4 hours later than observed. 
Furthermore, the decay phases across all six energy channels for both \textit{SOHO} and STB show a strong concordance between simulations and observations, within a factor of $\sim$2. 

\subsubsection{Energy Spectrum} \label{sec:Result4sep3_spec}

\begin{figure}[tp!]
\centering {\includegraphics[width=0.98\hsize]{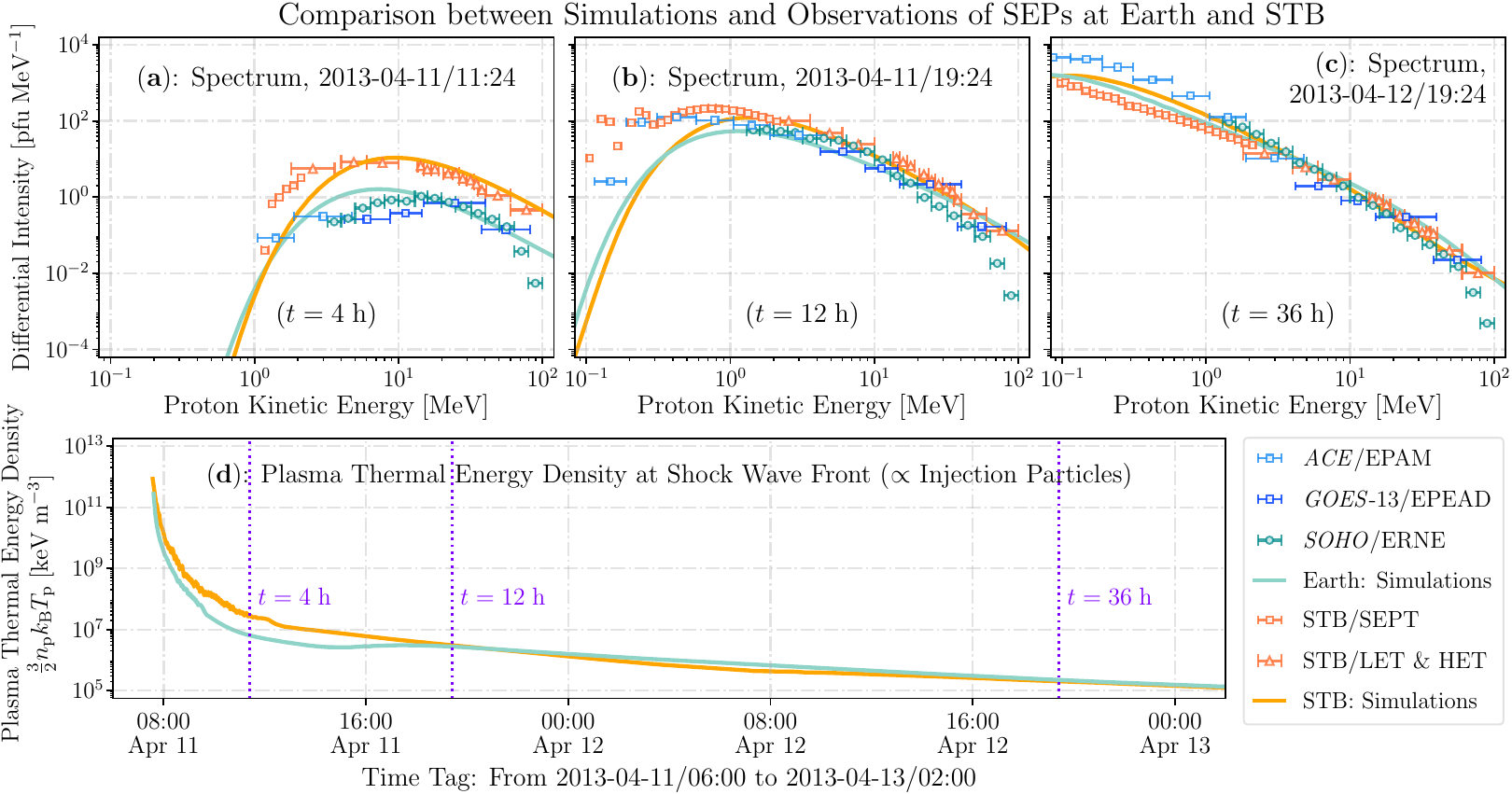}}
\caption{Comparison between simulations and observations of SEPs at Earth and STB. Panels (a)–(c) present the SEP spectra at Earth and STB at three time intervals after the CME flux rope eruption: (a) 4 hours, (b) 12 hours, and (c) 36 hours. In each panel, observational data are plotted as scattered points, while simulation results are shown as curves. Data from \textit{ACE}/EPAM are plotted as light blue squares. Data from \textit{GOES}-13/EPEAD are denoted by deep blue squares, with the so-called ``effective energies" calibrated by \cite{bruno2017calibration}. Measurements from \textit{SOHO}/ERNE are marked as green circles, and the simulated spectra at Earth are plotted in green. Similarly, orange squares represent data from STB/SEPT, orange triangles indicate data from STB/LET and HET instruments, and orange curves represent the simulated spectra at STB. Panel (d) shows the time profile of the ambient plasma thermal energy density at the shock wave front, using the same legends as in panels (a)–(c). Additionally, three purple vertical dotted lines indicate the moments of 4, 12 and 36 hours after the CME eruption, respectively.} \label{fig12:spectrum}
\end{figure}

% We further investigate the time-evolving energy spectrum. 
Panels (a)–(c) of Figure \ref{fig12:spectrum} depict the energy spectra at 4, 12 and 36 hours after the launch of the CME flux rope, respectively. In each panel, simulation results are plotted in curves, while observational data are shown as scattered points where the energy bin widths are indicated by horizontal bars. We subtract the background fluxes for all the particle measurements shown here. The \textit{SOHO}/ERNE data are marked as green circles, and the simulated spectra at Earth are plotted in green lines. Note that the \textit{SOHO}/ERNE data are subject to saturation effects, which may lead to inaccuracies in particle counts at high flux levels, especially at high energies \citep{miteva2018solar, miteva2020flux, kuhl2019revising}. Therefore, we also show particle measurements from the Energetic Proton, Electron and Alpha Detector \citep[EPEAD,][]{onsager1996operational, sellers1996design} on board \textit{GOES}-13, which are plotted in deep blue squares with energy bins calibrated by \cite{bruno2017calibration}. Measurements from STB/LET and HET instruments are marked as orange triangles, and the simulated spectra at STB are plotted in orange lines. At low-energy ranges, we include data from the Low-Energy Magnetic Spectrometer 120 (LEMS120) of the \textit{Electron, Proton, and Alpha Monitor} \citep[EPAM,][]{gold1998electron} on board \textit{ACE} for near-Earth observations, and data from the \textit{Solar Electron and Proton Telescope} \citep[SEPT,][]{muller2008solar} on board STB, marked as light blue squares and orange squares in Figure \ref{fig12:spectrum}, respectively. Note that the \textit{ACE}/EPAM/LEMS120 does not distinguish between ion species, and we assume that its measured intensities are dominated by protons and the contribution of heavier ions is small compared to that of protons \citep[e.g.,][]{marhavilas2015survey, lario2018flat}. The STB/SEPT has four direction-dependent channels (sunward, anti-sunward, north and south) and the observational data are averaged across these directional channels. 

Overall, the simulation results are comparable to the particle measurements, particular for energies above $1\;$MeV. Notably, the simulations capture the spectral shape during the onset phases and at later times with high fidelity. However, for low-energy protons, contamination from high-energy particles significantly impacts the \textit{ACE}/EPAM/LEMS120 and STB/SEPT measurements, especially during the onset phase of SEP events \citep[e.g.,][and references therein]{haggerty2006qualitative, malandraki2009energetic, marhavilas2015survey, morgado2015low, lario2018flat, brudern2022new}. As a result, only a limited number of energy channels in \textit{ACE}/EPAM/LEMS120 and STB/SEPT provide valid measurements at $t=4\;$h in Figure \ref{fig12:spectrum}(a), and we also see a fluctuating spectrum observed by STB/SEPT at $t=12\;$h in Figure \ref{fig12:spectrum}(b). 
Furthermore, from the time--intensity profiles in Figure \ref{fig11:dfluxtime}, we can find some short-term irregular structures at energies below 5 MeV in observations, which can also contribute to the differences in model--data comparison at low energies and at specific time in Figure \ref{fig12:spectrum}. 
% Furthermore, \textit{ACE}/EPAM/LEMS120 has its orientation at the specific time \citep{gold1998electron}, as its FOV is limited to particles arriving from certain directions, while our model averages out the directional dependence of particle distributions (cf. Eq.~(\ref{eqn:ParkerEqn})). This directional bias from \textit{ACE}/EPAM/LEMS120 also contributes to the model--data differences in Figure \ref{fig12:spectrum}. 
From the simulation perspective, the differences between the observed and simulated spectra at low energies (e.g., Figure \ref{fig12:spectrum}(b)(c)) underscore the challenges in complete accurate modeling the acceleration and transport of low-energy particles. 

Here, we discuss the spectrum differences between Earth and STB. 
In Figure \ref{fig12:spectrum}(a), the particle intensity at STB is about half an order of magnitude higher than at Earth, 4 hours after the flux rope eruption. The SEP intensities at these two locations become similar at later times, as illustrated in panels (b) and (c) of Figure \ref{fig12:spectrum}. To explore these variations in SEP flux levels, we examine the time-evolving properties on the shock surface. According to Eq.~(\ref{eqn:DistInj}), the number of particles injected at the shock wave front is proportional to the ambient plasma thermal energy density (per volume), calculated as $\frac{3}{2} n_\mathrm{p} k_\mathrm{B} T_\mathrm{p}$. 
Figure \ref{fig12:spectrum}(d) shows the time evolution of the plasma thermal energy at the shock wave front, with the same legends as those used in Figure \ref{fig12:spectrum}(a)–(c) and three purple vertical dotted lines indicating 4, 12 and 36 hours after the CME flux rope eruption, respectively. During the first few hours, the plasma thermal energy is slightly higher (by a factor of $\sim$3) at the cobpoint of STB compared to Earth, due to the properties of the flux rope and ambient solar wind. Later, the plasma thermal energy is similar at the two cobpoints. 
This tendency contributes to the differences at earlier times and similarities at later times regarding the magnitude of SEP fluxes observed at Earth and STB shown in Figure \ref{fig12:spectrum}(a)–(c). Since there is also a diffusion process, particle fluxes at later times are also affected by the earlier time to some extent. 

\begin{figure*}[tp!]
\centering {\includegraphics[width=0.75\hsize]{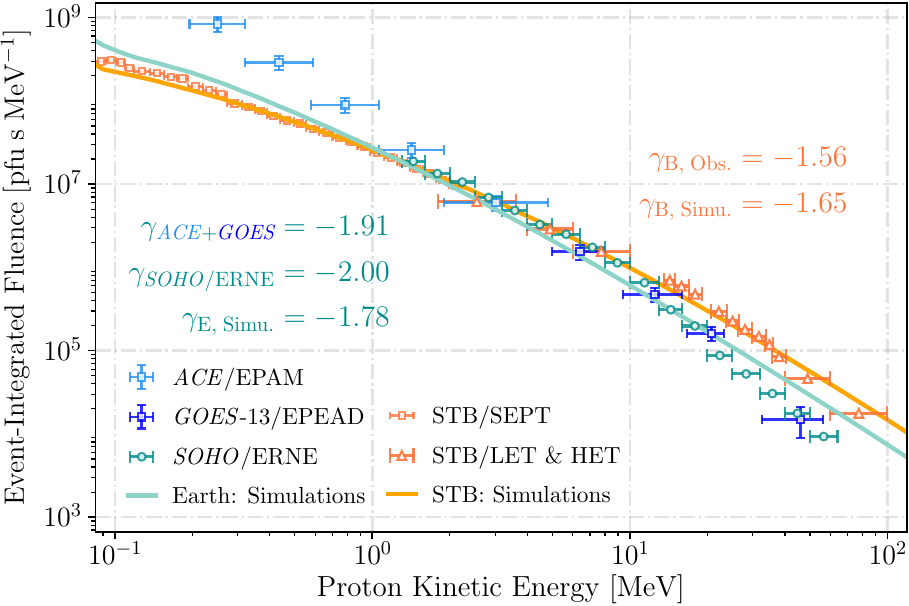}}
\caption{Event-integrated fluence spectrum at Earth and STB, with a similar legend style used in Figure \ref{fig12:spectrum} only except for the vertical bars in the observational data, if shown, indicating the uncertainties of particle measurements. Fluence data from \textit{ACE}/EPAM and \textit{GOES}-13/EPEAD are taken from \cite{bruno2021empirical}. Fitted spectral indices are included for both observations and simulations at Earth and STB in corresponding colors.} \label{fig13:fluence}
\end{figure*}

To evaluate the overall event-integrated fluence, we calculate the SEP intensities over time across multiple energy channels, as illustrated in Figure \ref{fig13:fluence}. The SEP intensities are integrated in the first 3 days of this event. Consistency of the fluence intensity and spectral shape can be found in Figure \ref{fig13:fluence}, especially for the fluence spectrum at STB and the part of $1\;$MeV at Earth. 
We also calculate the spectral index of the fluence spectrum in the energy range from $1.0$ to $50\;\mathrm{MeV}$ for both simulations and observations at Earth and STB. The spectral indices from simulations are consistent with the ones derived from observations. A slightly harder fluence spectrum at STB ($\gamma_\mathrm{B,\,Simu}=-1.65$) than at Earth ($\gamma_\mathrm{E,\,Simu}=-1.78$) is reproduced from our simulations. 
%although the model--data values for each spacecraft are different. %, probably indicating an overall less efficient particle acceleration or more energy loss at Earth than at STB in this event. Note that Figure \ref{fig13:fluence} is the event-integrated fluence spectrum; it reflects the efficiency of particle acceleration and/or energy loss during its transport, throughout the entire SEP event. 

In spite of the good agreement in model--data comparisons, we can tell noticeable differences at energies $\lesssim1$ MeV between simulations and observations in both time-evolving spectra shown in Figure \ref{fig12:spectrum}(a)–(c) and fluence spectra shown in Figure \ref{fig13:fluence}. As discussed in the text associated with Figure \ref{fig12:spectrum}, these differences between simulated and observed spectra arise from a combination of factors, including the background solar wind, CME propagation, shock properties, particle acceleration and transport processes, as well as the instrumental effects and some short-period irregularities at low energies (also see Figure \ref{fig11:dfluxtime}). Next, we will discuss how the MFP affects the particle acceleration and transport processes in simulations. 

\subsubsection{Influence of MFPs on SEP Acceleration and Transport} \label{sec:Result4sep5_mfp}
% MFP: Last but not least, we xxx
Based on the DSA mechanism, the diffusion coefficient plays a critical role in the acceleration and transport processes of energetic particles. In M-FLAMPA, only the parallel diffusion and MFP are considered (see Section \ref{sec:Method3.1}). In the following, we further investigate and discuss the influence of the parallel MFP estimation, which is derived from QLT but manipulated differently in the upstream and downstream shock regions (see more details in Section \ref{sec:Method3.3}). 

\begin{figure*}[tp!]
\centering {\includegraphics[width=0.9\hsize]{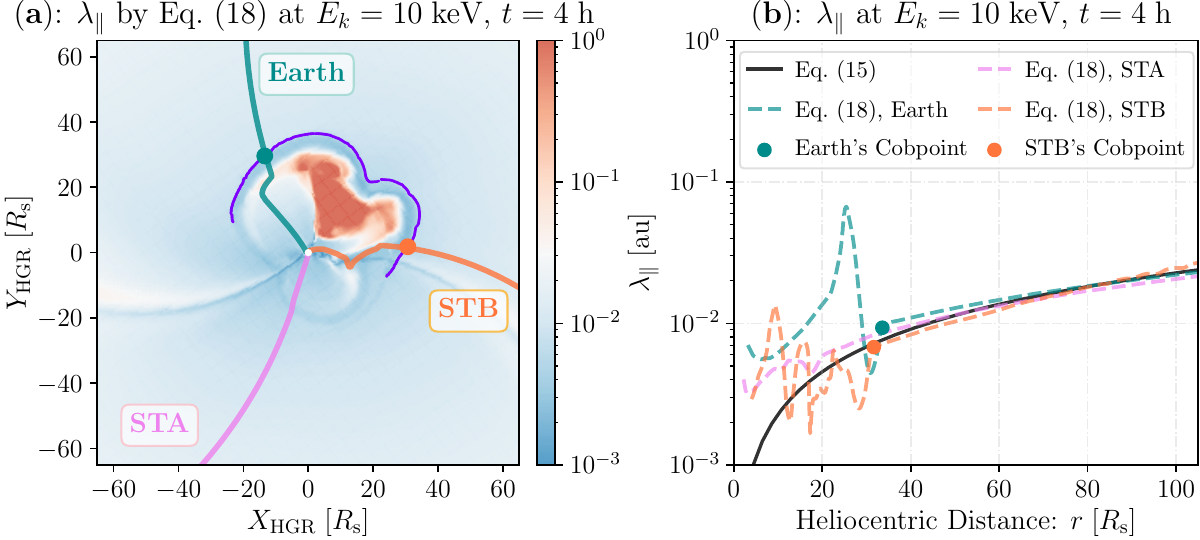}}
\caption{The parallel MFP $(\lambda_{\parallel})$ for the injected protons at 10 keV, 4 hours after the CME flux rope eruption. (a): $\lambda_{\parallel}$ in the solar equatorial plane ($Z_\mathrm{HGR}=0$, viewed from the north pole) using AWSoM-R modeled magnetic field turbulence and following Eq.~(\ref{eqn:lambdaxxDn}). The contour colors saturate for $\lambda_{\parallel}$ values beyond the scope from $10^{-3}$ to $10^0\;\mathrm{au}$. In panel (a), HGR coordinates are used, with the FOV spanning $-65\;R_\mathrm{s}\leqslant X_\mathrm{HGR},\, Y_\mathrm{HGR}\leqslant 65\;R_\mathrm{s}$. The white spot in the center represents the Sun ($1\;R_\mathrm{s}$). The shock front identified by the shock-capturing tool (see Section \ref{sec:Result3shk} and Figure \ref{fig9:z0Cut}(b)) is indicated by the purple markers. Magnetic field lines connecting to Earth, STA and STB are plotted in green, pink and orange, respectively. (b): The comparison of $\lambda_{\parallel}$ for $10\;$keV protons as a function of heliocentric distance. The black solid curve indicates $\lambda_{\parallel}$ calculated by Eq.~(\ref{eqn:lambdaxxUp}), while the dashed curves represent $\lambda_{\parallel}$ calculated by Eq.~(\ref{eqn:lambdaxxDn}) using time-accurate AWSoM-R parameters, plotted in the corresponding colors for each spacecraft as used in panel (a). Cobpoints associated with Earth and STB are shown in both panels. The color bar axis in panel (a) and the $y$-axis of panel (b) share the same label tag.} \label{fig14:lxxUpDn} 
\end{figure*}

% Figure MFP (Up & Dn)
In Figure \ref{fig14:lxxUpDn}, we plot the $\lambda_{\parallel}$ for $10\;$keV injected protons at 4 hours after the CME eruption. Figure \ref{fig14:lxxUpDn}(a) shows the $\lambda_{\parallel}$ calculated using Eq.~(\ref{eqn:lambdaxxDn}) in the equatorial plane, plotted with the shock front, the magnetic field lines connecting to Earth, STA and STB, as well as the cobpoints related to Earth and STB. 
In Figure \ref{fig14:lxxUpDn}(a), the regions with a dramatic increase of $\lambda_{\parallel}$ correspond to the disruptions of the flux rope to the background solar wind, and the regions with smaller $\lambda_{\parallel}$ ahead of the flux rope are where the particles are accelerated effectively. 
In Figure \ref{fig14:lxxUpDn}(b), we plot the $\lambda_{\parallel}$ at the injection energy of 10 keV for protons, calculated by Eqs.~(\ref{eqn:lambdaxxUp}) and (\ref{eqn:lambdaxxDn}) along the field lines connecting to Earth, STA and STB. In the region far upstream of the shock, i.e., in the background solar wind, we can see good agreement using both approaches. 
% The difference of $\lambda_{\parallel}$ from the low solar corona to the shock front is likely to be attributed to the distortion of the flux rope and the shock. It is understood that in the DSA mechanism, particles in the downstream region diffuse around the shock front due to magnetic turbulence. This slightly smaller estimation of downstream $\lambda_{\parallel}$ implies stronger scattering effects and can lead to higher acceleration efficiency of particles, producing higher intensities of high-energy particles as displayed in Figures \ref{fig12:spectrum} and \ref{fig13:fluence}. 

% Upstream MFP
Note that in the upstream region, the free parameter $\lambda_0$ in Eq.~(\ref{eqn:lambdaxxUp}) is chosen to be $0.3\;\mathrm{au}$ in order to match the $D_{\parallel}$ based on long-term \textit{PSP} solar wind turbulence observations \citep{chen2024parallel}, as shown in Figure \ref{fig2:DxxUp}. Even though the results in Figure \ref{fig2:DxxUp} are overall consistent, there are still discrepancies for $D_{\parallel}$, especially at a small heliocentric distance. 
% Note that the Alfv\'en wave turbulence produced in the vicinity of a shock-wave front have been demonstrated to have important consequences for SEP elemental abundance variations \citep{ng1999effect, tylka1999observations, sandroos2007simulation} and the evolution of SEP anisotropies \citep{reames2001angular, petrosian2012stochastic}. 
% In the downstream (post-shock) region, the parallel MFP is calculated using the Alfv\'en wave turbulence strength simulated by AWSoM-R. 
% Moreover, the upstream MFP can significantly influence the particle transport process, as reflected in the SEP time profiles and spectra \citep[e.g.,][]{qin2006effect, strauss2019solar, niemela2023advancing, zhong2024mean, wang2024statistical}. In the previous sections, we set the free parameter for the upstream MFP ($\lambda_0$ in Eq.~(\ref{eqn:lambdaxxUp})) as $0.3\;\mathrm{au}$. 
% Although the parallel diffusion coefficient with $\lambda_0=0.3\;\mathrm{au}$ in Equation \ref{eqn:calcDxx} based on QLT has been validated against Eq.~(\ref{eqn:DxxPSP}) taken from \cite{chen2024parallel} as shown in Figure \ref{fig2:DxxUp}, 
% Despite this free parameter validation, many assumptions about the dependence of $D_{\parallel}$ on the radial distance have been made as $D_{\parallel}\propto r^\alpha$ in many SEP transport models with $\alpha$ varying from 0 to 2 \citep[e.g.,][]{qin2004interplanetary, zhang2009propagation, droge2010anisotropic, giacalone2020solar}. 
For long-term solar wind magnetic field turbulence observations, although the results of \cite{chen2024parallel} are comparable to previous studies such as \cite{moussas1992mean} and \cite{erdHos2005situ}, there are still various factors that can influence the background turbulence strength and the estimation of $D_{\parallel}$, such as the IP transients \citep[e.g.,][and references therein]{desai2016large, pitvna2021turbulence} and switchbacks \citep[e.g.,][]{de2020switchbacks, shoda2021turbulent}. 
%As stated in \cite{chen2024parallel}, the manipulations for estimating $D_{\parallel}$ from turbulence observations can also be optimized by taking into account the magnetic field turbulence anisotropy \citep[e.g.,][]{bieber1994proton, matthaeus2003nonlinear, shalchi2009analytical, bandyopadhyay2021geometry}. 
As a result, the optimal value $\lambda_0$ for the upstream MFP may vary from event to event in modeling historical SEP events. 

\begin{figure*}[tp!]
\centering {\includegraphics[width=0.99\hsize]{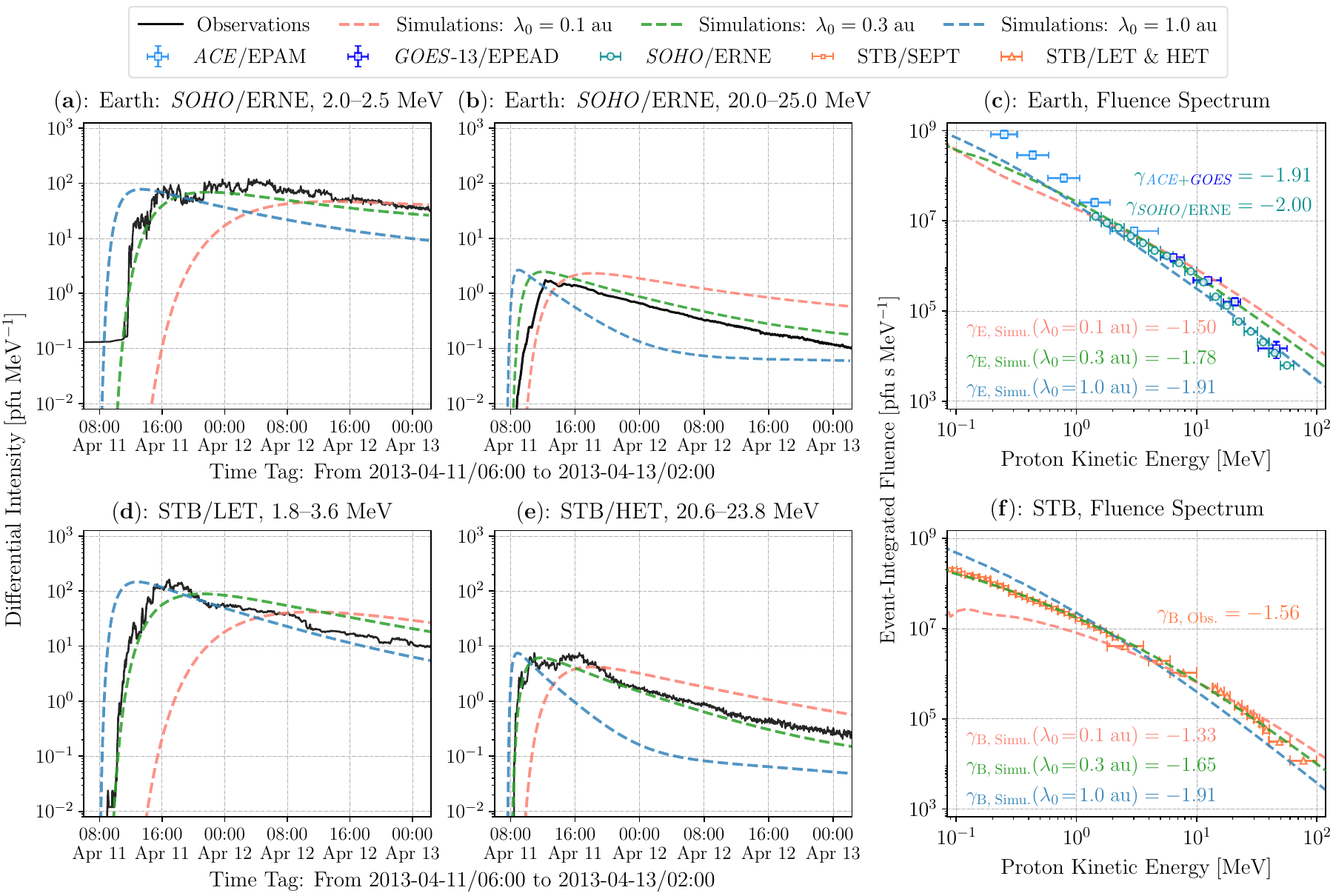}}
\caption{Calculated time--intensity profiles and energy spectra with different MFPs. (a)(b): The SEP time--intensity profiles at Earth in the energy ranges 2.0–2.5 MeV and 20.0–25.0 MeV, representing comparatively lower and higher energies, respectively. (c): Comparison of fluence spectra at Earth. (d)(e): The SEP time--intensity profiles at STB in the energy ranges 1.8–3.6 MeV and 20.6–23.8 MeV, corresponding to comparatively lower and higher energies, respectively. (f): Comparison of fluence spectra at STB. 
In each panel, simulation results are shown for three different MFP values, with $\lambda_0$ in Eq.~(\ref{eqn:lambdaxxUp}) set to 0.1, 0.3 and $1.0\;\mathrm{au}$, represented as red, green and blue dashed curves, respectively. 
Observational data are plotted as solid black lines for time series in panels (a), (b), (d) and (e), and marked as scattered points for the fluence spectrum in panels (c) and (f). The observational data in panels (c) and (f) follow the same legend style as used in Figure \ref{fig13:fluence}. 
Fitted spectral indices of the fluence spectrum are included for both observations and simulations using different MFPs at Earth and STB, displayed in the corresponding colors for the $\lambda_0$ value.} \label{fig15:diffmfp}
\end{figure*}

For this event, we also explore the variations in SEP time--intensity profiles and fluence spectra caused by different values of $\lambda_0$. We set $\lambda_0$ in Eq.~(\ref{eqn:lambdaxxUp}) to be 0.1, 0.3 and $1.0\;\mathrm{au}$, and present the time--intensity profiles and fluence spectra for both Earth and STB in Figure \ref{fig15:diffmfp}. 
Panels (a) and (c) show the time--intensity profiles for low-energy protons, with 2.0–2.5 MeV from \textit{SOHO}/ERNE and 1.8–3.6 MeV from STB/LET. Panels (b) and (d) correspond to the high-energy proton profiles, featuring 20.0–25.0 MeV in \textit{SOHO}/ERNE and 20.6–23.8 MeV in STB/HET. The fluence spectra and the fitted spectral indices are shown in panels (e) and (f) for Earth and STB, respectively. 
% In each panel, we show the observational data compared to the simulations for $\lambda_0$ values of 0.1, 0.3 and $1.0\;\mathrm{au}$, in which the simulation data are plotted in red, green and blue dashed curves, respectively. Particle measurements are plotted as black solid curves in the time--intensity profiles, and represented by scattered points following the legend style in Figure \ref{fig13:fluence}. Fluence spectral indices are included for both observations and simulations with different MFPs, written in the corresponding colors in panels (c) and (f). 

The SEP time--intensity profiles in Figure \ref{fig15:diffmfp}(a)(b)(d)(e) show that not only the absolute flux level, but also the onset and decay phases are sensitive to the MFP. Similarly, simulation results in Figure \ref{fig15:diffmfp}(c)(f) demonstrate that different MFPs lead to a softer or harder fluence spectrum at lower- and higher-energy ends. While the simulations with $\lambda_0=0.3\;\mathrm{au}$ achieve the best agreement with the observed time--intensity profiles, discrepancies still remain in the spectral index compared to observations at both Earth and STB. These comparisons reveal the fine tuning the upstream MFP parameter for this event in our model and also highlight the significance with respect to the transport process of SEPs. 
In fact, the transport of SEPs in the IP medium involves a range of different physical processes, including magnetic focusing, adiabatic cooling, drift, and parallel and perpendicular diffusion \citep{northrop1963adiabatic, roelof1969propagation, skilling1971cosmic, kota1997energy, prinsloo2019acceleration, wang2024statistical}. All of these processes depend intricately on the properties of the solar wind plasma speed and magnetic fields. The magnetic turbulence in the solar wind, for example, can influence the timing of the first arriving particles, the timing when the particle flux crosses a pre-set threshold \citep[e.g.,][]{qin2005model, wang2015simulations}, and also the event-integrated energy spectral index \citep[e.g.,][]{zhao2016double, zhao2017effects}. 
Discrepancies in model--data comparisons as shown in Sections \ref{sec:Result4sep2_time}–\ref{sec:Result4sep5_mfp} are consequences of the physical processes of SEPs mentioned above, as well as the uncertainties in modeling the background solar wind, CME flux rope and shock properties. 
%Our results in Figures \ref{fig2:DxxUp}, \ref{fig14:lxxUpDn} and \ref{fig15:diffmfp} emphasize the importance and necessity of transporting energetic particles within a solar wind solution derived from self-consistent MHD simulations, where the interplay of these processes can be accurately captured and studied. 

% \section{Discussions} \label{sec:discussion}

% The results presented in Section \ref{sec:Result} describe the background solar wind, CME flux rope, shock properties, as well as particle acceleration and transport processes. In what follows, we discuss each of these aspects from the numerical modeling perspective.

\section{Summary and Conclusions} \label{sec:Summary}

As SEPs can pose significant radiation hazards in space, it is crucial to understand their underlying physics including the particle acceleration and transport processes. Achieving this understanding requires not only a comprehensive analysis of observed SEP events but also the development of models and tools capable of capturing these complex processes for SEPs. %Numerical modeling based on first principles can offer valuable insights into SEPs, providing synthetic observables such as the WL images of the CME, SEP time profiles and spectra. 
In this work, we focus on the methodological advancements by developing a shock-capturing tool and implementing the Poisson bracket into the SWMF/SOFIE. To demonstrate the capability of the model development, we utilize them to investigate a historical SEP event on April 11, 2013, which is characterized by the absence of significant SEP fluxes at STA and faster SEP onset at STB compared to Earth. %, as demonstrated in Section \ref{sec:Event}. 

Our study begins with steady-state solar wind simulations driven by the stream-aligned AWSoM-R model, using the hourly updated GONG magnetogram. %, as processed in Section \ref{sec:Method1}. 
The simulations yield reasonable solar wind solutions, which show the magnetic field aligned with solar wind plasma streams and are validated against multi-point EUV observations shown in Figure \ref{fig4:backgroundsw}. %, as shown in Section \ref{sec:Result1sw}. 
Although the small-scale structures on the solar surface are not fully captured, the steady-state solutions show reasonably well open fluxes from the CHs and good agreement for the average brightness, scales and locations of large-scale structures. 
These solar wind solutions establish a suitable background through which CME and SEPs propagate. % suggest the readiness 

We then simulate the CME by placing a force-imbalanced GL flux rope with EEGGL-derived magnetic configurations within the source AR. %, as stated in Section \ref{sec:Method2}. 
The propagation of the CME flux rope is %shown in Section \ref{sec:Result2cme}, 
validated through the WL image comparisons (see Figure \ref{fig6:WLimage}) with multi-point observations from \textit{SDO}/AIA as well as COR1 on board STA and STB. As the CME propagates into the SC and IH domains, it interacts with the ambient solar wind, producing fast-mode shock fronts where particles are accelerated effectively. A shock-capturing tool has been developed to study the properties of CME-driven shocks starting from the low solar corona. %, as illustrated in Section \ref{sec:Result3shk}. 
Using the AMR technique and the speed jump ($\Delta U$ value) criterion, our shock-capturing tool shows the complex shock dynamics with high spatial resolutions. The shock surface extracted using the shock-capturing tool is not only asymmetric and non-uniform but also consists of three spheres (see Figure \ref{fig7:shocksurf}), primarily due to the non-uniform solar wind properties in the solar corona and the deformation of the flux rope. 

After identifying the shock front, we find that there is no magnetic connectivity to the shock for STA during this event, accounting for the absence of noticeable enhancements in SEP intensity at STA. Moreover, we plot the time-evolving shock properties in Figure \ref{fig8:shockprop}, demonstrating that the magnetic connectivity to the shock wave front is established earlier for STB than for Earth, along with a higher compression ratio at the onset phase. Our simulation results about the shock help understand how this complex shock surface affects the particle acceleration process and explain the observed differences of SEP behaviors, %as stated in Section \ref{sec:Event}, 
further underscoring the versatility and potential of the shock-capturing tool for simulating intricate CMEs. 

In order to accurately solve the kinetic equation governing the acceleration and transport processes of energetic particles, we have implemented the \cite{sokolov2023high} Poisson bracket scheme, which conserves the particle number and maintains the TVD property, into SOFIE/M-FLAMPA. We describe the formulation of the governing equation, the scheme implementation and the setup of the free parameters in detail. %in Section \ref{sec:Method3}. In Section \ref{sec:Result4sep},
Then, we show varieties of synthetic observables, including the SEP fluxes in plane cut, 2D particle intensity distribution in longitude and latitude on the $1\;$au sphere, time--intensity profiles and energy spectra, showcasing the integration of the Poisson bracket scheme and also providing insights into understanding the SEP behaviors. Moreover, we present the effects of the parallel MFP on the SEP time--intensity profiles and fluence spectra by choosing different upstream MFP parameters. Differences between the simulations and particle measurements arise from a combination of the instrumental effects, short-period structures, and uncertainties in modeling the background solar wind, CME propagation, shock properties, magnetic connectivity, and particle acceleration and transport processes. 

% Despite these advancements, challenges still remain with respect to the comparison between simulations and observations. In Section \ref{sec:discussion01_sw}, we discuss the mismatched results and uncertainties that may have potential impacts on SEP behaviors. In Section \ref{sec:discussion02_cmeshk}, we highlight the role of the flux rope initialization mechanism, the shock complexity, as well as their mutual effects on the determination of the scaling factor for SEP fluxes. Moreover, we examine the Alfv\'en wave turbulence obtained from AWSoM-R and the corresponding simulated parallel MFP in Section \ref{sec:discussion03_mfp}. We find that the overestimation of particle acceleration at high energies in our model is likely attributed to the underestimation of the parallel MFP in the downstream Alfv\'en wave turbulence, as well as the omission of finite acceleration time. We further study the variations of the SEP time--intensity profiles and fluence spectrum with different upstream MFPs. Note that the SEP onset phase contains competing processes between the continuous acceleration of protons and the diffusion process. The discussion in Section \ref{sec:discussion03_mfp} demonstrates the complexity of MFP studies and their implications on SEPs \citep[e.g.,][]{bieber1994proton, lang2024detailed}. 

In summary, this study represents an advancement of SWMF/SOFIE in SEP modeling by integrating newly developed methodologies and demonstrating their applications. The implementation of the Poisson bracket scheme within the SWMF enables high-resolution simulations for SEPs, and the shock-capturing tool facilitates understanding how the shock properties affect the particle acceleration process. These methodological developments, combined with a variety of synthetic observables, facilitate the elucidation of the underlying physics during the 2013 April 11 SEP event, enhancing our knowledge of SEPs about their acceleration and transport processes.  
% Looking ahead, the tools and methodologies presented here hold great promise to investigate key scientific questions \citep{anastasiadis2019solar, pevtsov2023long, guo2024particle} and support operational capabilities \citep{vourlidas2023nasa, georgoulis2024prediction, zheng2024overview}, ultimately enhancing our knowledge of heliophysical processes relevant to SEPs, improving radiation risk assessments, and contributing to safer space exploration in the future. 

\section*{Acknowledgments}

% The authors sincerely thank the anonymous reviewer for his/her time and effort in helping us improve this paper. 
This work is supported by NASA Living With a Star (LWS) Strategic Capability project under NASA grant 80NSSC22K0892 (SCEPTER), NASA Space Weather Center of Excellence program under award 80NSSC23M0191 (CLEAR), NASA grant 80NSSC21K1124, NSF ANSWERS grant GEO-2149771 and NASA LWS grant 80NSSC20K1778. 
The authors express their great gratitude to the \textit{SOHO} project, an international cooperation between ESA and NASA, and to the \textit{SDO}, \textit{STEREO}, \textit{ACE}, \textit{GOES} and \textit{PSP} teams. The authors acknowledge the development of StereoCAT\textsuperscript{\ref{url:STEREOCAT}} for CME analysis, as well as the DONKI\textsuperscript{\ref{url:DONKI}}, CDAW\textsuperscript{\ref{url:CDAW}} and OMNI\textsuperscript{\ref{url:OMNI}} datasets. The authors also thank the Virtual Solar Observatory\textsuperscript{\ref{url:vso}} (VSO) project at the National Solar Observatory (NSO), which serves as a research tool allowing scientists to search for the solar and heliospheric physics data. 
This work utilizes data from the NSO Integrated Synoptic Program, which is operated by the Association of Universities for Research in Astronomy, under a cooperative agreement with the National Science Foundation and with additional financial support from the National Oceanic and Atmospheric Administration, the National Aeronautics and Space Administration and the United States Air Force. The GONG\textsuperscript{\ref{url:gong}} network of instruments is hosted by the Big Bear Solar Observatory (USA), High Altitude Observatory (USA), Learmonth Solar Observatory (Australia), Udaipur Solar Observatory (India), Instituto de Astrof\'{i}sica de Canarias (Spain) and Cerro Tololo Inter-American Observatory (Chile). 
Computational resources supporting this work are provided by the NASA High-End Computing (HEC) Program\footnote{\url{https://www.nas.nasa.gov/hecc/}} through the NASA Advanced Supercomputing (NAS) Division at the Ames Research Center. 
Any opinions, findings, conclusions or recommendations expressed in this material are those of the authors and do not necessarily reflect the views of the National Aeronautics and Space Administration.

% Appendix
\setcounter{figure}{0}    
\renewcommand{\thefigure}{A\arabic{figure}}
\setcounter{table}{0}    
\renewcommand{\thetable}{A\arabic{table}}
\setcounter{equation}{0}    
\renewcommand{\theequation}{A\arabic{equation}}
\appendix

\section{Shock Formulation} \label{sec:appendA}

\subsection{Shock Normal and Shock Angle} \label{sec:appendA.1}

As the shock wave front is captured, the direction of the normal to the front, $\hat{\bs{n}}$, can be found using the continuity of the normal component of the magnetic field \cite[e.g., Eq.~(5.86) in][]{jackson1998book} upstream and downstream:
\begin{equation}
    \left(\bs{B}_\mathrm{down} - \bs{B}_\mathrm{up}\right)\cdot\hat{\bs{n}} = 0. \label{eqn:thetaBnCond}
\end{equation}
As far as it concerns the tangential components ($\bs{B}_\mathrm{t}$) of the magnetic field upstream and downstream, the \cite{rankine1870xv}-\cite{hugoniot1887propagation, hugoniot1889propagation} relationships as applied to MHD dictate that $\bs{B}_{t,\,\mathrm{up}}$ and $\bs{B}_{t,\,\mathrm{down}}$ are aligned with the jump in the tangential component of the plasma bulk velocity, $\left(\bs{u}_\mathrm{down} - \bs{u}_\mathrm{up}\right)_{t}$ in the MHD shock waves \cite[see, e.g., the introductory part of Chapter 72 in][]{landau2013electrodynamics}, giving:
\begin{equation}
    \left(\bs{u}_\mathrm{down} - \bs{u}_\mathrm{up}\right)_{t}
    \parallel \bs{B}_{t,\,\mathrm{up}} 
    \parallel \bs{B}_{t,\,\mathrm{down}}
    \parallel \left(\bs{B}_\mathrm{down} - \bs{B}_\mathrm{up}\right)_{t}
    \equiv \left(\bs{B}_\mathrm{down} - \bs{B}_\mathrm{up}\right), \label{eqn:calcBtPara}
\end{equation}
where the last identity accounts for Eq.~(\ref{eqn:thetaBnCond}) to express the alignment direction in terms of the easy-to-calculate difference in the magnetic field \citep[see also][]{lepping1971single, abraham1972determination}. 
Thus, eliminating the projection of the velocity jump, $\bs{u}_\mathrm{down} - \bs{u}_\mathrm{up}$ onto the direction of $\bs{B}_\mathrm{down} - \bs{B}_\mathrm{up}$ gives the jump in the normal velocity: 
\begin{equation}
    \left(\bs{u}_\mathrm{down} - \bs{u}_\mathrm{up}\right)_{n} = 
    \begin{cases}
        \bs{u}_\mathrm{down} - \bs{u}_\mathrm{up}, & \bs{B}_\mathrm{down} = \bs{B}_\mathrm{up}, \\
        \frac{\left[\left(\bs{B}_\mathrm{down} - \bs{B}_\mathrm{up}\right) \times \left(\bs{u}_\mathrm{down} - \bs{u}_\mathrm{up}\right)\right] \times \left(\bs{B}_\mathrm{down} - \bs{B}_\mathrm{up}\right)}{\left(\bs{B}_\mathrm{down} - \bs{B}_\mathrm{up}\right)^2}, & \bs{B}_\mathrm{down} \neq \bs{B}_\mathrm{up}.
    \end{cases} \label{eqn:calcUndnup}
\end{equation}
Finally, the direction of the unit normal to the front pointing from the shock downstream to upstream is given by:
\begin{equation}
    \hat{\bs{n}} = \frac{\left(\bs{u}_\mathrm{down} - \bs{u}_\mathrm{up}\right)_{n}}{\left| \left(\bs{u}_\mathrm{down} - \bs{u}_\mathrm{up}\right)_{n} \right|}. \label{eqn:calcshkn}
\end{equation}
The upstream shock angle, $\theta_{Bn}$, can be derived by measuring the angle between the upstream magnetic field and the shock normal \citep[see more analysis in, e.g.,][]{chao1984determining}:
\begin{equation}
    \theta_{Bn} = \arccos \frac{\left|\bs{B}_\mathrm{up}\cdot\hat{\bs{n}}\right|}{\left|\bs{B}_\mathrm{up}\right|}. \label{eqn:calcThetaBn}
\end{equation}

\subsection{Shock Speed and Fast-Mode Mach Number} \label{sec:appendA.2}

Based on the equation of continuity \citep[e.g., Chapter 1 of][]{landau1987fluid}, the mass flux remains continuous across the shock front:
\begin{equation}
    \rho_\mathrm{down}\left(U_{n,\,\mathrm{down}} - U_\mathrm{shock}\right) = \rho_\mathrm{up}\left(U_\mathrm{shock} - U_{n,\,\mathrm{up}}\right). \label{eqn:contmass}
\end{equation}
Here, $U_\mathrm{shock}$ denotes the shock speed, and $U_{n,\,\mathrm{down}} = \bs{u}_\mathrm{down} \cdot \hat{\bs{n}}$ and $U_{n,\,\mathrm{up}} = \bs{u}_\mathrm{up} \cdot \hat{\bs{n}}$ represent the downstream and upstream flow speed normal to the shock, respectively. The shock speed is then given as:
\begin{equation}
U_\mathrm{shock} = \frac{\rho_\mathrm{down}U_{n,\,\mathrm{down}} - \rho_\mathrm{up}U_{n,\,\mathrm{up}}}{\rho_\mathrm{down} - \rho_\mathrm{up}}, \label{eqn:calcUshk}
\end{equation}
as commonly employed in previous studies \citep[e.g.,][and references therein]{whang1996interplanetary, ding2022modeling, jin2022assessing}. With the shock normal in Eq.~(\ref{eqn:calcshkn}), the shock velocity can be expressed as:
\begin{equation}
    \bs{U}_\mathrm{shock} = U_\mathrm{shock}\, \hat{\bs{n}}. \label{eqn:calcUvec}
\end{equation}
Then, we calculate the fast-mode Mach number ($M_{\text{f}}$), corresponding to the fast magnetosonic wave, by:
\begin{equation}
    M_{\text{f}} = \frac{U_\mathrm{shock}}{V_\mathrm{fms}}, \label{eqn:calcMms}
\end{equation}
with 
\begin{equation}
    V_\mathrm{A} = \frac{B_\mathrm{up}}{\sqrt{\mu_0 \rho_\mathrm{up}}}, \, 
    c_\mathrm{s} = \sqrt{\Gamma \frac{P_\mathrm{up}}{\rho_\mathrm{up}}}, \, 
    V_{\text{fms}} = \sqrt{\frac{1}{2}\left( V_\mathrm{A}^2 + c_\mathrm{s}^2 \right) + \sqrt{\left( V_\mathrm{A}^2 + c_\mathrm{s}^2 \right)^2 - 4V_\mathrm{A}^2 c_\mathrm{s}^2 \cos^2\theta_{Bn}}}, \label{eqn:Vchar}
\end{equation}
where $V_\mathrm{A}$ denotes the upstream Alfv\'en speed, and $\mu_0$ is the vacuum permeability; $c_\mathrm{s}$ denotes the acoustic speed, $\Gamma=5/3$ is the ratio of specific heats, and $P_\mathrm{up}$ denotes the upstream ion thermal pressure; $V_\mathrm{fms}$ denotes the fast magnetosonic speed \citep[e.g., Chapter 69 of][]{landau2013electrodynamics}.

% \clearpage
\bibliography{ref_paper}{}

\begin{thebibliography}{}
\expandafter\ifx\csname natexlab\endcsname\relax\def\natexlab#1{#1}\fi
\providecommand{\url}[1]{\href{#1}{#1}}
\providecommand{\dodoi}[1]{doi:~\href{http://doi.org/#1}{\nolinkurl{#1}}}
\providecommand{\doeprint}[1]{\href{http://ascl.net/#1}{\nolinkurl{http://ascl.net/#1}}}
\providecommand{\doarXiv}[1]{\href{https://arxiv.org/abs/#1}{\nolinkurl{https://arxiv.org/abs/#1}}}

\bibitem[{Abraham-Shrauner(1972)}]{abraham1972determination}
Abraham-Shrauner, B. 1972, Journal of Geophysical Research, 77, 736,
  \dodoi{10.1029/JA077i004p00736}

\bibitem[{Altschuler \& Newkirk(1969)}]{altschuler1969magnetic}
Altschuler, M.~D., \& Newkirk, G. 1969, \solphys, 9, 131,
  \dodoi{10.1007/BF00145734}

\bibitem[{Alvarado-G{\'o}mez {et~al.}(2018)Alvarado-G{\'o}mez, Drake, Cohen,
  Moschou, \& Garraffo}]{alvarado2018suppression}
Alvarado-G{\'o}mez, J.~D., Drake, J.~J., Cohen, O., Moschou, S.~P., \&
  Garraffo, C. 2018, \apj, 862, 93, \dodoi{10.3847/1538-4357/aacb7f}

\bibitem[{Anastasiadis {et~al.}(2019)Anastasiadis, Lario, Papaioannou,
  Kouloumvakos, \& Vourlidas}]{anastasiadis2019solar}
Anastasiadis, A., Lario, D., Papaioannou, A., Kouloumvakos, A., \& Vourlidas,
  A. 2019, Philosophical Transactions of the Royal Society A, 377, 20180100,
  \dodoi{10.1098/rsta.2018.0100}

\bibitem[{Antiochos(2013)}]{antiochos2013helicity}
Antiochos, S.~K. 2013, \apj, 772, 72, \dodoi{10.1088/0004-637X/772/1/72}

\bibitem[{Antiochos {et~al.}(1999)Antiochos, DeVore, \&
  Klimchuk}]{antiochos1999model}
Antiochos, S.~K., DeVore, C., \& Klimchuk, J. 1999, \apj, 510, 485,
  \dodoi{10.1086/306563}

\bibitem[{Aran {et~al.}(2006)Aran, Sanahuja, \& Lario}]{aran2006solpenco}
Aran, A., Sanahuja, B., \& Lario, D. 2006, Advances in Space Research, 37,
  1240, \dodoi{10.1016/j.asr.2005.09.019}

\bibitem[{Armstrong {et~al.}(1985)Armstrong, Pesses, \&
  Decker}]{armstrong1985shock}
Armstrong, T.~P., Pesses, M.~E., \& Decker, R.~B. 1985, Collisionless Shocks in
  the Heliosphere: Reviews of Current Research, 35, 271,
  \dodoi{10.1029/GM035p0271}

\bibitem[{Axford {et~al.}(1977)Axford, Leer, \&
  Skadron}]{axford1977acceleration}
Axford, W., Leer, E., \& Skadron, G. 1977, in International Cosmic Ray
  Conference, Vol.~11, Springer, 132, \dodoi{10.1007/978-3-662-25523-0}

\bibitem[{Bain {et~al.}(2023)Bain, Copeland, Onsager, \&
  Steenburgh}]{bain2023noaa}
Bain, H.~M., Copeland, K., Onsager, T.~G., \& Steenburgh, R.~A. 2023, Space
  Weather, 21, e2022SW003346, \dodoi{10.1029/2022SW003346}

\bibitem[{Bale {et~al.}(2016)Bale, Goetz, Harvey, Turin, Bonnell, Dudok~de Wit,
  Ergun, MacDowall, Pulupa, Andr{\'e}, {et~al.}}]{bale2016fields}
Bale, S.~D., Goetz, K., Harvey, P., {et~al.} 2016, \ssr, 204, 49,
  \dodoi{10.1007/s11214-016-0244-5}

\bibitem[{Band {et~al.}(1993)Band, Matteson, Ford, Schaefer, Palmer, Teegarden,
  Cline, Briggs, Paciesas, Pendleton, {et~al.}}]{band1993batse}
Band, D., Matteson, J., Ford, L., {et~al.} 1993, \apj, 413, 281,
  \dodoi{10.1086/172995}

\bibitem[{Bell(1978{\natexlab{a}})}]{bell1978acceleration1}
Bell, A. 1978{\natexlab{a}}, Monthly Notices of the Royal Astronomical Society,
  182, 147, \dodoi{10.1093/mnras/182.2.147}

\bibitem[{Bell(1978{\natexlab{b}})}]{bell1978acceleration2}
---. 1978{\natexlab{b}}, Monthly Notices of the Royal Astronomical Society,
  182, 443, \dodoi{10.1093/mnras/182.3.443}

\bibitem[{Berger \& Colella(1989)}]{berger1989local}
Berger, M.~J., \& Colella, P. 1989, Journal of Computational Physics, 82, 64,
  \dodoi{10.1016/0021-9991(89)90035-1}

\bibitem[{Bertello {et~al.}(2014)Bertello, Pevtsov, Petrie, \&
  Keys}]{bertello2014uncertainties}
Bertello, L., Pevtsov, A., Petrie, G., \& Keys, D. 2014, \solphys, 289, 2419,
  \dodoi{10.1007/s11207-014-0480-3}

\bibitem[{Blandford \& Eichler(1987)}]{blandford1987particle}
Blandford, R., \& Eichler, D. 1987, Physics Reports, 154, 1,
  \dodoi{10.1016/0370-1573(87)90134-7}

\bibitem[{Blandford \& Ostriker(1978)}]{blandford1978particle}
Blandford, R.~D., \& Ostriker, J.~P. 1978, \apj, 221, L29,
  \dodoi{10.1086/182658}

\bibitem[{Borovikov {et~al.}(2019)Borovikov, Sokolov, Huang, Roussev, \&
  Gombosi}]{borovikov2019toward}
Borovikov, D., Sokolov, I., Huang, Z., Roussev, I., \& Gombosi, T. 2019, arXiv
  preprint arXiv:1911.10165, \dodoi{10.48550/arXiv.1911.10165}

\bibitem[{Borovikov {et~al.}(2018)Borovikov, Sokolov, Roussev, Taktakishvili,
  \& Gombosi}]{borovikov2018toward}
Borovikov, D., Sokolov, I., Roussev, I., Taktakishvili, A., \& Gombosi, T.
  2018, \apj, 864, 88, \dodoi{10.3847/1538-4357/aad68d}

\bibitem[{Borovikov {et~al.}(2017)Borovikov, Sokolov, Manchester, Jin, \&
  Gombosi}]{borovikov2017eruptive}
Borovikov, D., Sokolov, I.~V., Manchester, W.~B., Jin, M., \& Gombosi, T.~I.
  2017, Journal of Geophysical Research: Space Physics, 122, 7979,
  \dodoi{10.1002/2017JA024304}

\bibitem[{Borovikov {et~al.}(2015)Borovikov, Sokolov, \&
  T{\'o}th}]{borovikov2015efficient}
Borovikov, D., Sokolov, I.~V., \& T{\'o}th, G. 2015, Journal of Computational
  Physics, 297, 599, \dodoi{10.1016/j.jcp.2015.05.038}

\bibitem[{Brchnelova {et~al.}(2022)Brchnelova, Ku{\'z}ma, Perri, Lani, \&
  Poedts}]{brchnelova2022or}
Brchnelova, M., Ku{\'z}ma, B., Perri, B., Lani, A., \& Poedts, S. 2022, \apjs,
  263, 18, \dodoi{10.3847/1538-4365/ac8eb1}

\bibitem[{Br{\"u}dern {et~al.}(2022)Br{\"u}dern, Berger, Heber,
  Heidrich-Meisner, Klassen, Kollhoff, K{\"u}hl, Strauss, Wimmer-Schweingruber,
  \& Dresing}]{brudern2022new}
Br{\"u}dern, M., Berger, L., Heber, B., {et~al.} 2022, \aap, 663, A89,
  \dodoi{10.1051/0004-6361/202142761}

\bibitem[{{Brueckner} {et~al.}(1995){Brueckner}, {Howard}, {Koomen},
  {Korendyke}, {Michels}, {Moses}, {Socker}, {Dere}, {Lamy}, {Llebaria},
  {Bout}, {Schwenn}, {Simnett}, {Bedford}, \& {Eyles}}]{brueckner1995lasco}
{Brueckner}, G.~E., {Howard}, R.~A., {Koomen}, M.~J., {et~al.} 1995, \solphys,
  162, 357, \dodoi{10.1007/BF00733434}

\bibitem[{Bruno(2017)}]{bruno2017calibration}
Bruno, A. 2017, Space Weather, 15, 1191, \dodoi{10.1002/2017SW001672}

\bibitem[{Bruno \& Richardson(2021)}]{bruno2021empirical}
Bruno, A., \& Richardson, I.~G. 2021, \solphys, 296, 36,
  \dodoi{10.1007/s11207-021-01779-4}

\bibitem[{Bu{\v{c}}{\'\i}k(2020)}]{buvcik2020he3}
Bu{\v{c}}{\'\i}k, R. 2020, \ssr, 216, 24, \dodoi{10.1007/s11214-020-00650-5}

\bibitem[{Buzulukova \& Tsurutani(2022)}]{buzulukova2022space}
Buzulukova, N., \& Tsurutani, B. 2022, Frontiers in Astronomy and Space
  Sciences, 9, 1017103, \dodoi{10.3389/fspas.2022.1017103}

\bibitem[{Cane {et~al.}(2006)Cane, Mewaldt, Cohen, \&
  Von~Rosenvinge}]{cane2006role}
Cane, H., Mewaldt, R., Cohen, C., \& Von~Rosenvinge, T. 2006, Journal of
  Geophysical Research: Space Physics, 111, \dodoi{10.1029/2005JA011071}

\bibitem[{Chandran {et~al.}(2011)Chandran, Dennis, Quataert, \&
  Bale}]{chandran2011incorporating}
Chandran, B.~D., Dennis, T.~J., Quataert, E., \& Bale, S.~D. 2011, \apj, 743,
  197, \dodoi{10.1088/0004-637X/743/2/197}

\bibitem[{Chao \& Hsieh(1984)}]{chao1984determining}
Chao, J., \& Hsieh, K. 1984, Planetary and space science, 32, 641,
  \dodoi{10.1016/0032-0633(84)90115-6}

\bibitem[{Chen {et~al.}(2025)Chen, Sachdeva, Huang, van~der Holst,
  Manchester~IV, Jivani, Zou, Chen, Huan, \& Toth}]{chen2025decent}
Chen, H., Sachdeva, N., Huang, Z., {et~al.} 2025, Space Weather, 23,
  e2024SW004165, \dodoi{10.1029/2024SW004165}

\bibitem[{Chen(2011)}]{chen2011coronal}
Chen, P. 2011, Living Reviews in Solar Physics, 8, 1,
  \dodoi{10.12942/lrsp-2011-1}

\bibitem[{Chen {et~al.}(2024)Chen, Giacalone, Guo, \& Klein}]{chen2024parallel}
Chen, X., Giacalone, J., Guo, F., \& Klein, K.~G. 2024, \apj, 965, 61,
  \dodoi{10.3847/1538-4357/ad33c3}

\bibitem[{Chhiber {et~al.}(2021)Chhiber, Ruffolo, Matthaeus, Usmanov,
  Tooprakai, Chuychai, \& Goldstein}]{chhiber2021random}
Chhiber, R., Ruffolo, D., Matthaeus, W.~H., {et~al.} 2021, \apj, 908, 174,
  \dodoi{10.3847/1538-4357/abd7f0}

\bibitem[{Cliver {et~al.}(2022)Cliver, Schrijver, Shibata, \&
  Usoskin}]{cliver2022extreme}
Cliver, E.~W., Schrijver, C.~J., Shibata, K., \& Usoskin, I.~G. 2022, Living
  Reviews in Solar Physics, 19, 2, \dodoi{10.1007/s41116-022-00033-8}

\bibitem[{Cohen {et~al.}(2014)Cohen, Mason, Mewaldt, \&
  Wiedenbeck}]{cohen2014longitudinal}
Cohen, C., Mason, G., Mewaldt, R., \& Wiedenbeck, M. 2014, \apj, 793, 35,
  \dodoi{10.1088/0004-637X/793/1/35}

\bibitem[{Cohen {et~al.}(2005)Cohen, Stone, Mewaldt, Leske, Cummings, Mason,
  Desai, von Rosenvinge, \& Wiedenbeck}]{cohen2005heavy}
Cohen, C., Stone, E., Mewaldt, R., {et~al.} 2005, Journal of Geophysical
  Research: Space Physics, 110, \dodoi{10.1029/2005JA011004}

\bibitem[{Dahlin {et~al.}(2022)Dahlin, DeVore, \& Antiochos}]{dahlin2022stitch}
Dahlin, J.~T., DeVore, C.~R., \& Antiochos, S.~K. 2022, \apj, 941, 79,
  \dodoi{10.3847/1538-4357/ac9e5a}

\bibitem[{De~Pontieu {et~al.}(2007)De~Pontieu, McIntosh, Carlsson, Hansteen,
  Tarbell, Schrijver, Title, Shine, Tsuneta, Katsukawa,
  {et~al.}}]{de2007chromospheric}
De~Pontieu, B., McIntosh, S., Carlsson, M., {et~al.} 2007, Science, 318, 1574,
  \dodoi{10.1126/science.1151747}

\bibitem[{de~Wit {et~al.}(2020)de~Wit, Krasnoselskikh, Bale, Bonnell, Bowen,
  Chen, Froment, Goetz, Harvey, Jagarlamudi, {et~al.}}]{de2020switchbacks}
de~Wit, T.~D., Krasnoselskikh, V.~V., Bale, S.~D., {et~al.} 2020, \apjs, 246,
  39, \dodoi{10.3847/1538-4365/ab5853}

\bibitem[{Decker(1988)}]{decker1988computer}
Decker, R.~B. 1988, \ssr, 48, 195, \dodoi{10.1007/BF00226009}

\bibitem[{Delaunay(1934)}]{delaunay1934sphere}
Delaunay, B. 1934, Bull. Acad. Science USSR VII: Class. Sci. Mat. Nat., 793.
\newblock
  \url{https://www.mathnet.ru/links/95d7c8aa6111426eaf4114db10ced544/im4937.pdf}

\bibitem[{Desai \& Giacalone(2016)}]{desai2016large}
Desai, M., \& Giacalone, J. 2016, Living Reviews in Solar Physics, 13, 3,
  \dodoi{10.1007/s41116-016-0002-5}

\bibitem[{Ding {et~al.}(2015)Ding, Li, Le, Gu, \& Cao}]{ding2015seed}
Ding, L.-G., Li, G., Le, G.-M., Gu, B., \& Cao, X.-X. 2015, \apj, 812, 171,
  \dodoi{10.1088/0004-637X/812/2/171}

\bibitem[{Ding {et~al.}(2022)Ding, Wijsen, Li, \& Poedts}]{ding2022modeling}
Ding, Z., Wijsen, N., Li, G., \& Poedts, S. 2022, \aap, 668, A71,
  \dodoi{10.1051/0004-6361/202244732}

\bibitem[{Domingo {et~al.}(1995)Domingo, Fleck, \& Poland}]{domingo1995soho}
Domingo, V., Fleck, B., \& Poland, A.~I. 1995, \solphys, 162, 1,
  \dodoi{10.1007/BF00733425}

\bibitem[{Downs {et~al.}(2010)Downs, Roussev, van~der Holst, Lugaz, Sokolov, \&
  Gombosi}]{downs2010toward}
Downs, C., Roussev, I.~I., van~der Holst, B., {et~al.} 2010, \apj, 712, 1219,
  \dodoi{10.1088/0004-637X/712/2/1219}

\bibitem[{Downs {et~al.}(2021)Downs, Warmuth, Long, Bloomfield, Kwon, Veronig,
  Vourlidas, \& Vr{\v{s}}nak}]{downs2021validation}
Downs, C., Warmuth, A., Long, D.~M., {et~al.} 2021, \apj, 911, 118,
  \dodoi{10.3847/1538-4357/abea78}

\bibitem[{Dresing {et~al.}(2014)Dresing, G{\'o}mez-Herrero, Heber, Klassen,
  Malandraki, Dr{\"o}ge, \& Kartavykh}]{dresing2014statistical}
Dresing, N., G{\'o}mez-Herrero, R., Heber, B., {et~al.} 2014, \aap, 567, A27,
  \dodoi{10.1051/0004-6361/201423789}

\bibitem[{Dr{\"o}ge {et~al.}(2010)Dr{\"o}ge, Kartavykh, Klecker, \&
  Kovaltsov}]{droge2010anisotropic}
Dr{\"o}ge, W., Kartavykh, Y., Klecker, B., \& Kovaltsov, G. 2010, \apj, 709,
  912, \dodoi{10.1088/0004-637X/709/2/912}

\bibitem[{Drury(1983)}]{drury1983introduction}
Drury, L.~O. 1983, Reports on Progress in Physics, 46, 973,
  \dodoi{10.1088/0034-4885/46/8/002}

\bibitem[{Dumbovi{\'c} {et~al.}(2018)Dumbovi{\'c}, {\v{C}}alogovi{\'c},
  Vr{\v{s}}nak, Temmer, Mays, Veronig, \& Piantschitsch}]{dumbovic2018drag}
Dumbovi{\'c}, M., {\v{C}}alogovi{\'c}, J., Vr{\v{s}}nak, B., {et~al.} 2018,
  \apj, 854, 180, \dodoi{10.3847/1538-4357/aaaa66}

\bibitem[{{Earl}(1974)}]{earl1974diffuse}
{Earl}, J.~A. 1974, \apj, 193, 231, \dodoi{10.1086/153152}

\bibitem[{Ellison {et~al.}(1990)Ellison, Jones, \& Reynolds}]{ellison1990first}
Ellison, D.~C., Jones, F.~C., \& Reynolds, S.~P. 1990, \apj, 360, 702,
  \dodoi{10.1086/169156}

\bibitem[{Ellison \& Ramaty(1985)}]{ellison1985shock}
Ellison, D.~C., \& Ramaty, R. 1985, \apj, 298, 400, \dodoi{10.1086/163623}

\bibitem[{Engelbrecht(2019)}]{engelbrecht2019pitch}
Engelbrecht, N.~E. 2019, \apj, 880, 60, \dodoi{10.3847/1538-4357/ab2871}

\bibitem[{Erd{\H{o}}s \& Balogh(2005)}]{erdHos2005situ}
Erd{\H{o}}s, G., \& Balogh, A. 2005, Advances in Space Research, 35, 625,
  \dodoi{10.1016/j.asr.2005.02.048}

\bibitem[{Fermi(1949)}]{fermi1949origin}
Fermi, E. 1949, Physical Review, 75, 1169, \dodoi{10.1103/PhysRev.75.1169}

\bibitem[{Fisk \& Gloeckler(2006)}]{fisk2006common}
Fisk, L., \& Gloeckler, G. 2006, \apj, 640, L79, \dodoi{10.1086/503293}

\bibitem[{Fisk \& Gloeckler(2008)}]{fisk2008acceleration}
---. 2008, \apj, 686, 1466, \dodoi{10.1086/591543}

\bibitem[{Fisk \& Schwadron(2001)}]{fisk2001behavior}
Fisk, L., \& Schwadron, N. 2001, \apj, 560, 425, \dodoi{10.1086/322503}

\bibitem[{Fox {et~al.}(2016)Fox, Velli, Bale, Decker, Driesman, Howard, Kasper,
  Kinnison, Kusterer, Lario, {et~al.}}]{fox2016solar}
Fox, N., Velli, M., Bale, S., {et~al.} 2016, \ssr, 204, 7,
  \dodoi{10.1007/s11214-015-0211-6}

\bibitem[{Fulara {et~al.}(2019)Fulara, Chandra, Chen, Zhelyazkov, Srivastava,
  \& Uddin}]{fulara2019kinematics}
Fulara, A., Chandra, R., Chen, P., {et~al.} 2019, \solphys, 294, 1,
  \dodoi{10.1007/s11207-019-1445-3}

\bibitem[{Giacalone(2005{\natexlab{a}})}]{giacalone2005efficient}
Giacalone, J. 2005{\natexlab{a}}, \apj, 628, L37, \dodoi{10.1086/432510}

\bibitem[{Giacalone(2005{\natexlab{b}})}]{giacalone2005particle}
---. 2005{\natexlab{b}}, \apj, 624, 765, \dodoi{10.1086/429265}

\bibitem[{Giacalone {et~al.}(2000)Giacalone, Jokipii, \&
  Mazur}]{giacalone2000small}
Giacalone, J., Jokipii, J., \& Mazur, J. 2000, \apj, 532, L75,
  \dodoi{10.1086/312564}

\bibitem[{Giacalone \& K{\'o}ta(2007)}]{giacalone2007acceleration}
Giacalone, J., \& K{\'o}ta, J. 2007, Solar Dynamics and Its Effects on the
  Heliosphere and Earth, 277, \dodoi{10.1007/978-0-387-69532-7_19}

\bibitem[{Giacalone \& Neugebauer(2008)}]{giacalone2008energy}
Giacalone, J., \& Neugebauer, M. 2008, \apj, 673, 629, \dodoi{10.1086/524008}

\bibitem[{Gibson \& Low(1998)}]{gibson1998time}
Gibson, S.~E., \& Low, B. 1998, \apj, 493, 460, \dodoi{10.1086/305107}

\bibitem[{Gieseler {et~al.}(2023)Gieseler, Dresing, Palmroos, Freiherr~von
  Forstner, Price, Vainio, Kouloumvakos, Rodr{\'\i}guez-Garc{\'\i}a, Trotta,
  G{\'e}not, {et~al.}}]{gieseler2023solar}
Gieseler, J., Dresing, N., Palmroos, C., {et~al.} 2023, Frontiers in Astronomy
  and Space Sciences, 9, 1058810, \dodoi{10.3389/fspas.2022.1058810}

\bibitem[{Gloeckler(2003)}]{gloeckler2003ubiquitous}
Gloeckler, G. 2003, AIP Conference Proceedings, 679, 583,
  \dodoi{10.1063/1.1618663}

\bibitem[{{Gold} {et~al.}(1998){Gold}, {Krimigis}, {Hawkins}, {Haggerty},
  {Lohr}, {Fiore}, {Armstrong}, {Holland}, \& {Lanzerotti}}]{gold1998electron}
{Gold}, R.~E., {Krimigis}, S.~M., {Hawkins}, S.~E., I., {et~al.} 1998, \ssr,
  86, 541, \dodoi{10.1023/A:1005088115759}

\bibitem[{Gombosi(1998)}]{gombosi1998physics}
Gombosi, T.~I. 1998, Physics of the Space Environment (Cambridge University
  Press), 274--277, \dodoi{10.1017/CBO9780511529474}

\bibitem[{Gombosi {et~al.}(2003)Gombosi, De~Zeeuw, Powell, Ridley, Sokolov,
  Stout, \& T{\'o}th}]{gombosi2003adaptive}
Gombosi, T.~I., De~Zeeuw, D.~L., Powell, K.~G., {et~al.} 2003, Space Plasma
  Simulation, 247, \dodoi{10.1007/3-540-36530-3_12}

\bibitem[{Gombosi {et~al.}(2018)Gombosi, van~der Holst, Manchester, \&
  Sokolov}]{gombosi2018extended}
Gombosi, T.~I., van~der Holst, B., Manchester, W.~B., \& Sokolov, I.~V. 2018,
  Living Reviews in Solar Physics, 15, 1, \dodoi{10.1007/s41116-018-0014-4}

\bibitem[{Gombosi {et~al.}(2004)Gombosi, Powell, De~Zeeuw, Clauer, Hansen,
  Manchester, Ridley, Roussev, Sokolov, Stout, {et~al.}}]{gombosi2004solution}
Gombosi, T.~I., Powell, K.~G., De~Zeeuw, D.~L., {et~al.} 2004, Computing in
  Science \& Engineering, 6, 14, \dodoi{10.1109/MCISE.2004.1267603}

\bibitem[{Gombosi {et~al.}(2021)Gombosi, Chen, Glocer, Huang, Jia, Liemohn,
  Manchester, Pulkkinen, Sachdeva, Al~Shidi, {et~al.}}]{gombosi2021sustained}
Gombosi, T.~I., Chen, Y., Glocer, A., {et~al.} 2021, Journal of Space Weather
  and Space Climate, 11, 42, \dodoi{10.1051/swsc/2021020}

\bibitem[{Gopalswamy {et~al.}(2015)Gopalswamy, M{\"a}kel{\"a}, Yashiro, Xie,
  Akiyama, \& Thakur}]{gopalswamy2015high}
Gopalswamy, N., M{\"a}kel{\"a}, P., Yashiro, S., {et~al.} 2015, Journal of
  Physics: Conference Series, 642, 012012,
  \dodoi{10.1088/1742-6596/642/1/012012}

\bibitem[{Gopalswamy {et~al.}(2002)Gopalswamy, Yashiro, Micha{\l}ek, Kaiser,
  Howard, Reames, Leske, \& Von~Rosenvinge}]{gopalswamy2002interacting}
Gopalswamy, N., Yashiro, S., Micha{\l}ek, G., {et~al.} 2002, \apj, 572, L103,
  \dodoi{10.1086/341601}

\bibitem[{Gopalswamy {et~al.}(2009)Gopalswamy, Yashiro, Michalek, Stenborg,
  Vourlidas, Freeland, \& Howard}]{gopalswamy2009soho}
Gopalswamy, N., Yashiro, S., Michalek, G., {et~al.} 2009, Earth, Moon, and
  Planets, 104, 295, \dodoi{10.1007/s11038-008-9282-7}

\bibitem[{Guo {et~al.}(2021)Guo, Zeitlin, Wimmer-Schweingruber, Hassler,
  Ehresmann, Rafkin, Freiherr~von Forstner, Khaksarighiri, Liu, \&
  Wang}]{guo2021radiation}
Guo, J., Zeitlin, C., Wimmer-Schweingruber, R.~F., {et~al.} 2021, \aapr, 29, 1,
  \dodoi{10.1007/s00159-021-00136-5}

\bibitem[{Haggerty {et~al.}(2006)Haggerty, Roelof, Ho, \&
  Gold}]{haggerty2006qualitative}
Haggerty, D., Roelof, E., Ho, G., \& Gold, R. 2006, Advances in Space Research,
  38, 995, \dodoi{10.1016/j.asr.2005.08.030}

\bibitem[{Harten \& Clark(1995)}]{harten1995design}
Harten, R., \& Clark, K. 1995, \ssr, 71, 23, \dodoi{10.1007/BF00751324}

\bibitem[{Harvey {et~al.}(1996)Harvey, Hill, Hubbard, Kennedy, Leibacher,
  Pintar, Gilman, Noyes, Title, Toomre, {et~al.}}]{harvey1996global}
Harvey, J., Hill, F., Hubbard, R., {et~al.} 1996, Science, 272, 1284,
  \dodoi{10.1126/science.272.5266.1284}

\bibitem[{Hayes {et~al.}(2001)Hayes, Vourlidas, \& Howard}]{hayes2001deriving}
Hayes, A., Vourlidas, A., \& Howard, R. 2001, \apj, 548, 1081,
  \dodoi{10.1086/319029}

\bibitem[{Heras {et~al.}(1995)Heras, Sanahuja, Lario, Smith, Detman, \&
  Dryer}]{heras1995three}
Heras, A., Sanahuja, B., Lario, D., {et~al.} 1995, \apj, 445, 497,
  \dodoi{10.1086/175714}

\bibitem[{Hill(2018)}]{hill2018global}
Hill, F. 2018, Space Weather, 16, 1488, \dodoi{10.1029/2018SW002001}

\bibitem[{Hinterreiter {et~al.}(2019)Hinterreiter, Magdalenic, Temmer, Verbeke,
  Jebaraj, Samara, Asvestari, Poedts, Pomoell, Kilpua,
  {et~al.}}]{hinterreiter2019assessing}
Hinterreiter, J., Magdalenic, J., Temmer, M., {et~al.} 2019, \solphys, 294, 1,
  \dodoi{10.1007/s11207-019-1558-8}

\bibitem[{Hollweg(1986)}]{hollweg1986transition}
Hollweg, J.~V. 1986, Journal of Geophysical Research: Space Physics, 91, 4111,
  \dodoi{10.1029/JA091iA04p04111}

\bibitem[{Hoppock {et~al.}(2018)Hoppock, Chandran, Klein, Mallet, \&
  Verscharen}]{hoppock2018stochastic}
Hoppock, I.~W., Chandran, B.~D., Klein, K.~G., Mallet, A., \& Verscharen, D.
  2018, Journal of Plasma Physics, 84, 905840615,
  \dodoi{10.1017/S0022377818001277}

\bibitem[{Howard {et~al.}(2008)Howard, Moses, Vourlidas, Newmark, Socker,
  Plunkett, Korendyke, Cook, Hurley, Davila, {et~al.}}]{howard2008sun}
Howard, R.~A., Moses, J., Vourlidas, A., {et~al.} 2008, \ssr, 136, 67,
  \dodoi{10.1007/s11214-008-9341-4}

\bibitem[{Hu {et~al.}(2017)Hu, Li, Ao, Zank, \&
  Verkhoglyadova}]{hu2017modeling}
Hu, J., Li, G., Ao, X., Zank, G.~P., \& Verkhoglyadova, O. 2017, Journal of
  Geophysical Research: Space Physics, 122, 10, \dodoi{10.1002/2017JA024077}

\bibitem[{Huang {et~al.}(2024{\natexlab{a}})Huang, T{\'o}th, Huang, Sachdeva,
  van~der Holst, \& Manchester}]{huang2024adjusting}
Huang, Z., T{\'o}th, G., Huang, J., {et~al.} 2024{\natexlab{a}}, \apjl, 965,
  L1, \dodoi{10.3847/2041-8213/ad3547}

\bibitem[{Huang {et~al.}(2024{\natexlab{b}})Huang, T{\'o}th, Sachdeva, \&
  van~der Holst}]{huang2024solar}
Huang, Z., T{\'o}th, G., Sachdeva, N., \& van~der Holst, B. 2024{\natexlab{b}},
  \apj, 965, 1, \dodoi{10.3847/1538-4357/ad32ca}

\bibitem[{Hugoniot(1889{\natexlab{a}})}]{hugoniot1887propagation}
Hugoniot, H. 1889{\natexlab{a}}, Journal de l'\'{E}cole Polytechnique (French),
  57, 1.
\newblock \url{https://books.google.com/books?id=wccAAAAAYAAJ}

\bibitem[{Hugoniot(1889{\natexlab{b}})}]{hugoniot1889propagation}
---. 1889{\natexlab{b}}, Journal de l'\'{E}cole Polytechnique (French), 58, 1.
\newblock \url{https://gallica.bnf.fr/ark:/12148/bpt6k4337130/f11.item}

\bibitem[{Isenberg(1997)}]{isenberg1997hemispherical}
Isenberg, P.~A. 1997, Journal of Geophysical Research: Space Physics, 102,
  4719, \dodoi{10.1029/96JA03671}

\bibitem[{{Jackson}(1998)}]{jackson1998book}
{Jackson}, J.~D. 1998, {Classical Electrodynamics, 3rd Edition} (Wiley)

\bibitem[{Jian {et~al.}(2015)Jian, MacNeice, Taktakishvili, Odstrcil, Jackson,
  Yu, Riley, Sokolov, \& Evans}]{jian2015validation}
Jian, L., MacNeice, P., Taktakishvili, A., {et~al.} 2015, Space Weather, 13,
  316, \dodoi{10.1002/2015SW001174}

\bibitem[{Jin {et~al.}(2017{\natexlab{a}})Jin, Manchester, van~der Holst,
  Sokolov, T{\'o}th, Vourlidas, de~Koning, \& Gombosi}]{jin2017chromosphere}
Jin, M., Manchester, W., van~der Holst, B., {et~al.} 2017{\natexlab{a}}, \apj,
  834, 172, \dodoi{10.3847/1538-4357/834/2/172}

\bibitem[{Jin {et~al.}(2022)Jin, Nitta, \& Cohen}]{jin2022assessing}
Jin, M., Nitta, N.~V., \& Cohen, C.~M. 2022, Space Weather, 20, e2021SW002894,
  \dodoi{10.1029/2021SW002894}

\bibitem[{Jin {et~al.}(2016)Jin, Schrijver, Cheung, DeRosa, Nitta,
  {et~al.}}]{jin2016numerical}
Jin, M., Schrijver, C., Cheung, M., {et~al.} 2016, \apj, 820, 16,
  \dodoi{10.3847/0004-637X/820/1/16}

\bibitem[{Jin {et~al.}(2013)Jin, Manchester, Van Der~Holst, Oran, Sokolov,
  T{\'o}th, Liu, Sun, \& Gombosi}]{jin2013numerical}
Jin, M., Manchester, W., Van Der~Holst, B., {et~al.} 2013, \apj, 773, 50,
  \dodoi{10.1088/0004-637X/773/1/50}

\bibitem[{Jin {et~al.}(2017{\natexlab{b}})Jin, Manchester, van~der Holst,
  Sokolov, T{\'o}th, Mullinix, Taktakishvili, Chulaki, \&
  Gombosi}]{jin2017data}
Jin, M., Manchester, W., van~der Holst, B., {et~al.} 2017{\natexlab{b}}, \apj,
  834, 173, \dodoi{10.3847/1538-4357/834/2/173}

\bibitem[{Jokipii(1982)}]{jokipii1982particle}
Jokipii, J. 1982, \apj, 255, 716, \dodoi{10.1086/159870}

\bibitem[{Jokipii(1987)}]{jokipii1987rate}
---. 1987, \apj, 313, 842, \dodoi{10.1086/165022}

\bibitem[{Jokipii(1966)}]{jokipii1966cosmic}
Jokipii, J.~R. 1966, \apj, 146, 480, \dodoi{10.1086/148912}

\bibitem[{Jones \& Ellison(1991)}]{jones1991plasma}
Jones, F.~C., \& Ellison, D.~C. 1991, \ssr, 58, 259, \dodoi{10.1007/BF01206003}

\bibitem[{Joshi {et~al.}(2016)Joshi, Kushwaha, Veronig, Dhara, Shanmugaraju, \&
  Moon}]{joshi2016formation}
Joshi, B., Kushwaha, U., Veronig, A.~M., {et~al.} 2016, \apj, 834, 42,
  \dodoi{10.3847/1538-4357/834/1/42}

\bibitem[{Kahler {et~al.}(1978)Kahler, Hildner, \&
  Van~Hollebeke}]{kahler1978prompt}
Kahler, S., Hildner, E., \& Van~Hollebeke, M. 1978, \solphys, 57, 429,
  \dodoi{10.1007/BF00160116}

\bibitem[{Kahler \& Ling(2019)}]{kahler2019suprathermal}
Kahler, S., \& Ling, A. 2019, \apj, 872, 89, \dodoi{10.3847/1538-4357/aafb03}

\bibitem[{Kahler {et~al.}(1984)Kahler, Sheeley~Jr, Howard, Koomen, Michels,
  McGuire, Von~Rosenvinge, \& Reames}]{kahler1984associations}
Kahler, S., Sheeley~Jr, N., Howard, R., {et~al.} 1984, Journal of Geophysical
  Research: Space Physics, 89, 9683, \dodoi{10.1029/JA089iA11p09683}

\bibitem[{Kaiser {et~al.}(2008)Kaiser, Kucera, Davila, St~Cyr, Guhathakurta, \&
  Christian}]{kaiser2008stereo}
Kaiser, M.~L., Kucera, T., Davila, J., {et~al.} 2008, \ssr, 136, 5,
  \dodoi{10.1007/s11214-007-9277-0}

\bibitem[{Kasper {et~al.}(2016)Kasper, Abiad, Austin, Balat-Pichelin, Bale,
  Belcher, Berg, Bergner, Berthomier, Bookbinder, {et~al.}}]{kasper2016solar}
Kasper, J.~C., Abiad, R., Austin, G., {et~al.} 2016, \ssr, 204, 131,
  \dodoi{10.1007/s11214-015-0206-3}

\bibitem[{Kataoka {et~al.}(2009)Kataoka, Ebisuzaki, Kusano, Shiota, Inoue,
  Yamamoto, \& Tokumaru}]{kataoka2009three}
Kataoka, R., Ebisuzaki, T., Kusano, K., {et~al.} 2009, Journal of Geophysical
  Research: Space Physics, 114, \dodoi{10.1029/2009JA014167}

\bibitem[{Kecskem{\'e}ty {et~al.}(2009)Kecskem{\'e}ty, Daibog, Logachev, \&
  K{\'o}ta}]{kecskemety2009decay}
Kecskem{\'e}ty, K., Daibog, E., Logachev, Y.~I., \& K{\'o}ta, J. 2009, Journal
  of Geophysical Research: Space Physics, 114, \dodoi{10.1029/2008JA013730}

\bibitem[{Kennis {et~al.}(2024)Kennis, Perri, \& Poedts}]{kennis2024magnetic}
Kennis, S., Perri, B., \& Poedts, S. 2024, \aap, 691, A257,
  \dodoi{10.1051/0004-6361/202451005}

\bibitem[{Kilpua {et~al.}(2019)Kilpua, Lugaz, Mays, \&
  Temmer}]{kilpua2019forecasting}
Kilpua, E.~K., Lugaz, N., Mays, M.~L., \& Temmer, M. 2019, Space Weather, 17,
  498, \dodoi{10.1029/2018SW001944}

\bibitem[{King \& Papitashvili(2005)}]{king2005solar}
King, J., \& Papitashvili, N. 2005, Journal of Geophysical Research: Space
  Physics, 110, \dodoi{10.1029/2004JA010649}

\bibitem[{Kleimann {et~al.}(2022)Kleimann, Dialynas, Fraternale, Galli,
  Heerikhuisen, Izmodenov, Kornbleuth, Opher, \&
  Pogorelov}]{kleimann2022structure}
Kleimann, J., Dialynas, K., Fraternale, F., {et~al.} 2022, \ssr, 218, 36,
  \dodoi{10.1007/s11214-022-00902-6}

\bibitem[{Klein \& Dalla(2017)}]{klein2017acc}
Klein, K.-L., \& Dalla, S. 2017, \ssr, 212, 1107,
  \dodoi{10.1007/s11214-017-0382-4}

\bibitem[{{Kolmogorov}(1941)}]{kolmogorov1941local}
{Kolmogorov}, A. 1941, Akademiia Nauk SSSR Doklady, 30, 301,
  \dodoi{10.1098/rspa.1991.0075}

\bibitem[{Kong {et~al.}(2019)Kong, Guo, Chen, \&
  Giacalone}]{kong2019acceleration}
Kong, X., Guo, F., Chen, Y., \& Giacalone, J. 2019, \apj, 883, 49,
  \dodoi{10.3847/1538-4357/ab3848}

\bibitem[{K{\'o}ta(1997)}]{kota1997energy}
K{\'o}ta, J. 1997, in Proceedings of the 25th International Cosmic Ray
  Conference, Vol.~1, 213--216.
\newblock \url{https://articles.adsabs.harvard.edu/pdf/1997ICRC...25a.213K}

\bibitem[{K{\'o}ta(2000)}]{kota2000diffusion}
K{\'o}ta, J. 2000, Journal of Geophysical Research: Space Physics, 105, 2403,
  \dodoi{10.1029/1999JA900469}

\bibitem[{K{\'o}ta \& Jokipii(2004)}]{kota2004cosmic}
K{\'o}ta, J., \& Jokipii, J. 2004, AIP Conference Proceedings, 719, 272,
  \dodoi{10.1063/1.1809528}

\bibitem[{K{\'o}ta {et~al.}(2005)K{\'o}ta, Manchester, Jokipii, De~Zeeuw, \&
  Gombosi}]{kota2005simulation}
K{\'o}ta, J., Manchester, W., Jokipii, J., De~Zeeuw, D., \& Gombosi, T. 2005,
  AIP Conference Proceedings, 781, 201, \dodoi{10.1063/1.2032697}

\bibitem[{Kouloumvakos {et~al.}(2019)Kouloumvakos, Rouillard, Wu, Vainio,
  Vourlidas, Plotnikov, Afanasiev, \& {\"O}nel}]{kouloumvakos2019connecting}
Kouloumvakos, A., Rouillard, A.~P., Wu, Y., {et~al.} 2019, \apj, 876, 80,
  \dodoi{10.3847/1538-4357/ab15d7}

\bibitem[{Krivodonova \& Smirnov(2021)}]{krivodonova2021tvd}
Krivodonova, L., \& Smirnov, A. 2021, arXiv preprint arXiv:2110.00067,
  \dodoi{10.48550/arXiv.2110.00067}

\bibitem[{Krymskii(1977)}]{krymskii1977regular}
Krymskii, G. 1977, in Akademiia Nauk SSSR Doklady, Vol. 234, 1306--1308

\bibitem[{K{\"u}hl \& Heber(2019)}]{kuhl2019revising}
K{\"u}hl, P., \& Heber, B. 2019, Space Weather, 17, 84,
  \dodoi{10.1029/2018SW002114}

\bibitem[{Kwon \& Vourlidas(2017)}]{kwon2017investigating}
Kwon, R.-Y., \& Vourlidas, A. 2017, \apj, 836, 246,
  \dodoi{10.3847/1538-4357/aa5b92}

\bibitem[{Kwon {et~al.}(2014)Kwon, Zhang, \& Olmedo}]{kwon2014new}
Kwon, R.-Y., Zhang, J., \& Olmedo, O. 2014, \apj, 794, 148,
  \dodoi{10.1088/0004-637X/794/2/148}

\bibitem[{Laitinen {et~al.}(2013)Laitinen, Dalla, \&
  Marsh}]{laitinen2013energetic}
Laitinen, T., Dalla, S., \& Marsh, M. 2013, \apjl, 773, L29,
  \dodoi{10.1088/2041-8205/773/2/L29}

\bibitem[{Laitinen {et~al.}(2018)Laitinen, Effenberger, Kopp, \&
  Dalla}]{laitinen2018effect}
Laitinen, T., Effenberger, F., Kopp, A., \& Dalla, S. 2018, Journal of Space
  Weather and Space Climate, 8, A13, \dodoi{10.1051/swsc/2018001}

\bibitem[{Laitinen {et~al.}(2016)Laitinen, Kopp, Effenberger, Dalla, \&
  Marsh}]{laitinen2016solar}
Laitinen, T., Kopp, A., Effenberger, F., Dalla, S., \& Marsh, M. 2016, \aap,
  591, A18, \dodoi{10.1051/0004-6361/201527801}

\bibitem[{Landau \& Lifshitz(1960)}]{landau1959course}
Landau, L., \& Lifshitz, E. 1960, Course of Theoretical Physics: Mechanics,
  Vol.~1 (Pergamon Press)

\bibitem[{Landau {et~al.}(2013)Landau, Bell, Kearsley, Pitaevskii, Lifshitz, \&
  Sykes}]{landau2013electrodynamics}
Landau, L.~D., Bell, J.~S., Kearsley, M., {et~al.} 2013, {Electrodynamics of
  Continuous Media}, Vol.~8 (Elsevier)

\bibitem[{Landau \& Lifshitz(1987)}]{landau1987fluid}
Landau, L.~D., \& Lifshitz, E.~M. 1987, Fluid Mechanics, Vol.~6 (Elsevier)

\bibitem[{Lario {et~al.}(2019)Lario, Berger, Decker, Wimmer-Schweingruber,
  Wilson, Giacalone, \& Roelof}]{lario2019evolution}
Lario, D., Berger, L., Decker, R., {et~al.} 2019, \aj, 158, 12,
  \dodoi{10.3847/1538-3881/ab1e49}

\bibitem[{Lario {et~al.}(2018)Lario, Berger, LB~III, Decker, Haggerty, Roelof,
  Wimmer-Schweingruber, \& Giacalone}]{lario2018flat}
Lario, D., Berger, L., LB~III, W., {et~al.} 2018, Journal of Physics:
  Conference Series, 1100, 012014, \dodoi{10.1088/1742-6596/1100/1/012014}

\bibitem[{Lario {et~al.}(2006)Lario, Kallenrode, Decker, Roelof, Krimigis,
  Aran, \& Sanahuja}]{lario2006radial}
Lario, D., Kallenrode, M.-B., Decker, R., {et~al.} 2006, \apj, 653, 1531,
  \dodoi{10.1086/508982}

\bibitem[{Lario {et~al.}(2014)Lario, Raouafi, Kwon, Zhang, G{\'o}mez-Herrero,
  Dresing, \& Riley}]{lario2014solar}
Lario, D., Raouafi, N., Kwon, R.-Y., {et~al.} 2014, \apj, 797, 8,
  \dodoi{10.1088/0004-637X/797/1/8}

\bibitem[{Lario {et~al.}(1998)Lario, Sanahuja, \& Heras}]{lario1998energetic}
Lario, D., Sanahuja, B., \& Heras, A. 1998, \apj, 509, 415,
  \dodoi{10.1086/306461}

\bibitem[{Lario {et~al.}(2024)Lario, Balmaceda, G{\'o}mez-Herrero, Mason,
  Krupar, Mac~Cormack, Kouloumvakos, Cernuda, Collier, Richardson,
  {et~al.}}]{lario2024rapid}
Lario, D., Balmaceda, L.~A., G{\'o}mez-Herrero, R., {et~al.} 2024, \apj, 975,
  84, \dodoi{10.3847/1538-4357/ad6c47}

\bibitem[{Lee {et~al.}(2011)Lee, Luhmann, Hoeksema, Sun, Arge, \&
  de~Pater}]{lee2011coronal}
Lee, C., Luhmann, J., Hoeksema, J., {et~al.} 2011, \solphys, 269, 367,
  \dodoi{10.1007/s11207-010-9699-9}

\bibitem[{Lee \& Schachter(1980)}]{lee1980two}
Lee, D.-T., \& Schachter, B.~J. 1980, International Journal of Computer \&
  Information Sciences, 9, 219, \dodoi{10.1007/BF00977785}

\bibitem[{Lee(1982)}]{lee1982coupled}
Lee, M.~A. 1982, Journal of Geophysical Research: Space Physics, 87, 5063,
  \dodoi{10.1029/JA087iA07p05063}

\bibitem[{Lee(1983)}]{lee1983coupled}
---. 1983, Journal of Geophysical Research: Space Physics, 88, 6109,
  \dodoi{10.1029/JA088iA08p06109}

\bibitem[{Lee {et~al.}(2012)Lee, Mewaldt, \& Giacalone}]{lee2012shock}
Lee, M.~A., Mewaldt, R., \& Giacalone, J. 2012, \ssr, 173, 247,
  \dodoi{10.1007/s11214-012-9932-y}

\bibitem[{Lemen {et~al.}(2012)Lemen, Title, Akin, Boerner, Chou, Drake, Duncan,
  Edwards, Friedlaender, Heyman, {et~al.}}]{lemen2012atmospheric}
Lemen, J.~R., Title, A.~M., Akin, D.~J., {et~al.} 2012, \solphys, 275, 17,
  \dodoi{10.1007/s11207-011-9776-8}

\bibitem[{Lepping \& Argentiero(1971)}]{lepping1971single}
Lepping, R., \& Argentiero, P. 1971, Journal of Geophysical Research, 76, 4349,
  \dodoi{10.1029/JA076i019p04349}

\bibitem[{Li \& Lee(2015)}]{li2015scatter}
Li, G., \& Lee, M.~A. 2015, \apj, 810, 82, \dodoi{10.1088/0004-637X/810/1/82}

\bibitem[{Li {et~al.}(2012)Li, Moore, Mewaldt, Zhao, \& Labrador}]{li2012twin}
Li, G., Moore, R., Mewaldt, R., Zhao, L., \& Labrador, A. 2012, \ssr, 171, 141,
  \dodoi{10.1007/s11214-011-9823-7}

\bibitem[{Li \& Zank(2005)}]{li2005mixed}
Li, G., \& Zank, G. 2005, \grl, 32, \dodoi{10.1029/2004GL021250}

\bibitem[{Li {et~al.}(2003)Li, Zank, \& Rice}]{li2003energetic}
Li, G., Zank, G., \& Rice, W. 2003, Journal of Geophysical Research: Space
  Physics, 108, \dodoi{10.1029/2002JA009666}

\bibitem[{Li {et~al.}(2009)Li, Zank, Verkhoglyadova, Mewaldt, Cohen, Mason, \&
  Desai}]{li2009shock}
Li, G., Zank, G., Verkhoglyadova, O., {et~al.} 2009, \apj, 702, 998,
  \dodoi{10.1088/0004-637X/702/2/998}

\bibitem[{Linker {et~al.}(2019)Linker, Caplan, Schwadron, Gorby, Downs, Torok,
  Lionello, \& Wijaya}]{linker2019coupled}
Linker, J.~A., Caplan, R.~M., Schwadron, N., {et~al.} 2019, Journal of Physics:
  Conference Series, 1225, 012007, \dodoi{10.1088/1742-6596/1225/1/012007}

\bibitem[{Liouville(1838)}]{liouville1838note}
Liouville, J. 1838, Journal de math{\'e}matiques pures et appliqu{\'e}es, 3,
  342.
\newblock \url{http://www.numdam.org/item/JMPA_1838_1_3__342_0.pdf}

\bibitem[{Lugaz {et~al.}(2013)Lugaz, Farrugia, Manchester, \&
  Schwadron}]{lugaz2013interaction}
Lugaz, N., Farrugia, C.~J., Manchester, W.~B., \& Schwadron, N. 2013, \apj,
  778, 20, \dodoi{10.1088/0004-637X/778/1/20}

\bibitem[{Lugaz {et~al.}(2005{\natexlab{a}})Lugaz, Manchester, \&
  Gombosi}]{lugaz2005evolution}
Lugaz, N., Manchester, W.~B., \& Gombosi, T.~I. 2005{\natexlab{a}}, \apj, 627,
  1019, \dodoi{10.1086/430465}

\bibitem[{Lugaz {et~al.}(2005{\natexlab{b}})Lugaz, Manchester, \&
  Gombosi}]{lugaz2005numerical}
---. 2005{\natexlab{b}}, \apj, 634, 651, \dodoi{10.1086/491782}

\bibitem[{Lugaz {et~al.}(2007)Lugaz, Manchester, Roussev, T{\'o}th, \&
  Gombosi}]{lugaz2007numerical}
Lugaz, N., Manchester, W.~B., Roussev, I.~I., T{\'o}th, G., \& Gombosi, T.~I.
  2007, \apj, 659, 788, \dodoi{10.1086/512005}

\bibitem[{Luhmann {et~al.}(2007)Luhmann, Ledvina, Krauss-Varban, Odstrcil, \&
  Riley}]{luhmann2007heliospheric}
Luhmann, J., Ledvina, S., Krauss-Varban, D., Odstrcil, D., \& Riley, P. 2007,
  Advances in Space Research, 40, 295, \dodoi{10.1016/j.asr.2007.03.089}

\bibitem[{Luhmann {et~al.}(2020)Luhmann, Gopalswamy, Jian, \&
  Lugaz}]{luhmann2020icme}
Luhmann, J.~G., Gopalswamy, N., Jian, L., \& Lugaz, N. 2020, \solphys, 295, 61,
  \dodoi{10.1007/s11207-020-01624-0}

\bibitem[{MacNamara \& Strang(2016)}]{macnamara2016operator}
MacNamara, S., \& Strang, G. 2016, Splitting Methods in Communication, Imaging,
  Science, and Engineering, 95, \dodoi{10.1007/978-3-319-41589-5_3}

\bibitem[{Maguire {et~al.}(2020)Maguire, Carley, McCauley, \&
  Gallagher}]{maguire2020evolution}
Maguire, C.~A., Carley, E.~P., McCauley, J., \& Gallagher, P.~T. 2020, \aap,
  633, A56, \dodoi{10.1051/0004-6361/201936449}

\bibitem[{Malandraki {et~al.}(2009)Malandraki, Marsden, Lario, Tranquille,
  Heber, Mewaldt, Cohen, Lanzerotti, Forsyth, Elliott,
  {et~al.}}]{malandraki2009energetic}
Malandraki, O., Marsden, R., Lario, D., {et~al.} 2009, \apj, 704, 469,
  \dodoi{10.1088/0004-637X/704/1/469}

\bibitem[{Manchester {et~al.}(2004{\natexlab{a}})Manchester, Gombosi, DeZeeuw,
  \& Fan}]{manchester2004eruption}
Manchester, W., Gombosi, T., DeZeeuw, D., \& Fan, Y. 2004{\natexlab{a}}, \apj,
  610, 588, \dodoi{10.1086/421516}

\bibitem[{Manchester {et~al.}(2017)Manchester, Kilpua, Liu, Lugaz, Riley,
  T{\"o}r{\"o}k, \& Vr{\v{s}}nak}]{manchester2017physical}
Manchester, W., Kilpua, E.~K., Liu, Y.~D., {et~al.} 2017, \ssr, 212, 1159,
  \dodoi{10.1007/s11214-017-0394-0}

\bibitem[{Manchester {et~al.}(2014{\natexlab{a}})Manchester, Kozyra, Lepri, \&
  Lavraud}]{manchester2014simulation}
Manchester, W., Kozyra, J., Lepri, S., \& Lavraud, B. 2014{\natexlab{a}},
  Journal of Geophysical Research: Space Physics, 119, 5449,
  \dodoi{10.1002/2014JA019882}

\bibitem[{Manchester {et~al.}(2006)Manchester, Ridley, Gombosi, \&
  DeZeeuw}]{manchester2006modeling}
Manchester, W., Ridley, A., Gombosi, T., \& DeZeeuw, D. 2006, Advances in Space
  Research, 38, 253, \dodoi{10.1016/j.asr.2005.09.044}

\bibitem[{Manchester {et~al.}(2014{\natexlab{b}})Manchester, van~der Holst, \&
  Lavraud}]{manchester2014flux}
Manchester, W., van~der Holst, B., \& Lavraud, B. 2014{\natexlab{b}}, Plasma
  Physics and Controlled Fusion, 56, 064006,
  \dodoi{10.1088/0741-3335/56/6/064006}

\bibitem[{Manchester {et~al.}(2005)Manchester, Gombosi, De~Zeeuw, Sokolov,
  Roussev, Powell, K{\'o}ta, T{\'o}th, \& Zurbuchen}]{manchester2005coronal}
Manchester, W., Gombosi, T., De~Zeeuw, D., {et~al.} 2005, \apj, 622, 1225,
  \dodoi{10.1086/427768}

\bibitem[{Manchester {et~al.}(2004{\natexlab{b}})Manchester, Gombosi, Roussev,
  De~Zeeuw, Sokolov, Powell, T{\'o}th, \& Opher}]{manchester2004three}
Manchester, W.~B., Gombosi, T.~I., Roussev, I., {et~al.} 2004{\natexlab{b}},
  Journal of Geophysical Research: Space Physics, 109,
  \dodoi{10.1029/2002JA009672}

\bibitem[{Manchester {et~al.}(2004{\natexlab{c}})Manchester, Gombosi, Roussev,
  Ridley, De~Zeeuw, Sokolov, Powell, \& T{\'o}th}]{manchester2004modeling}
---. 2004{\natexlab{c}}, Journal of Geophysical Research: Space Physics, 109,
  \dodoi{10.1029/2003JA010150}

\bibitem[{Manchester {et~al.}(2008)Manchester, Vourlidas, T{\'o}th, Lugaz,
  Roussev, Sokolov, Gombosi, De~Zeeuw, \& Opher}]{manchester2008three}
Manchester, W.~B., Vourlidas, A., T{\'o}th, G., {et~al.} 2008, \apj, 684, 1448,
  \dodoi{10.1086/590231}

\bibitem[{Marhavilas {et~al.}(2015)Marhavilas, Malandraki, \&
  Anagnostopoulos}]{marhavilas2015survey}
Marhavilas, P., Malandraki, O., \& Anagnostopoulos, G. 2015, Planetary and
  Space Science, 117, 192, \dodoi{10.1016/j.pss.2015.06.010}

\bibitem[{Mason(2007)}]{mason2007he3}
Mason, G. 2007, \ssr, 130, 231, \dodoi{10.1007/s11214-007-9156-8}

\bibitem[{Mays {et~al.}(2015)Mays, Taktakishvili, Pulkkinen, MacNeice,
  Rast{\"a}tter, Odstrcil, Jian, Richardson, LaSota, Zheng,
  {et~al.}}]{mays2015ensemble}
Mays, M., Taktakishvili, A., Pulkkinen, A., {et~al.} 2015, \solphys, 290, 1775,
  \dodoi{10.1007/s11207-015-0692-1}

\bibitem[{Melrose \& Pope(1993)}]{melrose1993diffusive}
Melrose, D., \& Pope, M. 1993, \pasa, 10, 222,
  \dodoi{10.1017/S1323358000025716}

\bibitem[{Meng {et~al.}(2015)Meng, Van~der Holst, T{\'o}th, \&
  Gombosi}]{meng2015alfven}
Meng, X., Van~der Holst, B., T{\'o}th, G., \& Gombosi, T. 2015, Monthly Notices
  of the Royal Astronomical Society, 454, 3697, \dodoi{10.1093/mnras/stv2249}

\bibitem[{Menzel \& Purdom(1994)}]{menzel1994introducing}
Menzel, W.~P., \& Purdom, J.~F. 1994, Bulletin of the American Meteorological
  Society, 75, 757, \dodoi{10.1175/1520-0477(1994)075<0757:IGITFO>2.0.CO;2}

\bibitem[{Mewaldt {et~al.}(2005)Mewaldt, Cohen, Labrador, Leske, Mason, Desai,
  Looper, Mazur, Selesnick, \& Haggerty}]{mewaldt2005proton}
Mewaldt, R., Cohen, C., Labrador, A., {et~al.} 2005, Journal of Geophysical
  Research: Space Physics, 110, \dodoi{10.1029/2005JA011038}

\bibitem[{Mewaldt {et~al.}(2008)Mewaldt, Cohen, Cook, Cummings, Davis, Geier,
  Kecman, Klemic, Labrador, Leske, {et~al.}}]{mewaldt2008low}
Mewaldt, R.~A., Cohen, C., Cook, W., {et~al.} 2008, \ssr, 136, 285,
  \dodoi{10.1007/s11214-007-9288-x}

\bibitem[{Miki{\'c} \& Lee(2006)}]{mikic2006introduction}
Miki{\'c}, Z., \& Lee, M. 2006, \ssr, 123, 57,
  \dodoi{10.1007/s11214-006-9012-2}

\bibitem[{Millward {et~al.}(2013)Millward, Biesecker, Pizzo, \&
  De~Koning}]{millward2013operational}
Millward, G., Biesecker, D., Pizzo, V., \& De~Koning, C. 2013, Space Weather,
  11, 57, \dodoi{10.1002/swe.20024}

\bibitem[{Miroshnichenko(2018)}]{miroshnichenko2018retrospective}
Miroshnichenko, L.~I. 2018, Journal of Space Weather and Space Climate, 8, A52,
  \dodoi{10.1051/swsc/2018042}

\bibitem[{Miteva {et~al.}(2018)Miteva, Samwel, \&
  Costa-Duarte}]{miteva2018solar}
Miteva, R., Samwel, S., \& Costa-Duarte, M. 2018, Journal of Atmospheric and
  Solar-Terrestrial Physics, 180, 26, \dodoi{10.1016/j.jastp.2017.05.003}

\bibitem[{Miteva {et~al.}(2020)Miteva, Samwel, Zabunov, \&
  Dechev}]{miteva2020flux}
Miteva, R., Samwel, S.~W., Zabunov, S., \& Dechev, M. 2020, Bulgarian
  Astronomical Journal, 33, 99

\bibitem[{Morgado {et~al.}(2015)Morgado, Maia, Lanzerotti, Gon{\c{c}}alves, \&
  Patterson}]{morgado2015low}
Morgado, B., Maia, D. J.~F., Lanzerotti, L., Gon{\c{c}}alves, P., \& Patterson,
  J.~D. 2015, \aap, 577, A61, \dodoi{10.1051/0004-6361/201525960}

\bibitem[{Morgan {et~al.}(2006)Morgan, Habbal, \& Woo}]{morgan2006depiction}
Morgan, H., Habbal, S.~R., \& Woo, R. 2006, \solphys, 236, 263,
  \dodoi{10.1007/s11207-006-0113-6}

\bibitem[{Moussas {et~al.}(1992)Moussas, Quenby, Theodossiou-Ekaterinidi,
  Valdes-Galicia, Drillia, Roulias, \& Smith}]{moussas1992mean}
Moussas, X., Quenby, J., Theodossiou-Ekaterinidi, Z., {et~al.} 1992, \solphys,
  140, 161, \dodoi{10.1007/BF00148436}

\bibitem[{{M{\"u}ller-Mellin} {et~al.}(2008){M{\"u}ller-Mellin},
  {B{\"o}ttcher}, {Falenski}, {Rode}, {Duvet}, {Sanderson}, {Butler},
  {Johlander}, \& {Smit}}]{muller2008solar}
{M{\"u}ller-Mellin}, R., {B{\"o}ttcher}, S., {Falenski}, J., {et~al.} 2008,
  \ssr, 136, 363, \dodoi{10.1007/s11214-007-9204-4}

\bibitem[{Ng \& Reames(1994)}]{ng1994focused}
Ng, C., \& Reames, D. 1994, \apj, 424, 1032, \dodoi{10.1086/173954}

\bibitem[{Ng {et~al.}(2003)Ng, Reames, \& Tylka}]{ng2003modeling}
Ng, C., Reames, D., \& Tylka, A. 2003, \apj, 591, 461, \dodoi{10.1086/375293}

\bibitem[{Nikoli{\'c}(2019)}]{nikolic2019solutions}
Nikoli{\'c}, L. 2019, Space Weather, 17, 1293, \dodoi{10.1029/2019SW002205}

\bibitem[{Nitta {et~al.}(2006)Nitta, Reames, DeRosa, Liu, Yashiro, \&
  Gopalswamy}]{nitta2006solar}
Nitta, N.~V., Reames, D.~V., DeRosa, M.~L., {et~al.} 2006, \apj, 650, 438,
  \dodoi{10.1086/507442}

\bibitem[{{Northrop}(1963)}]{northrop1963adiabatic}
{Northrop}, T.~G. 1963, Reviews of Geophysics and Space Physics, 1, 283,
  \dodoi{10.1029/RG001i003p00283}

\bibitem[{Onsager {et~al.}(1996)Onsager, Grubb, Kunches, Matheson, Speich,
  Zwickl, \& Sauer}]{onsager1996operational}
Onsager, T., Grubb, R., Kunches, J., {et~al.} 1996, in GOES-8 and Beyond, Vol.
  2812, SPIE, 281--290, \dodoi{10.1117/12.254075}

\bibitem[{Ontiveros \& Vourlidas(2009)}]{ontiveros2009quantitative}
Ontiveros, V., \& Vourlidas, A. 2009, \apj, 693, 267,
  \dodoi{10.1088/0004-637X/693/1/267}

\bibitem[{Oran {et~al.}(2013)Oran, van~der Holst, Landi, Jin, Sokolov, \&
  Gombosi}]{oran2013global}
Oran, R., van~der Holst, B., Landi, E., {et~al.} 2013, \apj, 778, 176,
  \dodoi{10.1088/0004-637X/778/2/176}

\bibitem[{Paassilta {et~al.}(2018)Paassilta, Papaioannou, Dresing, Vainio,
  Valtonen, \& Heber}]{paassilta2018catalogue}
Paassilta, M., Papaioannou, A., Dresing, N., {et~al.} 2018, \solphys, 293, 1,
  \dodoi{10.1007/s11207-018-1284-7}

\bibitem[{Palmerio {et~al.}(2018)Palmerio, Kilpua, M{\"o}stl, Bothmer, James,
  Green, Isavnin, Davies, \& Harrison}]{palmerio2018coronal}
Palmerio, E., Kilpua, E.~K., M{\"o}stl, C., {et~al.} 2018, Space Weather, 16,
  442, \dodoi{10.1002/2017SW001767}

\bibitem[{Palmerio {et~al.}(2024)Palmerio, Luhmann, Mays, Caplan, Lario,
  Richardson, Whitman, Lee, S{\'a}nchez-Cano, Wijsen,
  {et~al.}}]{palmerio2024improved}
Palmerio, E., Luhmann, J.~G., Mays, M.~L., {et~al.} 2024, Journal of Space
  Weather and Space Climate, 14, 3,
  \dodoi{https://doi.org/10.1051/swsc/2024001}

\bibitem[{Pan {et~al.}(2022)Pan, Gou, \& Liu}]{pan2022sigmoid}
Pan, H., Gou, T., \& Liu, R. 2022, \apj, 937, 77,
  \dodoi{10.3847/1538-4357/ac8d64}

\bibitem[{Park {et~al.}(2015)Park, Innes, Bucik, Moon, \&
  Kahler}]{park2015study}
Park, J., Innes, D., Bucik, R., Moon, Y.-J., \& Kahler, S. 2015, \apj, 808, 3,
  \dodoi{10.1088/0004-637X/808/1/3}

\bibitem[{Park {et~al.}(2017)Park, Moon, \& Lee}]{park2017dependence}
Park, J., Moon, Y.-J., \& Lee, H. 2017, \apj, 844, 17,
  \dodoi{10.3847/1538-4357/aa794a}

\bibitem[{Parker(1958)}]{parker1958dynamics}
Parker, E.~N. 1958, \apj, 128, 664, \dodoi{10.1086/146579}

\bibitem[{Parker(1965)}]{parker1965passage}
---. 1965, Planetary and Space Science, 13, 9,
  \dodoi{10.1016/0032-0633(65)90131-5}

\bibitem[{{Pesnell} {et~al.}(2012){Pesnell}, {Thompson}, \&
  {Chamberlin}}]{pesnell2012solar}
{Pesnell}, W.~D., {Thompson}, B.~J., \& {Chamberlin}, P.~C. 2012, \solphys,
  275, 3, \dodoi{10.1007/s11207-011-9841-3}

\bibitem[{Petrie(2015)}]{petrie2015solar}
Petrie, G.~J. 2015, Living Reviews in Solar Physics, 12, 1,
  \dodoi{10.1007/lrsp-2015-5}

\bibitem[{Petrosian(2012)}]{petrosian2012stochastic}
Petrosian, V. 2012, \ssr, 173, 535, \dodoi{10.1007/s11214-012-9900-6}

\bibitem[{Pierrard \& Lazar(2010)}]{pierrard2010kappa}
Pierrard, V., \& Lazar, M. 2010, \solphys, 267, 153,
  \dodoi{10.1007/s11207-010-9640-2}

\bibitem[{Pit{\v{n}}a {et~al.}(2021)Pit{\v{n}}a, {\v{S}}afr{\'a}nkov{\'a},
  N{\v{e}}me{\v{c}}ek, {\v{D}}urovcov{\'a}, \& Kis}]{pitvna2021turbulence}
Pit{\v{n}}a, A., {\v{S}}afr{\'a}nkov{\'a}, J., N{\v{e}}me{\v{c}}ek, Z.,
  {\v{D}}urovcov{\'a}, T., \& Kis, A. 2021, Frontiers in Physics, 8, 626768,
  \dodoi{10.3389/fphy.2020.626768}

\bibitem[{Plotnikov {et~al.}(2017)Plotnikov, Rouillard, \&
  Share}]{plotnikov2017magnetic}
Plotnikov, I., Rouillard, A.~P., \& Share, G.~H. 2017, \aap, 608, A43,
  \dodoi{10.1051/0004-6361/201730804}

\bibitem[{Powell {et~al.}(1999)Powell, Roe, Linde, Gombosi, \&
  De~Zeeuw}]{powell1999solution}
Powell, K.~G., Roe, P.~L., Linde, T.~J., Gombosi, T.~I., \& De~Zeeuw, D.~L.
  1999, Journal of Computational Physics, 154, 284,
  \dodoi{10.1006/jcph.1999.6299}

\bibitem[{Prinsloo {et~al.}(2019)Prinsloo, Strauss, \&
  Le~Roux}]{prinsloo2019acceleration}
Prinsloo, P., Strauss, R., \& Le~Roux, J. 2019, \apj, 878, 144,
  \dodoi{10.3847/1538-4357/ab211b}

\bibitem[{Qin {et~al.}(2005)Qin, Zhang, Dwyer, Rassoul, \&
  Mason}]{qin2005model}
Qin, G., Zhang, M., Dwyer, J.~R., Rassoul, H.~K., \& Mason, G.~M. 2005, \apj,
  627, 562, \dodoi{10.1086/430136}

\bibitem[{Rankine(1870)}]{rankine1870xv}
Rankine, W. J.~M. 1870, Philosophical Transactions of the Royal Society of
  London, 277, \dodoi{10.1098/rstl.1870.0015}

\bibitem[{Reames(1999)}]{reames1999particle}
Reames, D.~V. 1999, {\ssr}, 90, 413, \dodoi{10.1023/A:1005105831781}

\bibitem[{Reames(2013)}]{reames2013two}
---. 2013, \ssr, 175, 53, \dodoi{10.1007/s11214-013-9958-9}

\bibitem[{Reames(2021)}]{reames2021solar}
---. 2021, {Solar Energetic Particles: A Modern Primer on Understanding
  Sources, Acceleration and Propagation} (Springer Nature),
  \dodoi{10.1007/978-3-030-66402-2}

\bibitem[{Reiss {et~al.}(2023)Reiss, Arge, Henney, Klimchuk, Linker, Muglach,
  Pevtsov, Pinto, \& Schonfeld}]{reiss2023progress}
Reiss, M.~A., Arge, C.~N., Henney, C.~J., {et~al.} 2023, Advances in Space
  Research, \dodoi{10.1016/j.asr.2023.08.039}

\bibitem[{Richardson {et~al.}(2018)Richardson, Mays, \&
  Thompson}]{richardson2018prediction}
Richardson, I., Mays, M., \& Thompson, B. 2018, Space Weather, 16, 1862,
  \dodoi{10.1029/2018SW002032}

\bibitem[{Richardson {et~al.}(2014)Richardson, Von~Rosenvinge, Cane, Christian,
  Cohen, Labrador, Leske, Mewaldt, Wiedenbeck, \& Stone}]{richardson201425}
Richardson, I., Von~Rosenvinge, T., Cane, H., {et~al.} 2014, Coronal
  Magnetometry, 437, \dodoi{10.1007/978-1-4939-2038-9_27}

\bibitem[{Robbrecht {et~al.}(2009)Robbrecht, Berghmans, \& Van~der
  Linden}]{robbrecht2009automated}
Robbrecht, E., Berghmans, D., \& Van~der Linden, R. 2009, \apj, 691, 1222,
  \dodoi{10.1088/0004-637X/691/2/1222}

\bibitem[{Roelof(1969)}]{roelof1969propagation}
Roelof, E. 1969, Lectures in High-Energy Astrophysics, 111.
\newblock \url{https://ntrs.nasa.gov/citations/19690020274}

\bibitem[{Rouillard {et~al.}(2011)Rouillard, Odstrc, Sheeley, Tylka, Vourlidas,
  Mason, Wu, Savani, Wood, Ng, {et~al.}}]{rouillard2011interpreting}
Rouillard, A., Odstrc, D., Sheeley, N., {et~al.} 2011, \apj, 735, 7,
  \dodoi{10.1088/0004-637X/735/1/7}

\bibitem[{Rouillard {et~al.}(2016)Rouillard, Plotnikov, Pinto, Tirole, Lavarra,
  Zucca, Vainio, Tylka, Vourlidas, De~Rosa, {et~al.}}]{rouillard2016deriving}
Rouillard, A.~P., Plotnikov, I., Pinto, R.~F., {et~al.} 2016, \apj, 833, 45,
  \dodoi{10.3847/1538-4357/833/1/45}

\bibitem[{Roussev {et~al.}(2003)Roussev, Forbes, Gombosi, Sokolov, DeZeeuw, \&
  Birn}]{roussev2003three}
Roussev, I.~I., Forbes, T.~G., Gombosi, T.~I., {et~al.} 2003, \apj, 588, L45,
  \dodoi{10.1086/375442}

\bibitem[{Sachdeva {et~al.}(2019)Sachdeva, van Der~Holst, Manchester, T{\'o}th,
  Chen, Lloveras, V{\'a}squez, Lamy, Wojak, Jackson,
  {et~al.}}]{sachdeva2019validation}
Sachdeva, N., van Der~Holst, B., Manchester, W.~B., {et~al.} 2019, \apj, 887,
  83, \dodoi{10.3847/1538-4357/ab4f5e}

\bibitem[{Sachdeva {et~al.}(2021)Sachdeva, T{\'o}th, Manchester, van~der Holst,
  Huang, Sokolov, Zhao, Al~Shidi, Chen, Gombosi,
  {et~al.}}]{sachdeva2021simulating}
Sachdeva, N., T{\'o}th, G., Manchester, W.~B., {et~al.} 2021, \apj, 923, 176,
  \dodoi{10.3847/1538-4357/ac307c}

\bibitem[{Sachdeva {et~al.}(2023)Sachdeva, Manchester, Sokolov, Huang, Pevtsov,
  Bertello, Pevtsov, T{\'o}th, van~der Holst, \& Henney}]{sachdeva2023solar}
Sachdeva, N., Manchester, W.~B., Sokolov, I., {et~al.} 2023, \apj, 952, 117,
  \dodoi{10.3847/1538-4357/acda87}

\bibitem[{Schatten {et~al.}(1969)Schatten, Wilcox, \& Ness}]{schatten1969model}
Schatten, K.~H., Wilcox, J.~M., \& Ness, N.~F. 1969, \solphys, 6, 442,
  \dodoi{10.1007/BF00146478}

\bibitem[{Sellers \& Hanser(1996)}]{sellers1996design}
Sellers, F.~B., \& Hanser, F.~A. 1996, in GOES-8 and Beyond, Vol. 2812, SPIE,
  353--364, \dodoi{10.1117/12.254083}

\bibitem[{Shalchi(2019)}]{shalchi2019field}
Shalchi, A. 2019, Advances in Space Research, 64, 2426,
  \dodoi{10.1016/j.asr.2019.03.005}

\bibitem[{Shalchi(2020)}]{shalchi2020perpendicular}
---. 2020, \ssr, 216, 23, \dodoi{10.1007/s11214-020-0644-4}

\bibitem[{Shalchi(2021)}]{shalchi2021field}
---. 2021, Physics of Plasmas, 28, \dodoi{10.1063/5.0061485}

\bibitem[{Shen {et~al.}(2011)Shen, Feng, Wu, Xiang, \& Song}]{shen2011three}
Shen, F., Feng, X., Wu, S., Xiang, C., \& Song, W. 2011, Journal of Geophysical
  Research: Space Physics, 116, \dodoi{10.1029/2010JA015809}

\bibitem[{Shi {et~al.}(2022)Shi, Manchester, Landi, van~der Holst, Szente,
  Chen, T{\'o}th, Bertello, \& Pevtsov}]{shi2022awsom}
Shi, T., Manchester, W., Landi, E., {et~al.} 2022, \apj, 928, 34,
  \dodoi{10.3847/1538-4357/ac52ab}

\bibitem[{Shiota \& Kataoka(2016)}]{shiota2016magnetohydrodynamic}
Shiota, D., \& Kataoka, R. 2016, Space Weather, 14, 56,
  \dodoi{10.1002/2015SW001308}

\bibitem[{Shoda {et~al.}(2021)Shoda, Chandran, \& Cranmer}]{shoda2021turbulent}
Shoda, M., Chandran, B.~D., \& Cranmer, S.~R. 2021, \apj, 915, 52,
  \dodoi{10.3847/1538-4357/abfdbc}

\bibitem[{Sime \& Hundhausen(1987)}]{sime1987coronal}
Sime, D., \& Hundhausen, A. 1987, Journal of Geophysical Research: Space
  Physics, 92, 1049, \dodoi{10.1029/JA092iA02p01049}

\bibitem[{Skilling(1971)}]{skilling1971cosmic}
Skilling, J. 1971, \apj, 170, 265, \dodoi{10.1086/151210}

\bibitem[{Smith \& Dryer(1990)}]{smith1990mhd}
Smith, Z., \& Dryer, M. 1990, \solphys, 129, 387, \dodoi{10.1007/BF00159049}

\bibitem[{Sokolov \& Gombosi(2023)}]{sokolov2023titov}
Sokolov, I.~V., \& Gombosi, T.~I. 2023, \apj, 955, 126,
  \dodoi{10.3847/1538-4357/aceef5}

\bibitem[{Sokolov {et~al.}(2006{\natexlab{a}})Sokolov, Powell, Gombosi, \&
  Roussev}]{sokolov2006tvd}
Sokolov, I.~V., Powell, K.~G., Gombosi, T.~I., \& Roussev, I.~I.
  2006{\natexlab{a}}, Journal of Computational Physics, 220, 1,
  \dodoi{10.1016/j.jcp.2006.07.021}

\bibitem[{Sokolov {et~al.}(2006{\natexlab{b}})Sokolov, Roussev, Fisk, Lee,
  Gombosi, \& Sakai}]{sokolov2006diffusive}
Sokolov, I.~V., Roussev, I., Fisk, L., {et~al.} 2006{\natexlab{b}}, \apj, 642,
  L81, \dodoi{10.1086/504406}

\bibitem[{Sokolov {et~al.}(2004)Sokolov, Roussev, Gombosi, Lee, K{\'o}ta,
  Forbes, Manchester, \& Sakai}]{sokolov2004new}
Sokolov, I.~V., Roussev, I., Gombosi, T., {et~al.} 2004, \apj, 616, L171,
  \dodoi{10.1086/426812}

\bibitem[{Sokolov {et~al.}(2009)Sokolov, Roussev, Skender, Gombosi, \&
  Usmanov}]{sokolov2009transport}
Sokolov, I.~V., Roussev, I.~I., Skender, M., Gombosi, T.~I., \& Usmanov, A.~V.
  2009, \apj, 696, 261, \dodoi{10.1088/0004-637X/696/1/261}

\bibitem[{Sokolov {et~al.}(2019)Sokolov, Sun, T\'oth, Huang, Kota, \&
  Gombosi}]{sokolov2019integral}
Sokolov, I.~V., Sun, H., T\'oth, G., {et~al.} 2019, arXiv preprint
  arXiv:1910.12636v1, \dodoi{10.48550/arXiv.1910.12636v1}

\bibitem[{Sokolov {et~al.}(2022)Sokolov, Zhao, \& Gombosi}]{sokolov2022stream}
Sokolov, I.~V., Zhao, L., \& Gombosi, T.~I. 2022, \apj, 926, 102,
  \dodoi{10.3847/1538-4357/ac400f}

\bibitem[{Sokolov {et~al.}(2013)Sokolov, van~der Holst, Oran, Downs, Roussev,
  Jin, Manchester, Evans, \& Gombosi}]{sokolov2013magnetohydrodynamic}
Sokolov, I.~V., van~der Holst, B., Oran, R., {et~al.} 2013, \apj, 764, 23,
  \dodoi{10.1088/0004-637X/764/1/23}

\bibitem[{Sokolov {et~al.}(2021)Sokolov, van~der Holst, Manchester, Ozturk,
  Szente, Taktakishvili, T{\'o}th, Jin, \& Gombosi}]{sokolov2021threaded}
Sokolov, I.~V., van~der Holst, B., Manchester, W.~B., {et~al.} 2021, \apj, 908,
  172, \dodoi{10.3847/1538-4357/abc000}

\bibitem[{Sokolov {et~al.}(2023)Sokolov, Sun, T{\'o}th, Huang, Tenishev, Zhao,
  Kota, Cohen, \& Gombosi}]{sokolov2023high}
Sokolov, I.~V., Sun, H., T{\'o}th, G., {et~al.} 2023, Journal of Computational
  Physics, 476, 111923, \dodoi{10.1016/j.jcp.2023.111923}

\bibitem[{Stone {et~al.}(1998)Stone, Frandsen, Mewaldt, Christian, Margolies,
  Ormes, \& Snow}]{stone1998advanced}
Stone, E.~C., Frandsen, A., Mewaldt, R., {et~al.} 1998, \ssr, 86, 1,
  \dodoi{10.1023/A:1005082526237}

\bibitem[{Strang(1968)}]{strang1968construction}
Strang, G. 1968, SIAM journal on numerical analysis, 5, 506,
  \dodoi{10.1137/0705041}

\bibitem[{Strauss \& Fichtner(2015)}]{strauss2015aspects}
Strauss, R., \& Fichtner, H. 2015, \apj, 801, 29,
  \dodoi{10.1088/0004-637X/801/1/29}

\bibitem[{Temmer {et~al.}(2009)Temmer, Preiss, \& Veronig}]{temmer2009cme}
Temmer, M., Preiss, S., \& Veronig, A. 2009, \solphys, 256, 183,
  \dodoi{10.1007/s11207-009-9336-7}

\bibitem[{Temmer {et~al.}(2023)Temmer, Scolini, Richardson, Heinemann, Paouris,
  Vourlidas, Bisi, Al-Haddad, Amerstorfer, Barnard, {et~al.}}]{temmer2023cme}
Temmer, M., Scolini, C., Richardson, I.~G., {et~al.} 2023, Advances in Space
  Research, \dodoi{10.1016/j.asr.2023.07.003}

\bibitem[{Tenishev {et~al.}(2022)Tenishev, Zhao, \&
  Sokolov}]{tenishev2022application}
Tenishev, V., Zhao, L., \& Sokolov, I. 2022, arXiv preprint arXiv:2209.09346,
  \dodoi{10.48550/arXiv.2209.09346}

\bibitem[{Thompson {et~al.}(2003)Thompson, Davila, Fisher, Orwig, Mentzell,
  Hetherington, Derro, Federline, Clark, Chen, {et~al.}}]{thompson2003cor1}
Thompson, W.~T., Davila, J.~M., Fisher, R.~R., {et~al.} 2003, in Innovative
  Telescopes and Instrumentation for Solar Astrophysics, Vol. 4853, SPIE,
  1--11, \dodoi{10.1117/12.460267}

\bibitem[{{Titov} \& {D{\'e}moulin}(1999)}]{titov1999basic}
{Titov}, V.~S., \& {D{\'e}moulin}, P. 1999, \aap, 351, 707

\bibitem[{Titov {et~al.}(2022)Titov, Downs, T{\"o}r{\"o}k, \&
  Linker}]{titov2022magnetogram}
Titov, V.~S., Downs, C., T{\"o}r{\"o}k, T., \& Linker, J.~A. 2022, \apj, 936,
  121, \dodoi{10.3847/1538-4357/ac874e}

\bibitem[{Titov {et~al.}(2014)Titov, T{\"o}r{\"o}k, Mikic, \&
  Linker}]{titov2014method}
Titov, V.~S., T{\"o}r{\"o}k, T., Mikic, Z., \& Linker, J.~A. 2014, \apj, 790,
  163, \dodoi{10.1088/0004-637X/790/2/163}

\bibitem[{T{\"o}r{\"o}k {et~al.}(2018)T{\"o}r{\"o}k, Downs, Linker, Lionello,
  Titov, Miki{\'c}, Riley, Caplan, \& Wijaya}]{torok2018sun}
T{\"o}r{\"o}k, T., Downs, C., Linker, J.~A., {et~al.} 2018, \apj, 856, 75,
  \dodoi{10.3847/1538-4357/aab36d}

\bibitem[{Torsti {et~al.}(1995)Torsti, Valtonen, Lumme, Peltonen, Eronen,
  Louhola, Riihonen, Schultz, Teittinen, Ahola, {et~al.}}]{torsti1995energetic}
Torsti, J., Valtonen, E., Lumme, M., {et~al.} 1995, \solphys, 162, 505,
  \dodoi{10.1007/BF00733438}

\bibitem[{T{\'o}th(2023)}]{toth2023total}
T{\'o}th, G. 2023, Journal of Computational Physics, 494, 112534,
  \dodoi{10.1016/j.jcp.2023.112534}

\bibitem[{T{\'o}th {et~al.}(2011)T{\'o}th, van~der Holst, \&
  Huang}]{toth2011obtaining}
T{\'o}th, G., van~der Holst, B., \& Huang, Z. 2011, \apj, 732, 102,
  \dodoi{10.1088/0004-637X/732/2/102}

\bibitem[{T{\'o}th {et~al.}(2005)T{\'o}th, Sokolov, Gombosi, Chesney, Clauer,
  De~Zeeuw, Hansen, Kane, Manchester, Oehmke, {et~al.}}]{toth2005space}
T{\'o}th, G., Sokolov, I.~V., Gombosi, T.~I., {et~al.} 2005, Journal of
  Geophysical Research: Space Physics, 110, \dodoi{10.1029/2005JA011126}

\bibitem[{T{\'o}th {et~al.}(2012)T{\'o}th, van~der Holst, Sokolov, De~Zeeuw,
  Gombosi, Fang, Manchester, Meng, Najib, Powell, {et~al.}}]{toth2012adaptive}
T{\'o}th, G., van~der Holst, B., Sokolov, I.~V., {et~al.} 2012, Journal of
  Computational Physics, 231, 870, \dodoi{10.1016/j.jcp.2011.02.006}

\bibitem[{Treumann(2009)}]{treumann2009fundamentals}
Treumann, R. 2009, \aapr, 17, 409, \dodoi{10.1007/s00159-009-0024-2}

\bibitem[{Tsurutani {et~al.}(2024)Tsurutani, Sen, Hajra, Lakhina, Horne, \&
  Hada}]{tsurutani2024review}
Tsurutani, B.~T., Sen, A., Hajra, R., {et~al.} 2024, Journal of Geophysical
  Research: Space Physics, 129, e2024JA032622, \dodoi{10.1029/2024JA032622}

\bibitem[{Tylka {et~al.}(2005)Tylka, Cohen, Dietrich, Lee, Maclennan, Mewaldt,
  Ng, \& Reames}]{tylka2005shock}
Tylka, A., Cohen, C., Dietrich, W., {et~al.} 2005, \apj, 625, 474,
  \dodoi{10.1086/429384}

\bibitem[{Vainio(2003)}]{vainio2003generation}
Vainio, R. 2003, \aap, 406, 735, \dodoi{10.1051/0004-6361:20030822}

\bibitem[{Valtonen {et~al.}(1997)Valtonen, Peltonen, Peltonen, Eronen, Hoisko,
  Louhola, Lumme, Nieminen, Riihonen, Teittinen,
  {et~al.}}]{valtonen1997energetic}
Valtonen, E., Peltonen, J., Peltonen, P., {et~al.} 1997, Nuclear Instruments
  and Methods in Physics Research Section A: Accelerators, Spectrometers,
  Detectors and Associated Equipment, 391, 249,
  \dodoi{10.1016/S0168-9002(97)00469-5}

\bibitem[{van~den Berg {et~al.}(2020)van~den Berg, Strauss, \&
  Effenberger}]{van2020primer}
van~den Berg, J., Strauss, D.~T., \& Effenberger, F. 2020, \ssr, 216, 146,
  \dodoi{10.1007/s11214-020-00771-x}

\bibitem[{van~der Holst {et~al.}(2010)van~der Holst, Manchester, Frazin,
  V{\'a}squez, T{\'o}th, \& Gombosi}]{van2010data}
van~der Holst, B., Manchester, W., Frazin, R., {et~al.} 2010, \apj, 725, 1373,
  \dodoi{10.1088/0004-637X/725/1/1373}

\bibitem[{van~der Holst {et~al.}(2019)van~der Holst, Manchester, Klein, \&
  Kasper}]{van2019predictions}
van~der Holst, B., Manchester, W., Klein, K., \& Kasper, J. 2019, \apjl, 872,
  L18, \dodoi{10.3847/2041-8213/ab04a5}

\bibitem[{van~der Holst {et~al.}(2009)van~der Holst, Manchester, Sokolov,
  T{\'o}th, Gombosi, DeZeeuw, \& Cohen}]{van2009breakout}
van~der Holst, B., Manchester, W., Sokolov, I., {et~al.} 2009, \apj, 693, 1178,
  \dodoi{10.1088/0004-637X/693/2/1178}

\bibitem[{van~der Holst {et~al.}(2014)van~der Holst, Sokolov, Meng, Jin,
  Manchester, T{\'o}th, \& Gombosi}]{van2014alfven}
van~der Holst, B., Sokolov, I.~V., Meng, X., {et~al.} 2014, \apj, 782, 81,
  \dodoi{10.1088/0004-637X/782/2/81}

\bibitem[{van~der Holst {et~al.}(2011)van~der Holst, T{\'o}th, Sokolov, Powell,
  Holloway, Myra, Stout, Adams, Morel, Karni, {et~al.}}]{van2011crash}
van~der Holst, B., T{\'o}th, G., Sokolov, I.~V., {et~al.} 2011, \apjs, 194, 23,
  \dodoi{10.1088/0067-0049/194/2/23}

\bibitem[{van~der Holst {et~al.}(2022)van~der Holst, Huang, Sachdeva, Kasper,
  Manchester, Borovikov, Chandran, Case, Korreck, Larson,
  {et~al.}}]{van2022improving}
van~der Holst, B., Huang, J., Sachdeva, N., {et~al.} 2022, \apj, 925, 146,
  \dodoi{10.3847/1538-4357/ac3d34}

\bibitem[{Vemareddy \& Mishra(2015)}]{vemareddy2015full}
Vemareddy, P., \& Mishra, W. 2015, \apj, 814, 59,
  \dodoi{10.1088/0004-637X/814/1/59}

\bibitem[{Verkhoglyadova {et~al.}(2015)Verkhoglyadova, Zank, \&
  Li}]{verkhoglyadova2015theoretical}
Verkhoglyadova, O.~P., Zank, G.~P., \& Li, G. 2015, Physics Reports, 557, 1,
  \dodoi{10.1016/j.physrep.2014.10.004}

\bibitem[{Von~Rosenvinge {et~al.}(2008)Von~Rosenvinge, Reames, Baker, Hawk,
  Nolan, Ryan, Shuman, Wortman, Mewaldt, Cummings, {et~al.}}]{von2008high}
Von~Rosenvinge, T., Reames, D., Baker, R., {et~al.} 2008, The STEREO Mission,
  391, \dodoi{10.1007/978-0-387-09649-0_14}

\bibitem[{Vourlidas {et~al.}(2003)Vourlidas, Wu, Wang, Subramanian, \&
  Howard}]{vourlidas2003direct}
Vourlidas, A., Wu, S., Wang, A., Subramanian, P., \& Howard, R. 2003, \apj,
  598, 1392, \dodoi{10.1086/379098}

\bibitem[{Wang \& Guo(2024)}]{wang2024statistical}
Wang, Y., \& Guo, J. 2024, \aap, 691, A54, \dodoi{10.1051/0004-6361/202450046}

\bibitem[{Wang \& Qin(2015)}]{wang2015simulations}
Wang, Y., \& Qin, G. 2015, \apj, 806, 252, \dodoi{10.1088/0004-637X/806/2/252}

\bibitem[{Webb \& Howard(2012)}]{webb2012coronal}
Webb, D.~F., \& Howard, T.~A. 2012, Living Reviews in Solar Physics, 9, 1,
  \dodoi{10.12942/lrsp-2012-3}

\bibitem[{Whang {et~al.}(1996)Whang, Zhou, Lepping, \&
  Ogilvie}]{whang1996interplanetary}
Whang, Y., Zhou, J., Lepping, R., \& Ogilvie, K. 1996, \grl, 23, 1239,
  \dodoi{10.1029/96GL01358}

\bibitem[{Whitman {et~al.}(2023)Whitman, Egeland, Richardson, Allison, Quinn,
  Barzilla, Kitiashvili, Sadykov, Bain, Dierckxsens,
  {et~al.}}]{whitman2023review}
Whitman, K., Egeland, R., Richardson, I.~G., {et~al.} 2023, Advances in Space
  Research, 72, 5161, \dodoi{10.1016/j.asr.2022.08.006}

\bibitem[{Wijsen {et~al.}(2019)Wijsen, Aran, Pomoell, \&
  Poedts}]{wijsen2019modelling}
Wijsen, N., Aran, A., Pomoell, J., \& Poedts, S. 2019, \aap, 622, A28,
  \dodoi{10.1051/0004-6361/201833958}

\bibitem[{Wijsen {et~al.}(2023)Wijsen, Li, Ding, Lario, Poedts, Filwett, Allen,
  \& Dayeh}]{wijsen2023seed}
Wijsen, N., Li, G., Ding, Z., {et~al.} 2023, Journal of Geophysical Research:
  Space Physics, 128, e2022JA031203, \dodoi{10.1029/2022JA031203}

\bibitem[{Wraback {et~al.}(2024)Wraback, Hoffmann, Manchester, Sokolov, van~der
  Holst, \& Carpenter}]{wraback2024simulating}
Wraback, E., Hoffmann, A., Manchester, W., {et~al.} 2024, \apj, 962, 182,
  \dodoi{10.3847/1538-4357/ad21fd}

\bibitem[{Wuelser {et~al.}(2004)Wuelser, Lemen, Tarbell, Wolfson, Cannon,
  Carpenter, Duncan, Gradwohl, Meyer, Moore, {et~al.}}]{wuelser2004euvi}
Wuelser, J.-P., Lemen, J.~R., Tarbell, T.~D., {et~al.} 2004, in Telescopes and
  instrumentation for solar astrophysics, Vol. 5171, SPIE, 111--122,
  \dodoi{10.1117/12.506877}

\bibitem[{Xie {et~al.}(2019)Xie, St.~Cyr, M{\"a}kel{\"a}, \&
  Gopalswamy}]{xie2019statistical}
Xie, H., St.~Cyr, O., M{\"a}kel{\"a}, P., \& Gopalswamy, N. 2019, Journal of
  Geophysical Research: Space Physics, 124, 6384, \dodoi{10.1029/2019JA026832}

\bibitem[{Yashiro {et~al.}(2004)Yashiro, Gopalswamy, Michalek, St.~Cyr,
  Plunkett, Rich, \& Howard}]{yashiro2004catalog}
Yashiro, S., Gopalswamy, N., Michalek, G., {et~al.} 2004, Journal of
  Geophysical Research: Space Physics, 109, \dodoi{10.1029/2003JA010282}

\bibitem[{Young {et~al.}(2021)Young, Schwadron, Gorby, Linker, Caplan, Downs,
  T{\"o}r{\"o}k, Riley, Lionello, Titov, {et~al.}}]{young2021energetic}
Young, M.~A., Schwadron, N.~A., Gorby, M., {et~al.} 2021, \apj, 909, 160,
  \dodoi{10.3847/1538-4357/abdf5f}

\bibitem[{Yu {et~al.}(2022)Yu, Kong, Guo, Liu, Jiang, Chen, \&
  Giacalone}]{yu2022double}
Yu, F., Kong, X., Guo, F., {et~al.} 2022, \apjl, 925, L13,
  \dodoi{10.3847/2041-8213/ac4cb3}

\bibitem[{Zakharov {et~al.}(2012)Zakharov, L'vov, \&
  Falkovich}]{zakharov2012kolmogorov}
Zakharov, V.~E., L'vov, V.~S., \& Falkovich, G. 2012, Kolmogorov Spectra of
  Turbulence I: Wave Turbulence (Springer Science \& Business Media),
  \dodoi{10.1007/978-3-642-50052-7}

\bibitem[{Zank {et~al.}(2014)Zank, Le~Roux, Webb, Dosch, \&
  Khabarova}]{zank2014particle}
Zank, G., Le~Roux, J., Webb, G., Dosch, A., \& Khabarova, O. 2014, \apj, 797,
  28, \dodoi{10.1088/0004-637X/797/1/28}

\bibitem[{Zank {et~al.}(2007)Zank, Li, \& Verkhoglyadova}]{zank2007particle}
Zank, G., Li, G., \& Verkhoglyadova, O. 2007, \ssr, 130, 255,
  \dodoi{10.1007/s11214-007-9214-2}

\bibitem[{Zank {et~al.}(2000)Zank, Rice, \& Wu}]{zank2000particle}
Zank, G., Rice, W., \& Wu, C. 2000, Journal of Geophysical Research: Space
  Physics, 105, 25079, \dodoi{10.1029/1999JA000455}

\bibitem[{Zhang {et~al.}(2023)Zhang, Cheng, Zhang, Riley, Kwon, Lario,
  Balmaceda, \& Pogorelov}]{zhang2023data}
Zhang, M., Cheng, L., Zhang, J., {et~al.} 2023, \apjs, 266, 35,
  \dodoi{10.3847/1538-4365/accb8e}

\bibitem[{Zhang {et~al.}(2009)Zhang, Qin, \& Rassoul}]{zhang2009propagation}
Zhang, M., Qin, G., \& Rassoul, H. 2009, \apj, 692, 109,
  \dodoi{10.1088/0004-637X/692/1/109}

\bibitem[{Zhang \& Zhao(2017)}]{zhang2017precipitation}
Zhang, M., \& Zhao, L. 2017, \apj, 846, 107, \dodoi{10.3847/1538-4357/aa86a8}

\bibitem[{Zhao {et~al.}(2019)Zhao, Li, Zhang, Wang, Moradi, \&
  Effenberger}]{zhao2019stat}
Zhao, L., Li, G., Zhang, M., {et~al.} 2019, \apj, 878, 107,
  \dodoi{10.3847/1538-4357/ab2041}

\bibitem[{Zhao {et~al.}(2016)Zhao, Zhang, \& Rassoul}]{zhao2016double}
Zhao, L., Zhang, M., \& Rassoul, H.~K. 2016, \apj, 821, 62,
  \dodoi{10.3847/0004-637X/821/1/62}

\bibitem[{Zhao {et~al.}(2017)Zhao, Zhang, \& Rassoul}]{zhao2017effects}
---. 2017, \apj, 836, 31, \dodoi{10.3847/1538-4357/836/1/31}

\bibitem[{Zhao {et~al.}(2024)Zhao, Sokolov, Gombosi, Lario, Whitman, Huang,
  T{\'o}th, Manchester, van~der Holst, Sachdeva, {et~al.}}]{zhao2024solar}
Zhao, L., Sokolov, I., Gombosi, T., {et~al.} 2024, Space Weather, 22,
  e2023SW003729, \dodoi{10.1029/2023SW003729}

\bibitem[{Zheng {et~al.}(2024)Zheng, Jun, Tu, Shprits, Kim, Matthi{\"a}, Meier,
  Tobiska, Miyoshi, Jordanova, {et~al.}}]{zheng2024overview}
Zheng, Y., Jun, I., Tu, W., {et~al.} 2024, Advances in Space Research,
  \dodoi{10.1016/j.asr.2024.05.017}

\bibitem[{Zhuang {et~al.}(2021)Zhuang, Lugaz, Gou, \&
  Ding}]{zhuang2021successive}
Zhuang, B., Lugaz, N., Gou, T., \& Ding, L. 2021, \apj, 921, 6,
  \dodoi{10.3847/1538-4357/ac17e9}

\bibitem[{Zucca {et~al.}(2018)Zucca, Morosan, Rouillard, Fallows, Gallagher,
  Magdalenic, Klein, Mann, Vocks, Carley, {et~al.}}]{zucca2018shock}
Zucca, P., Morosan, D.~E., Rouillard, A., {et~al.} 2018, \aap, 615, A89,
  \dodoi{10.1051/0004-6361/201732308}

\end{thebibliography}
\bibliographystyle{aasjournal}

\end{document}